\documentclass[iicol]{sn-jnl}
\usepackage{natbib}
\usepackage{geometry}                
\usepackage{amsmath}
\usepackage{amsfonts}
\usepackage{aas_macros}
\usepackage{amsmath}
\usepackage{amssymb}
\usepackage{bm}
\usepackage{dcolumn}
\usepackage[latin1]{inputenc}
\usepackage{rotating}
\usepackage{xspace} 
\usepackage{fancyhdr}
\usepackage{framed}
\usepackage{graphics}
\usepackage{xfrac}
\usepackage{wrapfig}
\usepackage{floatflt}
\usepackage[arrowdel]{physics}
\usepackage{changepage}
\usepackage{hyperref}
\usepackage{ulem}

\def\H{{\rm H}}
\def\HI{{\rm H\,I}}
\def\HII{{\rm H\,II}}
\def\He{{\rm He}}
\def\HeI{{\rm He\,I}}
\def\HeII{{\rm He\,II}}
\def\HeIII{{\rm He\,III}}
\def\nHI{{\rm HI}}
\def\nHII{{\rm HII}}
\def\nHeI{{\rm HeI}}
\def\nHeII{{\rm HeII}}
\def\nHeIII{{\rm HeIII}}

\def\fesc{f_{\rm esc}}
\def\dim#1{{\,\rm #1}}

\def\Lya{Lyman-$\alpha$}
\def\cc{\,{\rm cm^{-3}}}
\def\spose#1{\hbox to 0pt{#1\hss}}
\def\lta{\mathrel{\spose{\lower 3pt\hbox{$\mathchar"218$}}
     \raise 2.0pt\hbox{$\mathchar"13C$}}}
\def\gta{\mathrel{\spose{\lower 3pt\hbox{$\mathchar"218$}}
     \raise 2.0pt\hbox{$\mathchar"13E$}}}

\def\vec#1{\mathbf{#1}}

\begin{document}

\title[]{Modeling Cosmic Reionization}

\author*[1,2,3]{\fnm{Nickolay} Y.\ \sur{Gnedin}}\email{gnedin@fnal.gov}

\author[4]{\fnm{Piero} \sur{Madau}}\email{pmadau@ucsc.edu}

\affil*[1]{\orgdiv{Theoretical Physics Department}, \orgname{Fermi National Accelerator Laboratory}, \orgaddress{\city{Batavia}, \state{IL}, \postcode{60510}, \country{USA}}}

\affil[2]{\orgdiv{Kavli Institute for Cosmological Physics}, \orgname{The University of Chicago}, \orgaddress{\city{Chicago}, \postcode{60637}, \state{IL}, \country{USA}}}

\affil[3]{\orgdiv{Department of Astronomy \& Astrophysics}, \orgname{The University of Chicago}, \orgaddress{\city{Chicago}, \postcode{60637}, \state{IL}, \country{USA}}}

\affil[4]{\orgdiv{Department of Astronomy \& Astrophysics}, \orgname{University of California}, \orgaddress{\city{Santa Cruz}, \postcode{95064}, \state{CA}, \country{USA}}}

\abstract{
The transformation of cold neutral intergalactic hydrogen into a highly ionized warm plasma marks the end of the cosmic dark ages and the beginning of the age of galaxies. The details of this process reflect the nature of the early sources of radiation and heat, the statistical characteristics of the large-scale structure of the Universe, the thermodynamics and chemistry of  cosmic baryons, and the histories of star formation and black hole accretion. A number of massive data sets from new ground- and space-based instruments and facilities over the next decade are poised to revolutionize our understanding of primeval galaxies, the reionization photon budget, the physics of the intergalactic medium (IGM), and the fine-grained properties of hydrogen gas in the ``cosmic web". In this review we survey the physics and key aspects of reionization-era modeling and describe the diverse range of computational techniques and tools currently available in this field.
}


\maketitle
\pagestyle{plain}

\newpage

\setcounter{tocdepth}{3}
\tableofcontents


\section{Introduction}
\label{sec:intro}
About 380,000 years after the Big Bang the temperature of the Universe dropped for the first time below 
three thousand degrees Kelvin, the primordial plasma recombined to form neutral hydrogen atoms, matter decoupled from the photon field, and the radiation presently observed as the cosmic microwave background (CMB) was released. The Universe then entered a period devoid of any luminous sources cosmologists call the cosmic ``dark ages", quietly waiting for gravity to draw hydrogen gas into collapsing dark matter halos. The formation of the very first stars following gas condensation and cooling, some 200 million years after the Big Bang, marks the end of the dark ages and the beginning of the age of galaxies. These young stellar systems emitted energetic UV radiation capable of ionizing local ``bubbles" of hydrogen gas. As more and more stars 
lit up and the mean density of the diffuse intergalactic medium (IGM) -- the main reservoir of baryonic material at all epochs -- decreased, these \HII\ bubbles overlapped and progressively larger volumes became ionized and heated. Throughout this epoch of reionization and reheating, the all-pervading IGM became clumpy under the influence of gravity, and acted as both a source of fuel for the galaxy formation process, and as a sink for absorbing the products of star formation (metals, wind mechanical energy, and radiation).

The first theoretical calculations of inhomogeneous hydrogen reionization \citep{arons70,arons72} assumed a uniform IGM and estimated that quasar ionization zones would overlap by redshift $z\sim 3$. These timely studies made immediately clear that: 1) initially isolated cosmological \HII\ regions would grow in size around the first sources; 2) the thickness of ionization fronts -- of order the mean free path of Lyman-continuum (LyC) radiation in the neutral IGM, $\lambda_{\rm mfp}\simeq 0.27\,(1+z)^{-3}$ proper Mpc -- would be much smaller than the size of these \HII\ regions; 3) \HII\ regions would accumulate faster that they can recombine as the UV emissivity of the Universe rose; and 4) at the epoch when \HII\ overlap occurred, the photon mean free path and the intensity of the ionizing background would increase abruptly. 
Many of these salient features are seen in current state-of-the-art numerical simulations of this process. Today, however, we believe that hydrogen reionization ended some 0.9 Gyr after the Big Bang \citep{fan06} and was likely driven by star-forming galaxies rather than quasars \citep[e.g.,][]{shapiro87,madau99,gnedin00,Robertson2015}. We know that the IGM is clumpy and that the optically thick 
``Lyman-limit systems" (LLSs), high density regions that trace non-linear and collapsed structures, cause the mean free path to LyC radiation to remain smaller than the horizon even after overlap \citep[e.g.,][]{gnedin06,furlanetto09,worseck14}. And we argue about the fraction of ionizing photons that escape from individual galaxies into the IGM and the role played by energetic X-ray radiation in altering the thermal and ionization balance of the early IGM \citep[e.g.,][]{madau04,oh01,venkatesan01,madau_fragos17}. 

Modern observations of resonant absorption in the spectra of distant quasars show that reionization was nearly completed at $z\sim 6$ \citep{fan06,mcgreer15,becker15}. A very early onset of reionization is disfavoured by the final full-mission {\it Planck}  CMB anisotropy analysis, which prefers a late and fast transition from a neutral to an ionized Universe, and by a variety of astrophysical probes \citep{schroeder13, banados18, ouchi10, schenker14,  onorbe_ps17, villasenor22}. Despite a considerable community effort, however, 
many key aspects of this process, such as the very nature of the first sources of UV radiation, how they interacted with their 
environment, the back-reaction of reionization on infant galaxies and its imprint on their luminosity function and gas content,
the impact of patchy reionization on observables, and the thermodynamics of primordial baryonic gas, all remain highly uncertain. Even a complete knowledge of the sources and sinks of UV and X-ray light in the early Universe,
which astrophysicists by no means possess, would not automatically translate into a detailed understanding of the reionization process 
as a whole, as this ultimately requires extremely challenging cosmological numerical simulations that self-consistently couple all the relevant physical processes -- dark matter dynamics, gas dynamics, self-gravity, star formation/feedback, radiative transfer, non-equilibrium 
ionizations and recombinations, chemical enrichment, heating and cooling -- over a huge range of scales \citep[e.g.,][]{gnedin14,ocvirk16,pawlik17,finlator18,kannan22}.
Over the next decade, the large wavelength coverage, unique sensitivity, and manifold spectroscopic and imaging capabilities of the {James Webb Space Telescope} ({\it JWST}) will allow the discovery of star-forming galaxies out to redshift 15, the tracing of the evolution of the cosmic star-formation rate density  to the earliest times, the determination of the cross-correlation of galaxies with the opacity of the early IGM, the study of the highly ionized near-zones around $z>6$ quasars, and will provide tight constraints on both the reionization history and the  contribution of different sources to the ionizing photon budget of the Universe. These observations will be complemented by a number of massive datasets from several existing facilities, such as the Dark Energy Spectroscopic Instrument (DESI), the Subaru Hyper Suprime-Cam (HSC), the Extremely Large Telescope (ELT), the Atacama Large Millimeter Array (ALMA), and the Hydrogen Epoch of Reionization Array (HERA). Even more advanced observational capabilities will become available in the very near future, such as 30-meter class telescopes, the Euclid satellite, the Roman Space Telescope, intensity mapping instruments such as SPHEREx, CCat-p, TIME, CONCERTO, or COMAP, and, in a more distant future, the ultimate 21 cm machine, the Square Kilometre Array (SKA). It is generally believed that the coming years will mark the beginning of a golden age of early Universe and epoch of reionization science.

\begin{figure*}
\vspace{+0.2cm}
\centering
\includegraphics[width=\hsize]{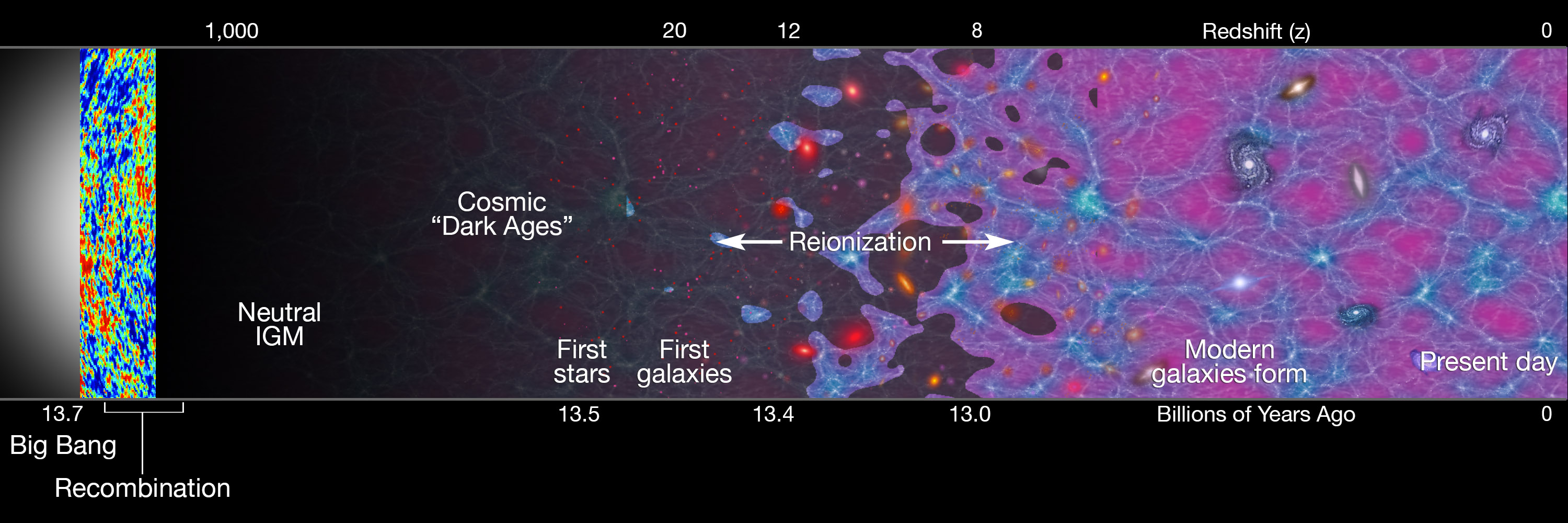}
\vspace{+0.0cm}
\caption{\footnotesize Cosmic Reionization Fundamentals. After recombination at $z\sim 1100$, hydrogen in the IGM remained neutral until the first stars and galaxies formed at $z\sim 15$. These primordial systems released energetic UV photons capable of ionizing 
local bubbles of hydrogen gas. As the abundance of these early galaxies increased, \HII\ bubbles overlapped and progressively larger volumes became ionized. The hydrogen  reionization process completed at $z\sim 6$, approximately 1 Gyr after the Big Bang.  Observations suggest that the double reionization of helium occurred at a 
later time \citep[e.g.,][]{Kriss2001,Shull2004,Zheng2004,Shull2010,worseck16}, when hard UV-emitting quasars and other active galactic nuclei (AGNs) became sufficiently abundant.
(From \citet{robertson10}; used by permission.)
}
\label{fig:reionization}
\end{figure*}

It seems then timely to survey in a ``living" document the key aspects of reionization-era modeling and the diverse range of computational techniques and tools currently available in this field, with the aim of providing a comprehensive road map for guiding and interpreting forthcoming observations of the early Universe. 
A note on terminology. In the remainder of this review, we call {\it analytical models of reionization} those based on a set of equations for some statistical description of the  process, such as the averaged neutral fraction, even in the absence of a closed form mathematical solution. By contrast, {\it numerical methods of reionization} are those that provide an actual realization of the hydrogen density field in a 3-dimensional volume, starting from initial Gaussian density fluctuations. A full hydrodynamical simulation is thus a numerical model, but so is the ``semi-numerical" realization of the \HII\ field provided by, say, the 21cmFast code \citep{Mesinger2011}. 

In the spirit of a ``living" article that is kept up-to-date, we have tried to structure this review in a way that maximizes the visibility of the latest findings for the reader, critically evaluate existing work, place it in a meaningful context, and suggest areas where more research and new results are needed.
Given the limited space available, it is impossible to provide here a thorough survey of the huge community effort behind reionization and IGM studies without leaving out significant contributions or whole subfields. For additional material, we therefore refer the reader to a number of recent complementary reviews on these topics  \citep{meiksin09,choudhury09,trac11,zaroubi13,lidz16,mcquinn16,dayal19,Robertson2021}. Unless otherwise stated, all results presented below assume a flat cosmology with parameters $(\Omega_M, \Omega_\Lambda, \Omega_b) = (0.3, 0.7, 0.045)$, a Hubble constant of $H_0=70\,$ km s$^{-1}$ Mpc$^{-1}$, and a primordial baryonic gas of total hydrogen and helium mass fractions $X=0.75$ and $Y=0.25$. We shall also use spectroscopic notation to indicate the ionization states -- \HI\ and \HII\ for neutral and ionized hydrogen, and \HeI, \HeII, and \HeIII\ for neutral,
singly ionized, and doubly ionized helium, respectively.

\section{The Physics of Intergalactic Gas}

The IGM is the storehouse of the bulk of cosmic baryons at high redshift. It forms a filamentary network that hosts the formation of galaxies and gives origin to a ``forest" of Lyman-$\alpha$ absorption lines, the signature of a highly fluctuating medium with temperatures characteristic of photoionized gas \citep[see, e.g.,][and references therein]{meiksin09,mcquinn16}. 
It is dark matter that provides the backbone of large-scale structure in the Universe, a web-like pattern present in embryonic form in the overdensity motif of the initial fluctuation field and only sharpened by nonlinear gravitational dynamics \citep{Bond1991}. The Lyman-$\alpha$ forest traces this underlying ``cosmic web" on scales and at redshifts that cannot be probed by any other observable. Because of its long cooling time, diffuse gas at $z\sim 2-5$ retains some memory of when and how it 
was reheated and reionized at $z\gtrsim 6$ \citep{miralda94}. The physics that governs the properties of the IGM throughout these epochs remains similar, as the evolving cosmic UV emissivity and the transfer of that
radiation through a medium made clumpy by gravity determine both the details of the reionization process and the thermodynamics of the forest.

\subsection{Photoionization of hydrogen and helium}

Photons with frequencies $\nu>\nu_\nHI$  will ionize intergalactic hydrogen at a rate per neutral hydrogen atom of
\begin{equation}
\Gamma_\nHI=4\pi\int_{\nu_\nHI}^\infty \frac{J_\nu}{h\nu}\,\sigma_\HI(\nu) d\nu, 
\label{eq:Gamma}
\end{equation}
where $\sigma_\HI(\nu)$ is the bound-free absorption cross-section from the ground state and $h\nu_\nHI=13.6$ eV is the ionization potential of the hydrogen atom. Here,  $J_\nu$ is the  angle-averaged specific intensity, 
 \begin{equation}
J_\nu(t,\vec{x})= \frac{1}{4\pi}\int d\Omega\, I_\nu(t,\vec{x},\vec{n}),
\label{eq:Jnu}
 \end{equation}
where $I_\nu(t,\vec{x},\vec{n})$ is the intensity of the radiation field at cosmic time $t$ and position $\vec{x}$, measured along the direction $\vec{n}$. Protons will radiatively capture free electrons at the rate per proton $n_e \alpha_\nHII(T)$, where 
\begin{equation}
\alpha^A_\nHII(T)\simeq 4.2\times 10^{-13}\,T_4^{-0.7}\,
{\rm cm^{3}\,s^{-1}}
\end{equation}
is the Case-A temperature-dependent radiative recombination coefficient onto all atomic levels of hydrogen, $T_4$ is the gas temperature in units of $10^4\,{\rm K}$, and 
$n_e$ is the proper electron density. 
There exist more accurate fits for $\alpha^A_\nHII$,
but this approximate form typically suffices for analytical calculations. Case-A recombination is the correct choice for modeling reionization, as H-ionizing photons emitted in radiative transitions from the continuum directly to the $1s$ level are typically redshifted below threshold before being absorbed by the IGM \citep{kaurov14}. In terms of the ionization fractions $x_\nHI=n_\nHI/n_\H$ and $x_\nHII=n_\nHII/n_\H$, the rate equation for low density photoionized gas can be then written as
\begin{equation}
\frac{dx_\nHI}{dt} =  -x_\nHI \Gamma_\nHI + n_ex_\nHII \alpha^A_\nHII,
\label{eq:hion}
\end{equation}
and similarly for the ionization states of helium $x_\nHeI=n_\nHeI/n_\He$, $x_\nHeII=n_\nHeII/n_\He$, and $x_\nHeIII=n_\nHeIII/n_\He$:
\begin{subequations}%
\begin{eqnarray}%
\frac{dx_\nHeI}{dt} & = & -x_\nHeI\Gamma_\nHeI+n_ex_\nHeII (\alpha^A_\nHeII\\
\nonumber & + & \alpha^D_\nHeII),\\
\frac{dx_\nHeII}{dt} & = & -x_\nHeII \Gamma_\nHeII + n_e x_\nHeIII \alpha^A_\nHeIII \\
\nonumber
& - & \frac{dx_\nHeI}{dt}, 
\end{eqnarray}%
\label{eq:heion}%
\end{subequations}%
where $d/dt$ is the Lagrangian derivative and the recombination coefficient of helium includes a \emph{dielectronic capture} term $\alpha^D_\nHeII$ from \HeII\ into \HeI\ \citep{aldrovandi73}. The rate equations above neglect collisional ionizations by thermal electrons and must be supplemented by three closure conditions for the conservation of charge and of the total abundances of hydrogen and helium. 

\subsection{Heating and cooling}

The temperature of an unshocked gas element of overdensity 
$\Delta \equiv \rho_b/ \langle \rho_b \rangle$, where  $\rho_b$ is the gas mass density, exposed to an ionizing UV photon field, obeys the equation \citep[e.g.][]{hui97}
\begin{equation}
\begin{split}
\frac{dT}{dt}= & -2HT + \frac{2T}{3\Delta}\frac{d\Delta}{dt} + \frac{T}{\mu}\frac{d\mu}{dt} \\ 
& + \frac{2\mu}{3k_B n}({\cal H}-\Lambda),
\label{eq:dT}
\end{split}
\end{equation}
where $H$ is the expansion rate, $\mu$ is the mean molecular weight, 
$n$ is the proper number density of  all species including electrons, $\Lambda$ is the total cooling rate per unit physical volume, and all other symbols have their usual meaning. The total photoheating rate ${\cal H}$ is summed over the species $i=\,$\HI, \HeI, and \HeII, ${\cal H}=\sum_i n_i H_i$, where
\begin{equation}
H_i = 4\pi \int_{\nu_i}^\infty \frac{J_\nu}{h\nu} (h\nu-h\nu_i)\,\sigma_i d\nu.
\label{eq:Heat}
\end{equation}
\begin{figure}
\centering
\includegraphics[width=\hsize]{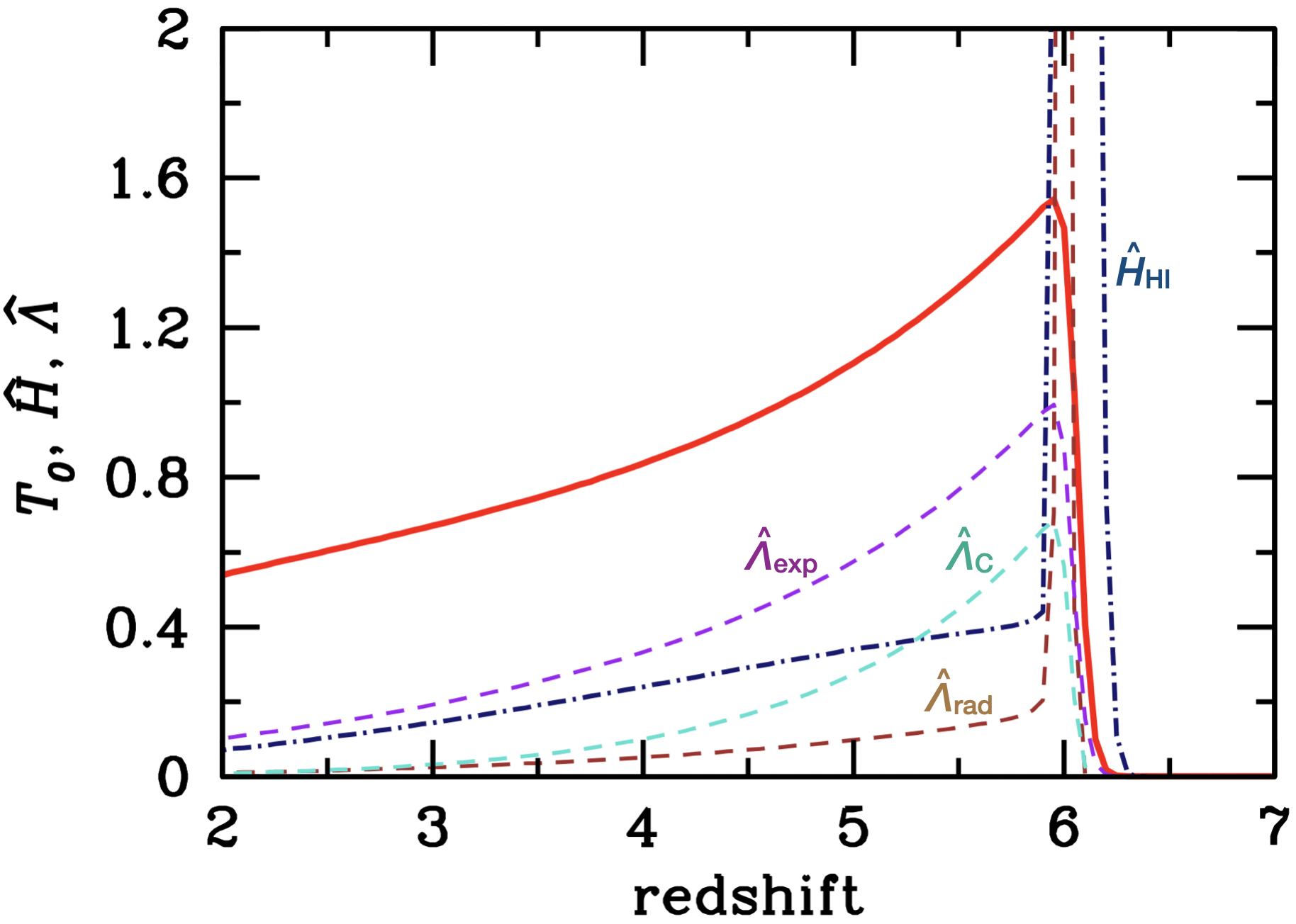}
\caption{Temperature evolution ($T_0$ in units of $10^4$ K, red curve) of an IGM gas parcel at the mean cosmic density ($\Delta=1$). In this example, a constant, power-law ($\alpha=1$) UV background flux, normalized to yield $\Gamma_\nHI=10^{-12}\, {\rm s^{-1}}$ and not energetic enough to 
photoionize \HeII, is 'switched on' at $6<z<6.5$. Also depicted are the hydrogen photoheating ($\hat{H}_\nHI$, dash-dotted blue curve), recombination ($\hat{\Lambda}_{\rm rad}$, dashed brown), Compton ($\hat{\Lambda}_C$, dashed turquiose), and expansion cooling ($\hat{\Lambda}_{\rm exp}$, dashed purple) rates \emph{per H-atom} (i.e. $\hat{H}_i\equiv n_i H_i/n_\H$ and 
$\hat{\Lambda}_i\equiv \Lambda_i/n_\H$), in units of $10^{-27.5}$ erg s$^{-1}$.
}
\label{fig:Heat}
\end{figure}
In equation (\ref{eq:dT}), the first two terms on the right-hand side account for adiabatic cooling, the third describes the change in the internal energy per particle resulting from varying  the total number of particles, and the fourth is the net heat gain or loss per unit volume from radiation processes. 
Suitable tabulations of the cooling rates can be found in \citet{hui97} and \citet{theuns98}. Equation (\ref{eq:dT}) must be integrated in conjunction with the rate equations (\ref{eq:hion}-\ref{eq:heion}) describing the evolving fractional abundances of the three species. Figure \ref{fig:Heat} shows the evolution of the temperature $T_0$ for an IGM gas element at the mean cosmic density ($\Delta=1$) photoionized by a constant  UV background that is suddenly 'switched on' between redshifts 6 and 6.5. Also depicted are the corresponding hydrogen photoheating, recombination, Compton, and expansion cooling rates per H-atom. In this example, the power-law ($\alpha=1$) radiation flux is not energetic enough to produce \HeIII. The gas temperature reaches a peak of about 15,000\,K as the IGM is flash photoionized and the abundances of \HI\ and \HeI\ quickly drop by 3 orders of magnitude. After reionization, $T$ is determined by photoionization heating, adiabatic cooling, and to a smaller extent by Compton cooling. Assuming a power-law background spectrum of the form $J_\nu=J_0 (\nu/\nu_\nHI)^{-\alpha}$, and approximating the photoionization cross-section as $\sigma_\nHI=\sigma_0 (\nu/\nu_\nHI)^{-3}$, we can rewrite the photoionization and photoheating rates of hydrogen as
\begin{equation}
\Gamma_\nHI=4\pi\int_{\nu_\nHI}^\infty \frac{J_\nu}{h\nu}\,\sigma_\HI(\nu) d\nu=\frac{4\pi \sigma_0 J_0}{h(3+\alpha)} 
\end{equation}
and 
\begin{eqnarray}
n_\nHI\,H_\nHI & = & 4\pi\,n_\nHI \int_{\nu_\nHI}^\infty \frac{J_\nu}{h\nu}\,(h\nu-h\nu_\nHI)\sigma_\HI(\nu) d\nu \nonumber \\
& = & \frac{4\pi n_\nHI\sigma_0 J_0\nu_\nHI}{(2+\alpha)(3+\alpha)}=\frac{h\nu_\nHI}{(2+\alpha)}\frac{n_\H}{t_{\rm rec}}.
\end{eqnarray}
Here, the last expression is valid in a highly ionized gas that satisfies the equilibrium condition
\begin{equation}
n_\nHI \Gamma_\nHI= \frac{n_\H}{t_{\rm rec}},
\end{equation}
where $t_{\rm rec}=(n_e \alpha^A_\nHII)^{-1}$ is the hydrogen recombination timescale. 
When this condition is satisfied, the photoheating rate does not depend on the amplitude of $J_\nu$ but only on its spectral shape, which opens up the possibility of constraining the nature of ionizing sources from measurements of the post-reionization temperature of the IGM.

\begin{figure}
\centering
\includegraphics[width=\hsize,clip]{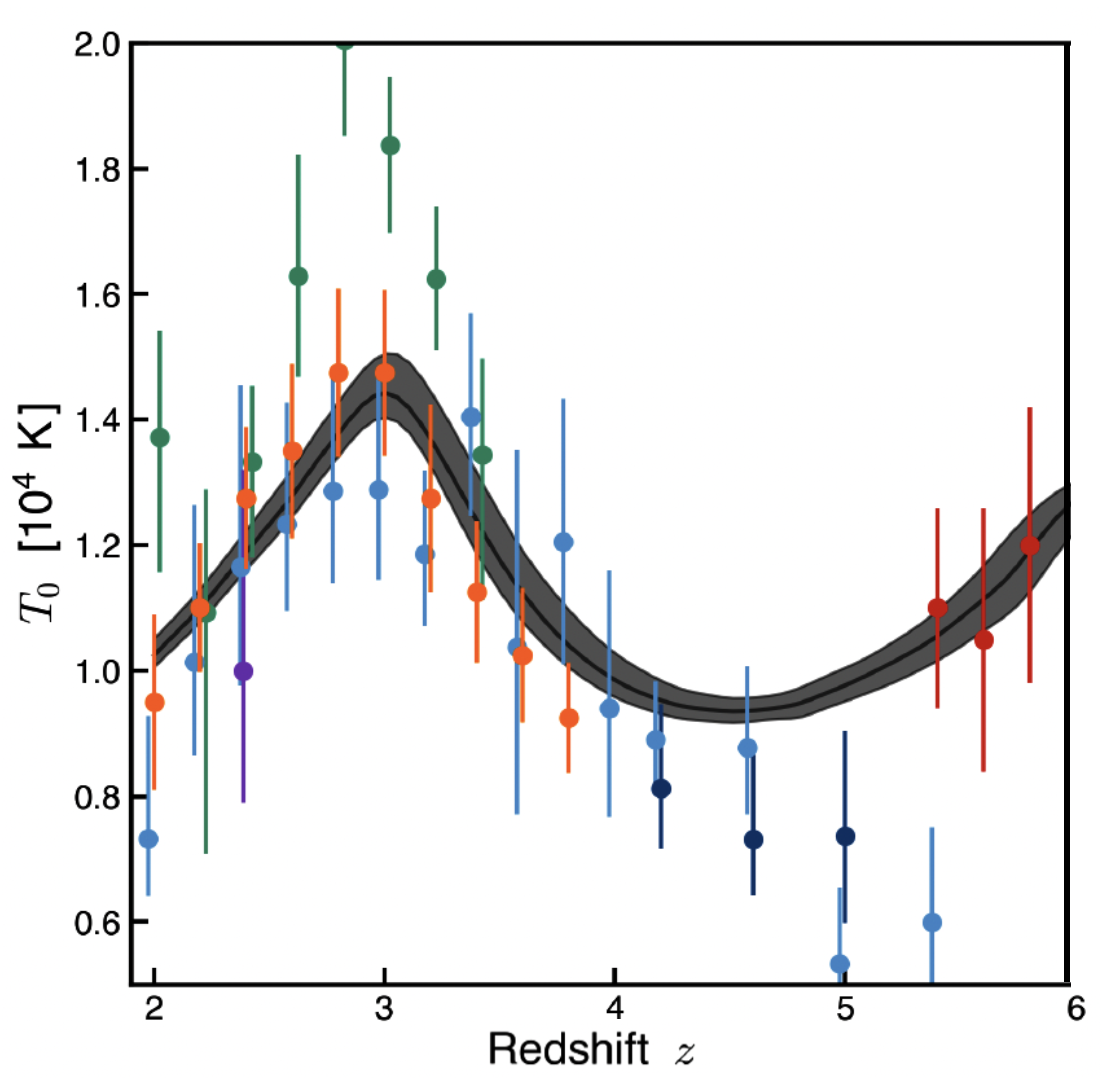}
\caption{Redshift evolution of the IGM temperature $T_0$ from the best-fit (black line) and 95\% confidence interval (gray band) model for the  UV background photoheating and photoionization rates of \citet{villasenor22}. The figure displays the epoch of cosmic expansion cooling at $z\lesssim 6$ following 
the reionization of hydrogen by massive stars. The onset of helium reionization by AGNs at $z\sim 4.5$ initiates a second epoch of heating that ends at $z\sim 3$ when \HeII\ reionization completes. A second epoch of cooling due to cosmic expansion then follows.  The data points with error bars show the values of $T_0$ inferred from observations of the Lyman-$\alpha$ forest by \citet{bolton14,hiss18,walther19,boera19,gaikwad20,gaikwad21}. The $\Lambda$CDM simulations used for statistical analysis assume a spatially homogeneous ionizing background.
(Image courtesy B.\ Villasenor.)}
\label{fig:T0}
\end{figure}
More realistically, it is believed that cosmic reionization was an extended process that started around $z\sim 10$, and that the cosmic web actually experienced two main reheating events. Figure \ref{fig:T0} shows the thermal history of the IGM at $z<6$ as recently inferred using a suite of more than 400 high-resolution cosmological hydrodynamics simulations of structure formation 
that self-consistently evolve a wide range of physically-motivated ionization and thermal histories of the IGM \citep{villasenor22}. A statistical comparison with the observed transmitted flux power spectrum of the hydrogen forest over the redshift range $2.2\le z\le 5$ and with the Lyman-$\alpha$ opacity of intergalactic \HeII\ over the redshift range $2.4\le z\le 2.9$  produces the best-fit temperature history depicted in the figure. A first peak in $T_0$, due to hydrogen reionization at $z\gtrsim 6$, is followed by an epoch of cooling due to adiabatic expansion until the onset of helium reionization from extreme UV radiation emitted by AGNs. The ionization of helium leads to a second increase of the temperature until \HeII\ is fully ionized ($z\simeq 3$), which is then followed by a second period of cooling from Hubble expansion.

\subsection{Secondary ionizations by X-rays}

It is generally recognized that energetic X-ray photons emitted by active galaxies and/or X-ray binaries may have significantly altered the thermal and ionization balance of the early neutral IGM. 
Since only highly energetic photons result in more than one ionization, this process is only important for radiation emitted by AGNs and X-ray binaries, while stars hardly emit any photons that are able to ionize hydrogen more than once.
Unlike stellar UV photons, X-ray photons can easily escape from their host galaxies, quickly establish a homogeneous X-ray background, make the low-density IGM warm and weakly ionized prior to the epoch of reionization breakthrough, set an entropy floor, reduce gas clumping, and promote gas-phase H$_2$ formation \citep[e.g.,][]{oh01,venkatesan01,madau04}. Photoionizations of hydrogen and helium by X-rays produce fast photoelectrons that can collisionally excite and ionize atoms before their energy is thermalized \citep[e.g.,][]{shull85}. To account for the effect of X-ray ``secondary'' ionizations, the standard photoionization rates of the neutral species \HI\ and \HeI\ must be ``enhanced'' as \citep[see][]{ricotti02a,kuhlen05,madau_fragos17}
\begin{align}
\widetilde \Gamma_\nHI & = \Gamma_\nHI + \frac{f_{\rm ion}^\nHI}{h\nu_\nHI}\,\sum_i \frac{n_i}{n_\nHI} H_i,\\
\widetilde \Gamma_\nHeI & = \Gamma_\nHeI + \frac{f_{\rm ion}^\nHeI}{h\nu_\nHeI}\,\sum_i \frac{n_i}{n_\nHeI} H_i.
\label{eq:sec}
\end{align}
Here, the sum is taken over $i=\,$\HI, \HeI, and \HeII, and $H_i$
are the standard photoheating rates per atom (eq. \ref{eq:Heat}). The above expressions are valid in the high-energy limit, in which the fractions of the primary electron energy that
go into secondary ionizations of \HI\ and \HeI, $f_{\rm ion}^\nHI$ and $f_{\rm ion}^\nHeI$, are independent of the electron energy \citep{shull85}. Each correction term for secondary ionizations is proportional to $n_i$, the number density of the species $i$ that is the source of primary photoelectrons. Energy losses from electron impact excitation and ionization of \HeII\ are small and can be neglected, i.e. one can safely assume $\widetilde \Gamma_\nHeII=\Gamma_\nHeII$. Collisional excitations of helium also generate line photons
that can ionize atomic hydrogen, but this contribution amounts to less than 5\% of the direct ionizations by secondaries.
Electrons with energies $<10.2$ eV are unable to interact with any atoms or ions, so all of the energy is deposited as heat. 
Since a fraction of the energy of the primary photoelectron goes into secondary ionizations and excitations, all photoheating rates must be reduced accordingly by the factor $f_{\rm heat}$,
\begin{equation}
\widetilde H_i = f_{\rm heat}\, H_i.
\end{equation}
Thus, the reduced energy deposition rate from species $i$ is related to the excess photoionization rate of species $j$ by the factor
($f_{\rm heat}h\nu_j/f_{\rm ion}^j$).

Monte Carlo simulations have been run to determine the fractions  $f_{\rm ion}^\nHI, f_{\rm ion}^\nHeI$, and $f_{\rm heat}$ as a function of electron energy and
hydrogen ionized fraction $x_\nHII$ \citep{shull85,valdes08,furlanetto10}.\footnote{The energy deposition fractions are insensitive to the total hydrogen density, which enters
only through the Coulomb logarithm in the electron scattering cross-section \citep{shull85}.}\, Here, we use the fitting functions provided by \citet{shull85}, which apply 
in the limit $E\gtrsim 0.3$ keV, 
\begin{align}
f_{\rm ion}^\nHI & = 0.3908 (1-x_\nHII^{0.4092})^{1.7592},\nonumber \\
f_{\rm ion}^\nHeI & =  0.0554 (1-x_\nHII^{0.4614})^{1.6660},\\
f_{\rm heat} & =  1-(1-x_\nHII^{0.2663})^{1.3163}.\nonumber
\end{align}
For photon energies above $0.2\,$keV, the ratio of the \HI/\HeI\ photoionization cross sections (see Section~2.6) drops below 4 per cent: since the primordial ratio of helium to hydrogen is about 8 per cent, the photoionization and heating rates are dominated by helium absorption, which exceeds the hydrogen contribution by $\sim 2:1$. While \HeI\ is the main source of hot primary photoelectrons, however, it is \HI\ that is more abundant and therefore undergoes the bulk of secondary ionizations. A primary nonthermal photoelectron of energy $E= 0.6\,$ keV in a medium with ionization fraction $x_\nHII=0.01$ creates about a dozen secondary
ionizations of hydrogen (corresponding to a fraction $f_{\rm ion}^\nHI=0.29$ of its initial energy), 
and only one secondary ionization of \HeI\ 
($f_{\rm ion}^\nHeI=0.045$), therefore depositing in the process 34\% of its initial energy as secondary ionizations, 37\% as heat
via Coulomb collisions with thermal electrons, and the rest as secondary excitations. Once the IGM ionized fraction increases to $x_\nHII\gtrsim 0.1$,
the number of secondary ionizations drops ($f_{\rm ion}^\nHI+f_{\rm ion}^\nHeI\lesssim 0.19$), and the bulk of the primary's energy goes into heat ($f_{\rm heat}\gtrsim 0.64$). 

\subsection{Radiative transfer in an expanding universe}

Ignoring scatterings into and out of the beam, the evolution of the specific intensity $I_\nu(t,\bf{x},\bf{n})$ in an expanding Universe is described by the radiative transfer equation \citep[e.g.,][]{peebles71,gnedin97,abel99,petkova09}: 
 \begin{equation}
 \begin{split}
\frac{1}{c}\frac{\partial I_\nu}{\partial t} +  \frac{\vec{n}}{a} \cdot 
\frac{\partial I_\nu}{\partial \vec{x}} & - \frac{H(t)}{c}\left(\nu \frac{\partial I_\nu}{\partial \nu} -3I_{\nu}\right) \\
& = -\kappa_\nu I_\nu + S_\nu,
\label{eq:rt}
\end{split}
\end{equation}
where $\bf{x}$ is the comoving spatial coordinate, $\bf{n}$ is a unit vector along the direction of propagation of the ray, $S_\nu$ is the source term or emission coefficent (with units erg s$^{-1}$ cm$^{-3}$
Hz$^{-1}$ sr$^{-1}$), and $k_\nu$ is the absorption coefficient -- the product of an atomic absorption cross-section and the number density of atoms that can interact with photons of frequency $\nu$. Equation (\ref{eq:rt}) is essentially the Boltzmann equation for the phase-space density $I_\nu/\nu^3$, with sources and sinks on the RHS. It can be recognized as the classical equation of radiative transfer in a static medium with two modifications: the denominator $a(t)=1/(1+z)$ in the second term, which accounts for the changes in path length along the ray due to cosmic expansion, and the third and fourth terms, which account for cosmological redshift and dilution of intensity. Comparing the third with the second term, one recovers the classical transfer equation in the limit when the scale of interest $L$ is much smaller than the horizon size, $L \ll c/H(t)$. 
Applying the space- and direction-averaging operator $\langle ~\rangle$ to Equation (\ref{eq:rt}), we obtain the following equation for the mean specific background intensity $\bar J_\nu\equiv \langle J_\nu\rangle$:
\begin{equation}
\frac{\partial \bar J_\nu}{\partial t} - H\left(\nu \frac{\partial  \bar J_\nu}{\partial \nu} 
-3 \bar J_{\nu} \right) = -c \bar \kappa_\nu\bar  J_\nu
+ \frac{c}{4\pi} \epsilon_\nu,
\label{eq:tra2}
\end{equation}
where $\epsilon_\nu=4\pi \langle S_\nu\rangle$ is the mean proper volume
emissivity. The above equation admits an integral solution
\begin{equation}
\bar J_\nu(t) = \frac{c}{4\pi}\int_0^{t}\, {dt'}  \epsilon_{\nu'}(t') \left[\frac{a(t')}{a(t)}\right]^3 e^{-\bar\tau(\nu,t',t)},
\label{Jnu}
\end{equation}
where $\nu'=\nu a(t)/a(t')$,
\begin{equation}
\bar\tau(\nu,t',t)=c\int\limits_{t'}^{t} dt'' \bar\kappa_{\nu''}(t''),
\end{equation}
and $\nu''=\nu a(t)/a(t'')$. A packet of photons will travel a proper mean free path $\lambda_{\rm mfp}(\nu)=1/\bar\kappa_\nu$ before 
suffering a $1/e$ attenuation.

The mean opacity $\bar \kappa_\nu$ in Equation (\ref{eq:tra2}) is defined as $\bar \kappa_\nu\equiv \langle \kappa_\nu I_\nu\rangle/ \bar J_\nu$ \citep{gnedin97}. This is not the \emph{space average} of the absorption coefficient, since it is weighted in the radiative transfer equation  by the local value of the specific intensity  $I_\nu$. The relation $\bar \kappa_\nu=\langle \kappa_\nu\rangle$ holds only in the limit of a uniform and isotropic ionizing background, or when $\kappa_\nu$ and $I_\nu$ are independent random variables. The former is a reasonable approximation during the late stages of reionization, when ionized  bubbles expand into low-density regions under the collective influence of many UV sources \citep{iliev06}, the photon mean free path is almost insensitive to the strength of the local radiation field, and variations in the LyC  opacity are relatively modest. The assumption of a nearly uniform UV background, however, breaks down during the early stages of reionization, when rare peaks in the density field that contain more absorbers as well as more sources ionize first, and the photon mean free path is comparable to or smaller than the average source separation. A detailed modeling of the spatial correlations between sources and sinks of ionizing radiation can only be achieved using radiative transfer simulations or semi-numerical schemes, which we describe in the rest of this review.

When the photon mean free path is much smaller than the horizon size, cosmological effects such as source evolution and frequency shifts can be neglected. In this local source 
$\bar\tau\rightarrow \infty$ 
approximation, the mean specific intensity relaxes to
\begin{equation}
4\pi \bar J_\nu \simeq \frac{\epsilon_\nu}{\bar\kappa_\nu},
\\
\label{eq:sourcefunction}
\end{equation}
and only emitters within the volume defined by an absorption length
$1/\bar\kappa_\nu$ contribute to the background radiation field \citep{zuo93,madau99}. While, for a given ionizing photon emissivity, the local source approximation can 
overestimate the background intensity close to the hydrogen Lyman edge at $z\sim 2$ 
(this is because a significant fraction of the emitted LyC photons at these epochs gets redshifted beyond the 
Lyman limit and does not contribute to the ionizing flux), it quickly approaches the exact value of $\bar J_\nu$ at redshifts $z\gtrsim 5$ \citep{becker13}. 

\subsection{The source term}

Both because of finite resolution and neglected physics, it is customary to separate the transport of radiation in the IGM (and in the circumgalactic medium or CGM) from that in the interstellar medium (ISM) of the source population by introducing an escape fraction parameter $f_{\rm esc}$ -- the fraction of LyC photons that escape the star forming or nuclear activity regions and leak into the IGM. The escape of ionizing radiation from each individual galaxy varies with propagation direction, time, and scale and  depends on the detailed spatial distribution of neutral gas, stars, and dust in the ISM as well as the clumpiness of halo gas. It is often convenient to sweep these complexities into a single parameter, which should be thought of as a global average over direction and (some of) host galaxy properties. If we denote then with $\phi(L_\nu,t)dL_\nu$ the source rest-frame UV luminosity function (LF), i.e. the comoving space density of emitters with monochromatic luminosities in the range ($L_\nu, L_\nu+dL_\nu$), the mean proper ionizing emissivity can be written as 
\begin{equation}
\epsilon_\nu(t)=a(t)^{-3}\int_0^\infty f_{\rm esc, \nu}(t) \phi(L_\nu,t)\,dL_\nu.
\end{equation}
\begin{figure}
\centering
\includegraphics[width=\hsize,clip]{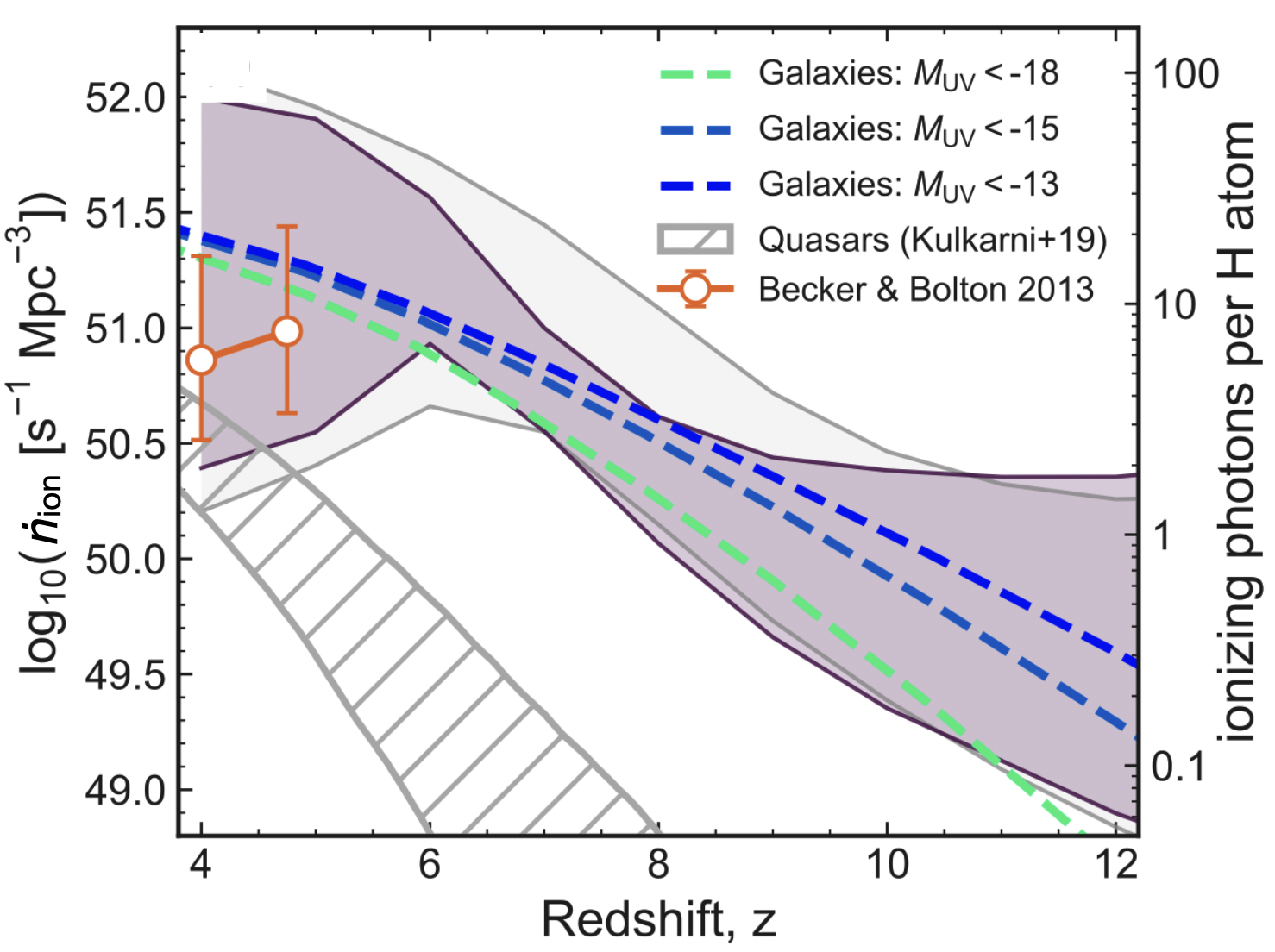}
\caption{Redshift evolution of the mean emission rate of hydrogen ionizing photons into the IGM inferred from measurement of the evolving galaxy (assuming a constant $f_{\rm esc}=0.2$) and quasar ($f_{\rm esc}=1$) LFs. To highlights uncertainties in the faint-end cut-off of the galaxy LF, the three dashed curves show $\dot n_{\rm ion}$ for galaxies brighter than $M_{\rm UV}=-18, -15,$ and $-13$ mag. Similarly, the hatched region encompasses the range of values obtained by integrating the LF down to $M_{\rm UV}<-18$ or $M_{\rm UV}<-21$ mag. The shaded regions show the 68 per cent confidence regions for $\dot n_{\rm ion}$ inferred using different observational constraints on the reionization timeline:
1) the CMB electron scattering optical depth 
and the Lyman-$\alpha$ and Lyman-$\beta$ forest dark pixel fraction (gray shade); and 2) same as 1) including additional measurements of the hydrogen neutral fraction from the Lyman-$\alpha$ damping wings observed in quasar spectra (purple shade). (From \citealt{mason-naidu19}, used by permission.)
}
\label{fig:Emiss}
\end{figure}
The rest-frame UV ($1500\,$\AA) LF of galaxies is well measured down to $M_{\rm UV} =-15$ mag \citep{Bouwens2022}, and has a steep  faint-end slope, $\phi(L_\nu)\propto L_\nu^{\alpha}$ with $\alpha\sim -2$. A slope exactly equal to $-2$ implies an equal contribution to the radiation emissivity per unit logarithmic interval of galaxy luminosity. Thus, each unit magnitude interval between the  exponential cutoff of the Schechter function $M^\star \approx -20$ mag  and $-15$ mag contributes equally about 20\% of the total UV photon output. Unless such a steep faint-end slope 
continues down to extremely faint magnitudes of $-10$ mag or so, the implication is that we may have already observed the bulk of the galaxies responsible for the reionization of cosmic hydrogen. A challenge to this line of argument may come from the fact that  rest-frame UV luminosities do not directly translate into ionizing luminosities -- the two are related by the relative escape fraction,
\begin{equation}
L_{\rm ion} = \frac{f_{\rm esc, ion}}{f_{\rm esc, 1500\AA}} R L_{\rm 1500\AA},
\end{equation}
where $R$ is ratio of the intrinsic, non-attenuated ionizing and 1500\,\AA\ luminosities. Here, 
$f_{\rm esc, 1500\AA}$ depends primarily on the dust abundance and spatial distribution, while 
$f_{\rm esc, ion}$ is thought to depend primarily on the distribution of neutral gas from the scales of  the disk ISM to those characterizing the structure of molecular clouds (and, perhaps, on the abundance of ``runaway'' massive stars that travel through interstellar space with anomalously high velocities) \citep{Gnedin2008,fesc1,fesc2,fesc3,fesc4,fesc5,fesc6,fesc8,fesc9,fesc10,fesc11,fesc12,fesc13,fesc14,fesc15,fesc15a,fesc15b,fesc16,fesc17,fesc18,fesc19,fesc20,fesc21,fesc22}. Simulations of radiative feedback in molecular clouds support the view that the escape fractions are most sensitive to the short temporal and spatial scale processes in molecular clouds rather than the global properties of the host galaxies \citep{fesc15a,fesc15b,fescKOa,fescKOb,fescKOc}. If that conclusion was fundamentally incorrect, variations of the escape fractions could impact the contribution of different galaxy types to the total ionizing photon budget. While the escape fraction is a function of frequency or a spectral band, in order to simplify notation we shall use below a common shorthand $f_{\rm esc}$ to indicate the escape fraction of ionizing radiation, i.e.\ hereafter $f_{\rm esc} \equiv f_{\rm esc, ion}$.

Galaxy-dominated scenarios that satisfy many of the observational constraints can be constructed if one extrapolates the UV LF
to sufficiently faint magnitudes ($M_{\rm UV} \sim -13$) and  assumes an escape fraction of $f_{\rm esc}\sim$ 10\% at $z\gtrsim 6$
\citep[e.g.,][]{madau99,robertson10,finkelstein12,haardt12,Shull2012,kuhlen12,Robertson2015,bouwens15}. In AGNs, $f_{\rm esc}$ is observed to be between 44 and 100\% \citep{grazian18}. There is a vast literature on measuring escape fractions from star-forming galaxies, and surveying it is beyond the scope of this review. In general, however, population-average escape fractions from galaxies of around 10\% may not be unreasonable. Another often-used quantity is 
\begin{equation}
\dot n_{\rm ion}(t)=a(t)^3\int_{\nu_i}^\infty \frac{\epsilon_\nu(t)}{h\nu}\,d\nu,
\end{equation}
the mean emission rate into the IGM of ionizing photons for species $i=\,$\HI, \HeI, and \HeII\ per unit comoving volume -- not to be confused with the \emph{production rate}, as the latter does not account for local absorption and hence does not depend on the escape fraction. Figure \ref{fig:Emiss} (from \citealt{mason-naidu19}) shows the uncertain H-ionizing photon production rate $\dot n_{\rm ion}$ inferred from measurements of the evolving galaxy and quasar LFs. The shadings mark the 68 per cent confidence regions for $\dot n_{\rm ion}$ inferred using a number of observational constraints on the reionization timeline. 

Recombination radiation from a clumpy, ionized IGM can also provide a non-negligible contribution \citep[up to tens of percent,][]{haardt12} to the ionizing photon field. The main processes that contribute to the H-ionizing emissivity include recombinations from the continuum to the ground state of \HI, \HeI, and \HeII, as well as \HeII\ Balmer, two-photon, and \Lya\ emission \citep{haardt96,sokasian03,haardt12}. The mean proper emissivity from a generic recombination process of species $i$ can be written as $\epsilon_\nu=h\nu\phi(\nu)\langle \alpha_i n_e n_{i}\rangle$, where $\phi(\nu)$ is the normalized emission profile and $\alpha_i$ is the relevant recombination coefficient. The emission profile of free-bound recombination radiation can be computed via the Milne detailed-balance relation, which relates the velocity-dependent recombination cross section to the photoionization cross section, while a delta-function profile is sufficient for bound-bound transitions.

\subsection{Continuum absorption}

In a \emph{homogeneous} primordial IGM, the mean free path for LyC absorption of a photon of frequency $\nu$ at redshift $z$ is given by
\begin{equation}
\lambda_{\rm mfp}(\nu,z)=\left(\sum\limits_i n_i\sigma_i\right)^{-1},
\label{eq:mfp}
\end{equation}
where $n_i$ and $\sigma_i$ are the proper number densities and photoionization cross-sections measured at $z$ and $\nu$ of species  $i=\,$\HI, \HeI, and \HeII. Accurate fits to the bound-free absorption cross-sections from the ground state for many astrophysically important atoms and ions are given by \citet{verner96}. There is also an analytical formula for hydrogen (and, by extension, for any hydrogenic atom) that is exact in the Bohr atom approximation \citep{osterbrock06}:
\begin{equation}
\sigma_\nHI(\nu)=\sigma_0\,(\nu_\nHI/\nu)^4\frac{e^{(4-4\tan^{-1}x/x)}}{1-e^{-2\pi/x}},
\end{equation}
where $\sigma_0=6.30\times 10^{-18}\,{\rm cm}^2$ and $x\equiv \sqrt{(\nu/\nu_\nHI)-1}$.
The cross-sections drop off rapidly with increasing photon energy, $\sigma(\nu) \propto \nu^{-3.5}$ in the X-ray regime, for all atoms and ions, so X-ray photons can penetrate significantly further into the IGM than those with UV energies close to the ionization thresholds. For crude estimates, the effective bound-free cross-section {per hydrogen atom} in a \emph{neutral} IGM can be approximated as $\sigma_{\rm bf}\equiv 
\sigma_\nHI+f_{\rm He}\sigma_\nHeI\simeq (3.3\times 10^{-23}\,{\rm cm^2})\, (h\nu/{\rm keV})^{-3.2}$ to an 
accuracy of better than 20\% in the photon energy range $0.07<h\nu<8$ keV, where $f_{\rm He}\equiv Y/(4X)=0.083$ is the helium to hydrogen ratio by number. Over this energy range, helium dominates photoabsorption over hydrogen by a factor of 1.4-2.6. The mean free path for photoelectric absorption in a neutral IGM is then
\begin{equation}
\lambda_{\rm mfp}(\nu,z)\simeq 52{\rm Mpc}\,\left(\frac{h\nu}{{\rm keV}}\right)^{3.2}\,\left(\frac{1+z}{10}\right)^{-3}
\label{eq:mfpN}
\end{equation}
in physical units. Because of the strong energy dependence of the cross-section, photons with $h\nu\ll 1$ keV are absorbed closer to the source galaxies, giving origin to a fluctuating  UV/soft X-ray background. More energetic photons instead permeate the Universe more uniformly, heating the low-density IGM far away from galaxies. The pre-reionization Universe is optically thin to all radiation for which $\lambda_{\rm mfp}>c/H$, i.e. photons above $\sim (1.6\,{\rm keV})\,[(1 + z)/10]^{0.47}$ have a low probabilitiy of absorption across a Hubble volume \citep{mcquinn12}.

The IGM, however, is known to be very clumpy. Its diffuse, close-to-mean-density component is observed to be highly photoionized at $z\lesssim 6$, with only a small residual amount of neutral hydrogen set by the balance between radiative recombinations and photoionizations from a nearly uniform UV radiation background, and provides negligibly small  \HI\ photoelectric absorption. The continuum optical depth is instead dominated by the  LLSs, high density regions in the outskirts of galaxies \citep{fumagalli11,shen13} that occupy a small portion of the volume and are able to keep a significant fraction of their hydrogen in neutral form. It is the LLS opacity that causes the mean free path of LyC radiation to remain relatively small even after overlap. At $z<6$, the mean free path of ionizing photons in the IGM can be measured directly from the shape of the transmitted flux profile blueward of the Lyman limit in the stacked  spectra of high-redshift quasars \citep{prochaska09, worseck14,becker21}. 

\subsection{A clumpy IGM} \label{sec:PDF}

The IGM is highly inhomogeneous owing to the nonlinear evolution of density fluctuations  over time. Inhomogeneities are key to understanding absorption features in  the spectra of high-redshift sources, increase the effective  recombination rate of the IGM, slow down the reionization  process, and cause the mean free path of LyC photons to remain sub-horizon even after the epoch of overlap. A statistical description of the baryonic density field is provided by the volume-weighted probability density distribution (PDF) of IGM gas $P_V(\Delta)$, normalized according to $\int_0^\infty P_V(\Delta) d\Delta=1$ \citep{miralda00}. 

\begin{figure}
\centering
\includegraphics[width=\hsize]{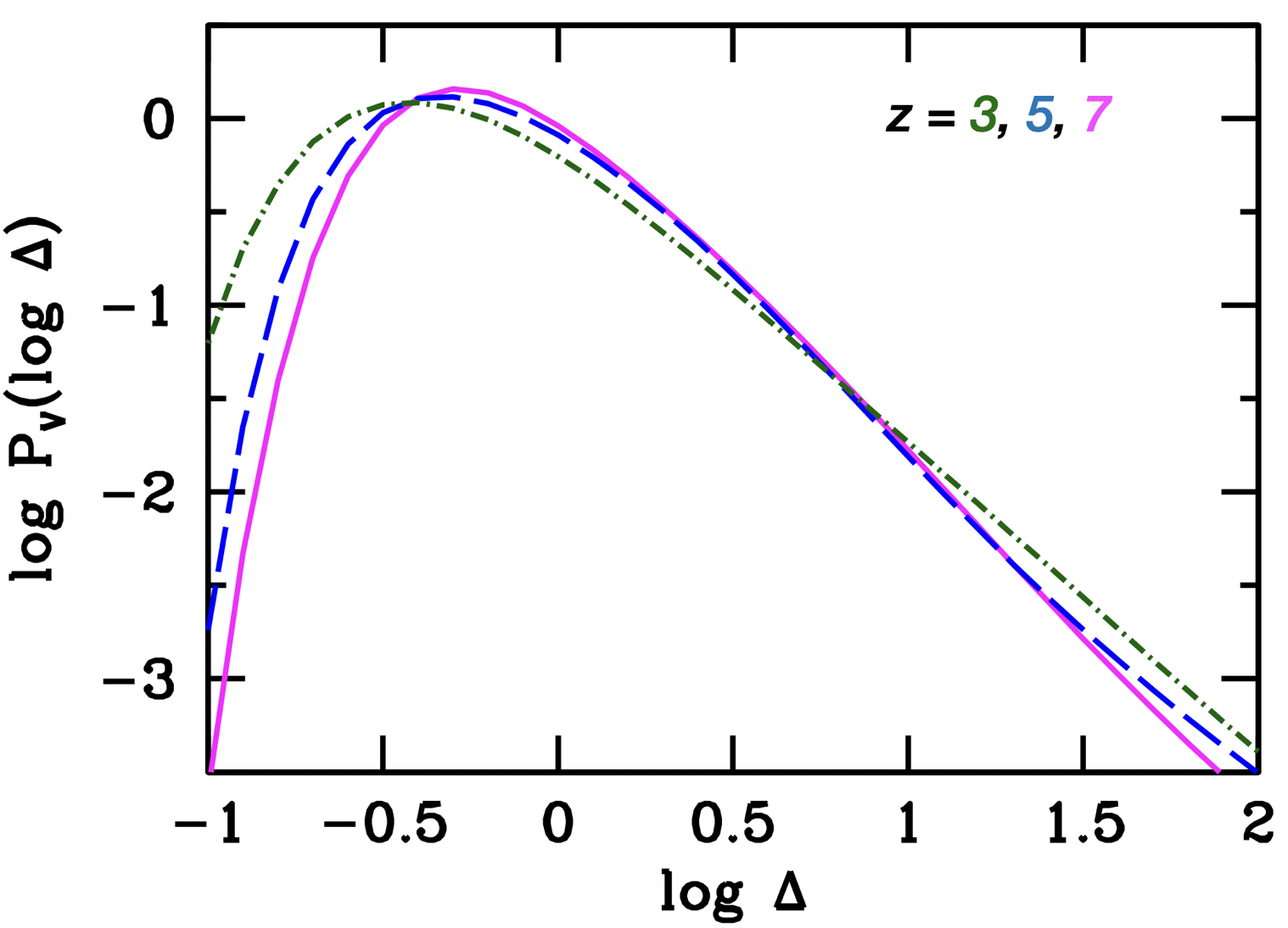}
\caption{\small Volume-weighted PDF $P_V(\log \Delta)$ of the baryon overdensity $\Delta$ from cosmological  hydrodynamics simulations in which a uniform UV background is switched on at $z=9$ \citep{bolton09}. The curves shows the PDF at 
redshifts $z=7$ (magenta solid line), $z=5$ (blue dashed line), and $z=3$ (green dot-dashed line). Reionization photoevaporates gas in low-mass halos, steepening the PDF at $ \log\Delta \gtrsim 1$ and redistributing baryons back into the $\log\Delta \lesssim 1$ diffuse IGM reservoir. 
}
\label{fig:PDF}
\end{figure}
Figure \ref{fig:PDF} shows the PDF per unit $\log\Delta$ computed from a large set of cosmological smoothed particle hydrodynamics (SPH) simulations in which a uniform UV background is switched on at $z = 9$ \citep{bolton09}. The density PDF can be accurately parameterized by the linear rms density fluctuation and the local slope of the density power spectrum at the scales of interest \citep{Chen2022}; that reference also provides fitting formulas that can be used in practical calculations.

As shown by \citet{pawlik09}, the presence of an  ionizing background has a strong impact on the PDF. In the absence of photoheating, gravitational collapse proceeds unimpeded to the smallest scales set by the CMB decoupling, causing a flattening of the PDF for overdensities $1\lesssim \log\Delta \lesssim 2$, a deficit of gas in the diffuse IGM ($\log\Delta \lesssim 1$), and shifting its maximum to lower overdensities. Reionization photoevaporates gas in low-mass halos, reducing the clustering at $\sim 100$ kpc scales and steepening the PDF at $ \log\Delta \gtrsim 1$ and redistributing baryons back into the $\log\Delta \lesssim 1$ diffuse reservoir. The details are sensitive to the timing of reionization, the photoheating energy input per baryon, and radiative transfer effects like the self-shielding of gas inside halos. 

\begin{figure*}[th]
\centering
\includegraphics[width=\textwidth]{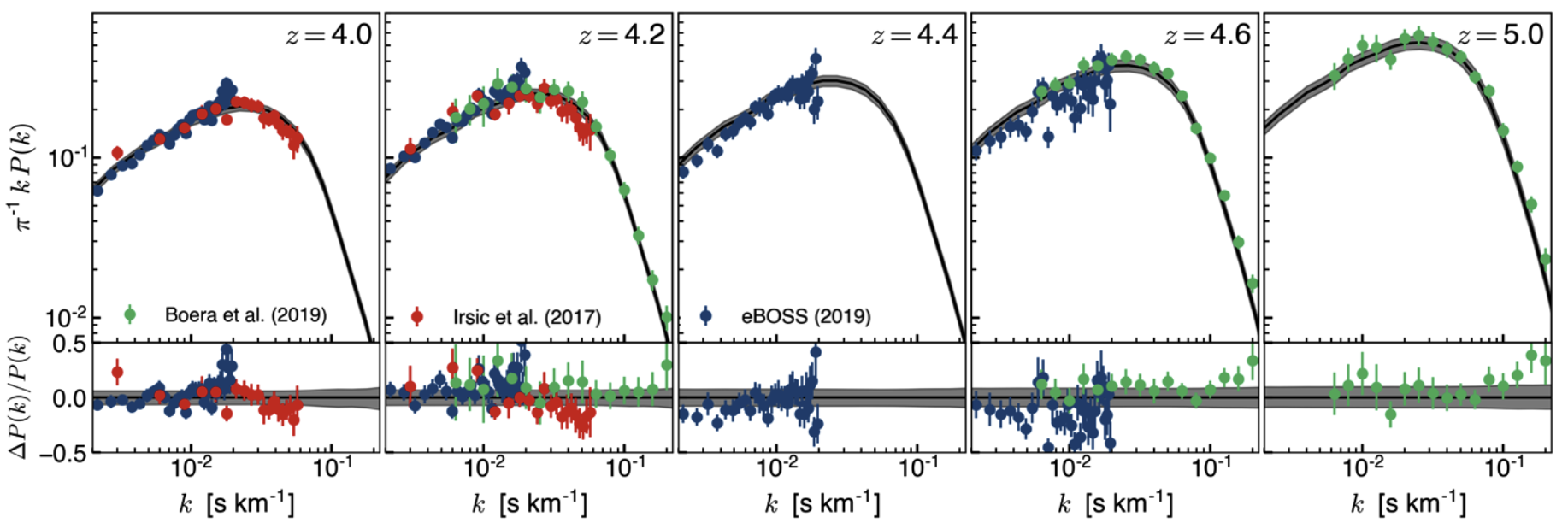}
\caption{Dimensionless transmitted flux power spectra of the \Lya\  forest at redshifts $z = 4-5$. The data points with error bars are from \citet{irsic17,chabanier19,boera19}. Also shown are the maximum likelihood fit (black line) and 95\% confidence interval (gray band) of \citet{villasenor22}, computed from hundreds of cosmological hydrodynamics simulations with varying photoionization and thermal histories of the IGM. Note that the velocity scale corresponding to a given comoving spatial scale changes with redshift.
The fractional differences between the observations and the best-fit model are shown in the bottom part of each panel. (Used by permission from B.\ Villasenor.)}
\label{fig:Pk}
\end{figure*}

More detailed information on the clumpiness of the IGM can be extracted by second-order statistics like the 1D flux power spectrum $P_F(k)$, which measures correlations of fluctuations in the transmitted flux fraction $F$ (the observed quantity) along the line of sight to distant quasars \citep[e.g.][]{croft98,mcdonald00,irsic17,chabanier19,boera19}. These are caused by variations in the \Lya\ scattering optical depth $\tau=-\ln F$, and therefore probe the gas density, temperature, and velocity field of neutral hydrogen. After measuring the distance between pixels in units of the local velocity scale $v$ and defining  the flux contrast $\delta_F(v)=F(v)/\langle F\rangle -1$, where $\langle F
\rangle$ is the mean transmitted flux, one can decompose each absorption spectrum into Fourier modes $\tilde \delta_F(k)$. Their variance as a function of the Fourier wavenumber $k=2\pi/v$ is the flux power spectrum over the velocity interval $\Delta v$, 
\begin{equation}
P(k)=\Delta v\langle \tilde \delta_F(k)^2\rangle,
\label{eq:Pk}
\end{equation}
which is commonly expressed in terms of the dimensionless quantity $\Delta_F^2(k)=kP(k)/\pi$. Figure \ref{fig:Pk} shows measurements of  $P(k)$ over the redshift range $4\le z\le 5$ from observations by eBOSS \citep{chabanier19}, \textit{Keck}, and the \textit{VLT} telescopes \citep{irsic17,boera19}. The figure also shows the best-fit $P(k)$ from the cosmic photoionization and photoheating history of \citet{villasenor22} (see also  Fig. \ref{fig:T0}). The flux power spectrum contains information that are encoded across different velocity scales. While on scales larger than $\sim 100\,$ km s$^{-1}$ $P(k)$ is sensitive to the ionization fraction of hydrogen in the IGM, on smaller scales the power spectrum is suppressed by the Doppler broadening of absorption features, the pressure smoothing of gas density  fluctuations, the smearing effect of peculiar velocities or, alternatively, by the free streaming of dark matter particles in the early Universe \citep[e.g.][]{viel13, garzilli19}.

\section{Numerical Methods}

Numerical algorithms for modeling the dynamics of both collisionless dark matter particles and baryons are well established and mature, and we do not discuss them here
\citep[see][for recent reviews that cover this topic in depth]{Teyssier2015, Springel2016}. 
We also do not survey models for star formation and stellar feedback, as these vary widely between different reionization simulation projects. Reionization simulations largely adopt recipes developed for and used in mainstream galaxy formation simulations. Recent reviews by \citet{Teyssier2019} and \citet{Vogelsberger2020} provide a general description of most of these recipes.

\subsection{Radiative transfer}
\label{sec:rtalgos}

Radiative transfer algorithms solve for the evolution of the radiation field, taking into account emission, absorption, and scattering processes. In general, the evolution is described by the differential equation
(\ref{eq:rt}) for the specific intensity $I_\nu$, which is a function of seven variables: 3D position, 2D angular coordinates, time, and frequency. Because of the high dimensionality of the problem, radiative transfer is usually the most computationally expensive component of a reionization simulation. In order to implement a radiative transfer scheme into cosmological simulations with a mass and spatial resolution comparable to those featured in modern galaxy formation runs, a radiative transfer algorithm should scale close to linearly with the number of resolution elements, just like $N$-body and hydrodynamics algorithms. To satisfy this requirement, some degree of physical approximation and computational optimization is needed. 

Existing algorithms for solving the radiative transfer equation (\ref{eq:rt}) can be broadly divided into two categories: ``ray tracing'' (we also include Monte-Carlo methods in this category) and ``moments methods''. It is interesting to note that only the moments methods are being used in the largest fully-coupled simulations, and so it appears, at least superficially, that ray-tracing techniques may be gradually falling out of use.

\subsubsection{Ray tracing}
\label{sec:ray}

Ray tracing is a domain-specific jargon for the standard method of characteristics \citep[e.g.][]{courant2008methods} for solving partial differential equations. The key advantage of using the method of characteristics in radiative transfer calculations is that photons move along geodesics (i.e.\ straight lines in the Newtonian limit); therefore one does not need to solve for the shape of characteristic curves, but only for the time dependence of the radiation intensity along each characteristic.

Even with this simplification, ray-tracing techniques suffer from unfavorable scaling. In the simplest case of $N_s$ sources on a uniform grid of size $N_g^3$, and requiring that each grid cell has at least one ray from each source passing through it, the number of rays required is $N_s \times 6 \times N_g^2$. Given that each ray, on average, passes through $N_g/2$ cells, the total number of operations then scales as $N_s \times N_g^3$. In a large-box simulation, the number of sources may be substantial. Moreover, at fixed mass resolution, galaxies of a specified mass contain the same number of resolution elements, and the number of sources $N_s$ at a given time is then a fixed (small) fraction of the total number of resolution elements. Therefore, for a set of simulations having fixed mass resolution but in progressively larger volumes, the cost of the simplest implementation of ray tracing scales as $N_g^6$ (the same scaling as a direct summation gravity solver),  which is prohibitively expensive for any realistic simulation.

A significant effort has gone  into reducing this unfavorable scaling. \citet{Abel2002} noticed that if $6N_g^2$ rays are cast to cover all cells at the faces of the computational volume, multiple rays will pass through each cell close to the sources. They used the HEALPix subdivision of the sphere \citep{healpix}
to implement a  more efficient adaptive ray-tracing scheme, where the angular resolution continually increases farther away from sources. For a given source, a small number of parent rays are cast and travel a short distance before they split into four daughter rays. Successive generations of splitting continue downstream. This algorithm, however, still scales as $N_s \times N_g^3$ with a smaller coefficient in front. Other variations of ray-tracing methods include grouping near sources in a tree-like fashion \citep{Razoumov2005}, merging closely aligned rays \citep{Trac2007}, or limiting the splitting of rays \citep{McQuinn2007} in the \citet{Abel2002} scheme. All these approaches break the unfavorable $N_s \times N_g^3$ scaling at the expense of angular resolution. The effect of such resolution loss on the accuracy of the solution has not been thoroughly studied. 

There exist several variations of the ray-tracing method, e.g. ``short characteristics'' and ``long characteristics'', which we do not cover here in detail as the differences between them are not fundamental (i.e.\ they do not eliminate the unfavorable scaling with $N_s \times N_g^3$). The primary difference between the
``short'' and ``long characteristics'' approaches is that in the latter the rays originate from the source, while in the former the rays are centered on each target cell -- such constrast is analogous to the difference between separate methods for solving hydrodynamics on the grid, like a Riemann solver versus a ``Total Variation Diminishing" (TVD) scheme.

A somewhat more distinct flavor of ray tracing is a class of Monte-Carlo methods. In this approach a source emits a number of ``photon packets" in random directions, therefore randomly sampling a subset of all possible rays. As the simulation progresses, all possible directions get eventually sampled, but only at a finite temporal cadence. These schemes, therefore, are less suitable for modeling rapidly varying sources.

\subsubsection{Moments methods}

The moments method converts the radiative transfer equation (\ref{eq:rt}) into a hierarchy of moments of the radiation intensity $I_\nu$,
\begin{subequations}%
\label{eq:rtmoms}%
\begin{multline}%
    \frac{1}{c}\frac{\partial J_\nu}{\partial t} + \frac{\partial F^j_\nu}{a\partial x^j} - \frac{H}{c} \left(\nu\frac{\partial J_\nu}{\partial \nu} - 3 J_\nu\right) = \\
    -\kappa_\nu J_\nu + \tilde S_\nu,
\end{multline}%
\begin{multline}%
    \frac{1}{c}\frac{\partial F^i_\nu}{\partial t} + \frac{\partial P^{ij}_\nu}{a\partial x^j} - \frac{H}{c} \left(\nu\frac{\partial F^i_\nu}{\partial \nu} - 3 F^i_\nu\right) = \\
    -\kappa_\nu F^i_\nu + Q^i_\nu, 
\end{multline}%
\end{subequations}
where $J_\nu$ is the mean specific intensity defined in equation (\ref{eq:Jnu}), $F^i_\nu = \int d\Omega\,n^i I_\nu$ is the photon flux, $\tilde S_\nu$ is the angle-averaged radiation source function $S_\nu$ from Equation (\ref{eq:rt}), and $Q^i_\nu$ is the flux source term. The second moment $P^{ij}_\nu\equiv \int d\Omega\,n^in^j I_\nu$ is the radiation pressure tensor, and its trace is equal to $J_\nu$. The radiation Eddington tensor,  defined as
\begin{equation}
    h^{ij}_\nu \equiv \frac{P^{ij}_\nu}{J_\nu},
    \label{eq:hij}
\end{equation}
is dimensionless and has a unit trace. 
In the general case the Eddington tensor or the radiation pressure tensor cannot be specified from the photon energy density and flux alone. One can write down an equation for $P^{ij}_\nu$ similar to the above partial differential equations for $J_\nu$ and $F^i_\nu$, but that equation would depend on the third moment of $I_\nu$. The hierarchy of moments is therefore not closed. 

In direct analogy with gas dynamics, for scales much larger than the photon mean free path and when photons preferentially scatter rather than being absorbed, one can adopt a ``diffusion'' approximation under which $h^{ij}_\nu = \delta^{ij}/3$. In modeling reionization this limit is never achieved, however, and this approximation is not useful.
Another interesting limiting case is when the gas is optically thin ($k_\nu=0$). Equations (\ref{eq:rtmoms}) then have an exact quadrature solution, which in the simplified case of isotropic sources ($\tilde S_\nu=S_\nu$ and $Q^i_\nu = 0$) takes the familiar $1/(4\pi r^2)$ inverse-square law form for the flux,
\begin{subequations}
\label{eq:rtot}
\begin{align}
    J_\nu(\vec{x}) & = \frac{a}{4\pi} \int d^3x_1 S_\nu(\vec{x}_1) \frac{1}{(\vec{x}-\vec{x}_1)^2}, \\
    F^i_\nu(\vec{x}) & = \frac{a}{4\pi} \int d^3x_1 S_\nu(\vec{x}_1) \frac{x^i-x^i_1}{\|\vec{x}-\vec{x}_1\|^3}, \\
    P^{ij}_\nu(\vec{x}) & = \frac{a}{4\pi} \int d^3x_1 S_\nu(\vec{x}_1) \frac{(x^i-x^i_1)(x^j-x^j_1)}{(\vec{x}-\vec{x}_1)^4}.
\end{align}
\end{subequations}
%
Since the Eddington tensor cannot be computed exactly within the scope of the moments method, approximate schemes have been commonly used. Of these the two most widely used are the ``M1 closure'' and ``Optically Thin Variable Eddington Tensor'' (OTVET) approximations.

The M1 closure was first introduced by \citet{Levermore1984} with the following ansatz for the Eddington tensor (dropping the frequency dependence for brevity):
\[
  h^{ij} = \frac{1-\alpha}{2}\delta^{ij} + \frac{3\alpha-1}{2}n^in^j,
\]
where $n^i=F^i/\|\vec{F}\|$ is the unit vector in the direction of the flux propagation, 
\[
  \alpha = \frac{3+4f^2}{5+2\sqrt{4-3f^2}},
\]
and $f=\|\vec{F}\|/J$. In the vicinity of a single strong source $f$ is close to 1, $\alpha$ is also close to 1, and $h^{ij}\approx n^in^j$, which is the correct solution in this case. In the opposite limit when multiple sources contribute more-or-less equally to the radiation field at a given location, then $f\ll1$, $\alpha\approx 1/3$, and $h^{ij}\approx \delta^{ij}/3$ as in the diffusion limit. The functional form of $\alpha(f)$ then smoothly interpolates between the two regimes.

In the OTVET approximation, first introduced by \citet{Gnedin2001}, the Eddington tensor is computed as the ratio of the {\it optically thin} radiation pressure  $P^{ij}_\nu$ and mean intensity $J_\nu$ from equations (\ref{eq:rtot}). OTVET then uses this ``optically thin'' tensor in the full, ``optically thick'' equations for the radiation mean intensity and flux. 
In the vicinity of a strong source (or anywhere in the simulated volume in the case of a single source), OTVET produces the same (exact) answer as the M1 closure, $h^{ij} = n^in^j$. 
In a general location with several sources of radiation, OTVET accounts for their spatial location and clustering, but it may do so incorrectly in the presence of luminous sources embedded in very optically thick regions -- in that case OTVET would still include their contribution to the Eddington tensor. 

It is important to keep in mind that Eqs.~(\ref{eq:rtmoms}) are written in conservative form:
in the absence of absorption the radiation density and flux are conserved, and in the presence of absorptions they are reduced proportionately. Thus, errors in the Eddington tensor do \emph{not} violate photon density or flux conservation, but rather result in the photon flux being advected in the wrong direction. For example, if the recombination term in equation (\ref{eq:dqdt}) is not large, the overall reionization history is not affected by errors in the Eddington tensor (provided that the LyC mean free path is short, as it is the case during reionization) since the number of ionizations is captured correctly by equations (\ref{eq:rtmoms}). For the same reason these equations automatically yield the correct speeds for expanding ionization fronts.
In the presence of large gas density fluctuations and significant radiative recombinations, the different approximations for computing the Eddington tensor may produce somewhat different results, as the photon flux may be incorrectly advected into dense regions in one scheme (where it is absorbed by gas that later recombines) or into lower density regions in the other, where it  results in persistent photoionizations. A detailed comparison between M1 closure and OTVET has never been performed, however.

An interesting variation of the moments methods was developed by \citet{Finlator2009}, who used ray tracing to compute the Eddington tensor accurately, without any approximation. This approach, however, combines the computational expense of both the ray-tracing and the moments schemes. For comparison, \citet{finlator18} 
carried out radiative transfer simulations on a  uniform $64^3$ grid, while modern reionization simulations with the moments methods, such as CROC and THESAN, typically use $2000^3$ resolution elements and achieve a dynamic range with AMR or moving mesh in excess of $10^5$.

\subsubsection{Reduced speed of light approximation}

So far we have only discussed  techniques for addressing the spatial dependence of the  radiative transfer equation. Its time dependence presents an additional challenge, since photons travel at the speed of light, which is much larger than any other relevant speed in a typical reionization simulation. In any explicit scheme, the time-step will be limited by an appropriate Courant--Friedrichs--Lewy stability condition \citep{cfl28} to a value that is about $v_{\max}/c$ times smaller than the hydrodynamic time-step (here $v_{\max}$ is the maximum gas velocity in the simulation, including the sound speed). For reionization simulations this ratio can be as small as $10^{-3}$. Clearly this is challenging, as it is computationally expensive (although not impossible) to make 1{,}000 radiative transfer time steps for every hydrodynamic time step in a realistic simulation.

The time dependent term in the radiative transfer equation (\ref{eq:rt}) is only crucial for capturing ``light fronts'' --  waves in the radiation field that propagate with the speed of light just like waves in gas propagate with the speed of sound. Presently it is unclear if capturing such waves is ever important.\footnote{Note that radiation sources can also change rapidly. This, however, is presumably captured by the simulation with, for example, a hydrodynamic solver or a star formation algorithm.}
For example, ionization fronts in the general IGM typically propagate at speeds of order a few thousand km/s. Capturing these ionization fronts is sufficient for following the overall distribution of ionized bubbles, and if that is the primary goal of a reionization simulation, capturing light fronts may be an overkill.
Following this logic, \citet{Gnedin2001} introduced the ``Reduced Speed of Light'' (RSL) approximation, which replaces the speed of light $c$ in the first term of equation (\ref{eq:rt}) with an  ``effective'' speed of light $\hat{c}$,
\[
  \frac{1}{c}\frac{\partial J_\nu}{\partial t} \rightarrow \frac{1}{\hat{c}}\frac{\partial J_\nu}{\partial t}.
\]
For example, taking $\hat{c}=c/10$ would still be sufficient for capturing ionization fronts propagating at 3{,}000\,km/s with 10\% precision, while increasing the time-step in the radiative transfer solver by a factor of 10. Obviously, such a modification would not be appropriate when light fronts need to be tracked or when I-fronts move at a significant fraction of the speed of light. However, this is not expected to be the case during reionization \citep{Shapiro2006}.

Unfortunately, the RSL approximation is a delicate numerical tool, and there exist multiple examples of its incorrect use. Reducing the speed of light in the transfer equation for the specific intensity of the cosmic UV background results in the incorrect redshifting of photon frequencies, so the RSL approximation can only be used when computing the contribution from sources located at distances smaller than the cosmic horizon. The radiation field must then be represented as the sum of two components, one from local sources and the other from ``distant'' ones -- sources
outside the simulation box.  \emph{In simulations that model the radiation field with just one spatial component, the RSL approximation should never be used!}
The second pitfall is that only the speed of light appearing in the first term of equation (\ref{eq:rt}) can be reduced. The speed of light enters explicitly or implicitly in many physical expressions, and thus great care must be taken in deciding where it can and where it cannot be modified. A detailed description of the RSL approximation and how it should be implemented can be found in \citet{Gnedin2016}.

\subsection{Thermodynamics}

The cooling and heating processes that control the thermodynamics of  cosmic gas are well known, but complex. For simple geometries they can be modeled in great detail with the ``industry-standard'' Cloudy software \citep{Cloudy}. In realistic simulations, however, it is still too computationally expensive to run a code like Cloudy for every resolution element at every step, and hence one has to resort to some reasonable approximations.

The complexity of the problem is highlighted by considering the dependence of the cooling function $\Lambda$ and the heating function ${\cal H}$ in the temperature equation
(\ref{eq:dT}). In addition to obvious arguments such as gas temperature $T$ and total hydrogen density $n_\H$, in the most general case the cooling and heating functions depend on a large number of other parameters that describe the fractional abundance for each atomic, ionic, and molecular species at each of their quantum level. If all species can be assumed to be in local thermodynamic equilibrium (almost always a safe assumption) \emph{and} to be close enough to ionization equilibrium (sometimes a safe assumptions, see discussion below), and the abundance pattern for heavy elements is assumed to be the same everywhere in the simulation volume (e.g., to be solar), then the cooling and heating functions become just functions of $T$, $n_\H$, the gas metallicity $Z$, and the radiation field $J_\nu$,
\begin{equation}
  \{{\cal H},\Lambda\} = \{{\cal H},\Lambda\}(T,n_\H,Z,J_\nu).
  \label{eq:cfsim}
\end{equation}
Even under these simplifying assumptions equation (\ref{eq:cfsim}) is too complex, since its RHS is not a function but rather a \emph{functional}, as it depends on the specific mean intensity $J_\nu$.

One approach to account for this dependence is to use a spatially uniform radiation spectrum. 
For example, the spectrum of the evolving cosmic radiation background has been used in several galaxy formation simulations
\citep{Kravtsov2002,Wiersma2009,Vogelsberger2013,illust,EAGLE} -- so the dependence of the heating and cooling functions on the radiation spectrum is reduced to a dependence on cosmic time only.
The main limitation of such an approach is that the resultant cooling and heating functions are only valid in the general IGM, but become inaccurate near and inside galaxies, where accounting for gas thermodynamics is especially important.

An alternative simplification is to only treat hydrogen and helium accurately. In that limit not only the cooling and heating rates can be computed accurately, but even the assumption of ionization equilibrium can be relaxed -- for modeling reionization this is particularly relevant, as equilibrium is not maintained at ionization fronts \citep{gnedin00,Aubert2008,gnedin14,thesan1}. The metal-dependent part of the cooling and heating functions is then added to the cooling and heating processes from hydrogen and helium with some approximation --  for example, using a uniform cosmic radiation background or an approximation that parameterizes the effect of an arbitrary radiation field with several photoionization rates \citep{Gnedin2012}.

The assumption of ionization equilibrium for all metal ions may not hold in all relevant physical regimes. \citet{Richings2016} explored departures from ionization equilibrium in galaxy formation simulations with a full time-dependent modeling of multiple metallic species, and found some differences. It is currently unknown if these differences have any systematic effects on gas dynamics on large scales.

\section{A General Overview of the Reionization Process}

The reionization of intergalactic hydrogen during the first billion years of cosmic history involves the early production, propagation, and absorption of LyC radiation. It is, in essence,  a relatively simple story of UV sources and sinks. While it is generally agreed that star-forming galaxies and AGNs are the best candidates for providing the bulk of the photoionizing power across cosmic time, the apparent contrast between the sharp decline in the number density of bright quasars at $z\gtrsim 3$ \citep[e.g.][]{kulkarni19} and the relatively shallower decline in the UV luminosity density from galaxies \citep[e.g.][]{madau14} has led to the predominant theory that the bulk of the photon budget for hydrogen reionization came from massive stars. There also remains some room for a contribution from more exotic possibilities such as dark matter annihilation or decay \citep{mapelli06,liu16}, primordial globular clusters \citep{ricotti02,ma21}, mini- \citep{madau04} and micro-quasars \citep{mirabel11}. The basic energy source in all these scenarios is often rather different -- rest-mass and nuclear binding energy in the case of particle annihilation/decay and stellar nucleosynthesis, and gravitational binding energy in the case of accretion onto black holes. And various reionization sources will be characterized by very distinct radiative efficiencies, spectra, abundances, and spatial distributions. 

The sinks of LyC photons during reionization come essentially in two flavors: hydrogen atoms in the diffuse IGM (both pristine and previously photoionized and recombined) and the LLSs (we use this term here to also include the rarer and denser \HI\ Damped Lyman-$\alpha$ absorbers). 
The separation between the two is, admittedly, somewhat artificial, as there is just one continuous \HI\ density field in the Universe. Nevertheless, the LLSs are well-defined observationally (they are caused by structures with integrated \HI\ column densities large enough to produce optically thick absorption at the Lyman edge) after the reionization of the diffuse IGM is completed, and therefore one can extrapolate their abundance into the reionization epoch and draw a distinction between these absorbers and hydrogen atoms
closer to the mean density. The key physical difference is that LLSs trace high density regions in the outskirts of galaxies \citep{fumagalli11,shen13} that occupy a small portion of the volume and are able to keep a significant fraction of their hydrogen in neutral form for a Hubble time. 
By contrast, a comparable amount of hydrogen gas in the diffuse IGM is photoionized almost instantly.\footnote{For example, 
a region of physical size 100 kpc at the mean density contains as much gas as a 10 kpc LLS with overdensity $\Delta=1,000$. A typical ionization front moving with 3,000 km s$^{-1}$ would ionize such diffuse region in 30 Myr.}

\begin{figure}
\includegraphics[width=\hsize]{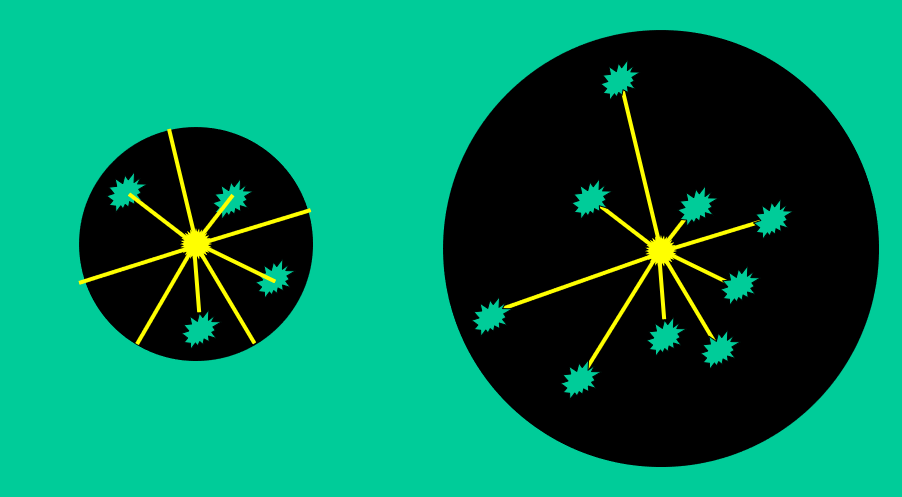}
\caption{A cartoon illustration of the relationship between the sizes of \HII\ bubbles and the LLSs they contain. The green color marks neutral gas, while ionized gas is in black. The small ionized bubble on the left keeps expanding, as many LyC photons (yellow rays, emitted from the source at the center) are absorbed by the surrounding neutral gas rather than by the few LLSs 
it contains. The large bubble on the right has instead stopped expanding, as it contains many LSSs that 
determine the mean free path of LyC radiation.}
\label{fig:sketch}
\end{figure}

On quite general grounds, one expects that a fraction of the ionizing radiation emitted by individual sources will leak into intergalactic space  and carve out cosmological \HII\ regions (more commonly colloquially called ``ionized bubbles'') in their immediate vicinity. These isolated ionized bubbles will expand in an otherwise largely neutral phase as more LyC photons are produced, and will start overlapping -- in a ``swiss-cheese'' reionization topology -- when their typical size becomes comparable to the mean distance between highly clustered groups of sources; for galaxies this would be comparable to their correlation length, about 5 Mpc \citep{Barone-Nugent2014}.
Merged bubbles will contain more radiation sources, and the ionization fronts delimiting their boundaries will expand faster as a consequence. A faster expanding 
bubble will rapidly engulf more sources and expand even faster, in a runaway process that continues until the mean free path of LyC photons becomes effectively 
controlled by the LLSs rather than by the typical size of \HII\ bubbles (see Fig. \ref{fig:sketch}). There must therefore be a short time interval when the mean free path increases sharply (and so does the UV background intensity) causing a sudden drop in the neutral fraction. This unambiguous ``epoch of overlap'' can then be defined mathematically as the moment when the LyC mean free path increases at the fastest rate \citep{Gnedin2004}, a physically-motivated definition that is preferable to some arbitrary chosen thresholds for the mass- or volume-weighted hydrogen neutral fractions.

\begin{figure}
\includegraphics[width=\hsize]{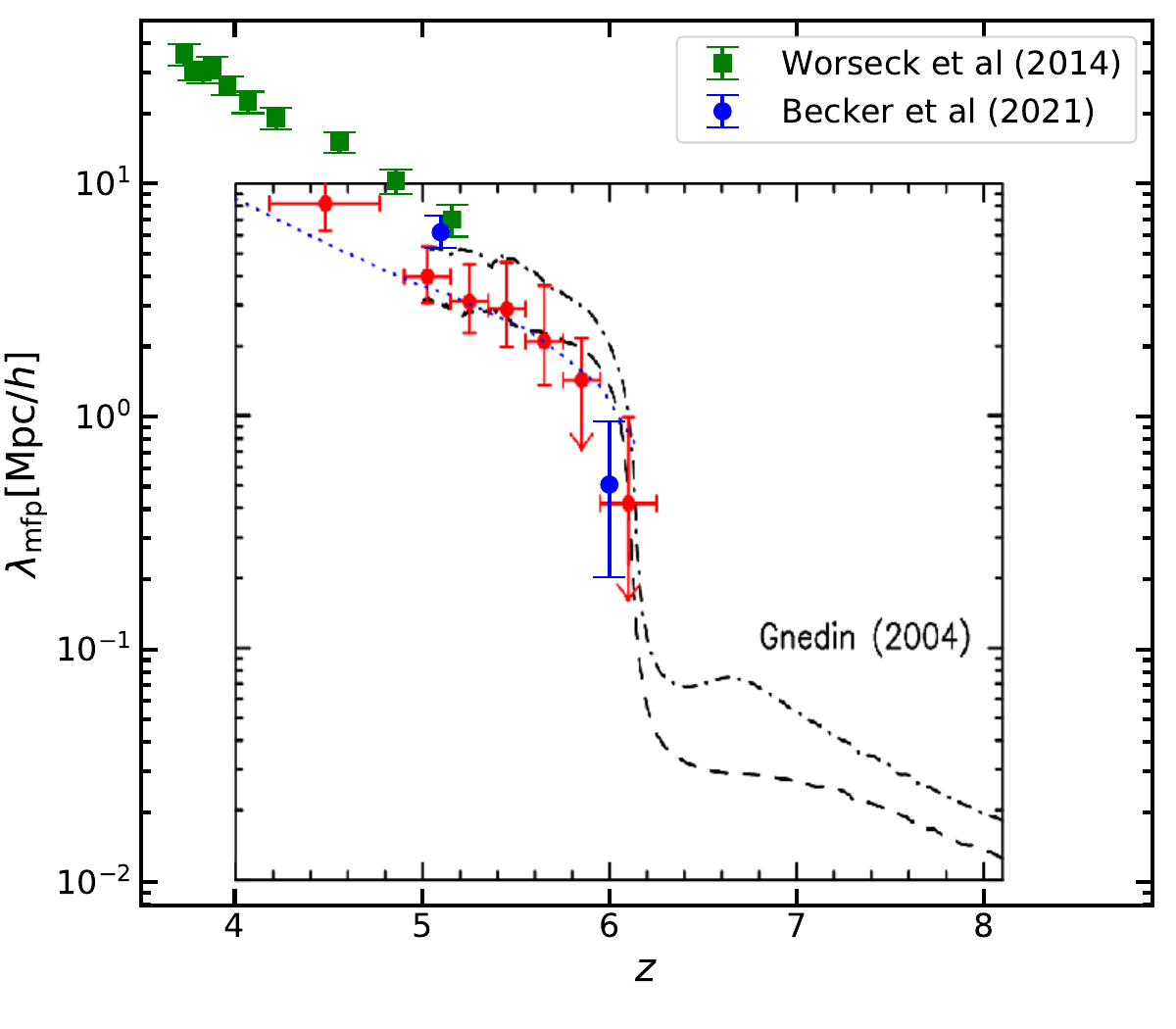}
\caption{The redshift-dependent $\lambda_{\rm mfp}$ of H-ionizing radiation measured from quasar spectra. Figure 8 from \citet{fan06} is overplotted with more recent observational determinations from \citet[][green]{worseck14} and \citet[][blue]{becker21}. The plot from \citet{fan06} also shows the evolution of the mean free path in the early simulations of \citet{Gnedin2004}. Note how the hightest-redshift value lie well below extrapolations from lower redshifts, a rapid decrease in the mean free path that is  qualitatively consistent with models wherein reionization ends at $z\sim 6$. In the units on the $y-$axis, the symbol $h$ denotes the Hubble constant in units of 100 km s$^{-1}$ Mpc$^{-1}$. (Used by permission from Xiaohui Fan.)
}
\label{fig:mfp}
\end{figure}

Under some assumptions, the LyC mean free path can be inferred from the absorption spectra of high redshift quasars. Two recent measurements are shown in Fig.~\ref{fig:mfp} together with the first determinations  by \citet{fan06}. While more observations are needed to fully constrain the evolution of the IGM opacity at these redshifts, the existing data already hint at the rapid decrease in the mean free path expected for 
a epoch of overlap at $z\sim 6$. There is one exception to the above scenario, and that is when the number density of LLSs is so high that the growth of \HII\ regions stops before overlap and overlap never actually occurs. The process of subsequent reionization (and whether it completes at all) is then controlled entirely by the evolution of the LLS  abundance. As the observational data in Fig.~\ref{fig:mfp} indicate, this situation is not realized in our Universe.

\section{Analytical Models of Reionization}
To zeroth order, tracing the history of cosmic reionization is mainly a photon counting exercise, where the growth rate of \HII\ regions is equal to the rate at which ionizing photons are produced minus the rate at which they are consumed by radiative recombinations. 
Most of the existing analytical tools, while appearing rather dissimilar from each other, apply in essence 
the same bookkeeping algorithm to different spatial scales. For example, one of the long running historical 
debates in the topology of reionization is whether such process proceeds ``inside-out'' (high density regions 
are ionized first, and lower-density regions follow, \citealt{Furlanetto2004})  or ``outside-in'' (lower density regions are ionized  first, \citealt{miralda00}). While the two approaches have been often presented as mutually exclusive, we shall show in this  section that they are, in fact, just two different stages of the same ``inside-out-back in'' process.

\subsection{The ``reionization equation''}

Let us start then with counting the number of hydrogen ionizing photons present in an arbitrary \emph{comoving} patch of the Universe (without 
regard to its size, shape, gas density, or other physical properties). This patch is fully ionized when the number of photons with energies above 13.6 eV, $N_{\gamma}$, either emitted inside the region or arriving into it from external sources, is equal or exceeds the number of hydrogen atoms $N_\H$ in the patch, $N_{\gamma} \geq N_{\H}$. If this condition does not hold the patch is only partially ionized, with $N_{\HII} < N_{\H}$ and
\[
N_{\gamma} = N_{\HII}.
\]
This is not enough, however, since atoms can recombine following the absorption of  UV radiation, and must be photoionized again. Hence, our ``master equation'' should also include photon losses from radiative recombinations,
\begin{equation}
  N_{\gamma}(t) = N_\HII(t) + \int_0^t N_\HII(t^\prime) \frac{dt^\prime}{t_{\rm rec}(t^\prime)},
  \label{eq:master}
\end{equation}
where $t_{\rm rec}$ is the recombination time inside our patch. Reionization is complete when $N_{\HII} = N_{\H}$ in every region of the Universe that is close (in some particular sense) to the mean cosmic density. Averaging now our master equation over a representative large comoving volume of the Universe $V$, denoting this average with angle brackets, and differentiating with respect to time, one gets
\begin{equation}
  \frac{1}{V}\frac{d}{dt}\langle N_{\gamma}\rangle = \frac{1}{V} \frac{d}{dt}\langle N_\HII\rangle + \frac{1}{V}\left\langle \frac{N_\HII}{t_{\rm rec}}\right\rangle.
\end{equation}
For historical reasons, the universe-average ionized fraction is often labeled with the symbol $Q$,
\[
  Q \equiv \lim_{V\rightarrow\infty} \frac{N_\HII}{N_\H} = \frac{\langle n_\HII\rangle}{\langle n_\H\rangle}.
\]
Rearranging terms, one recovers the original ``reionization equation'' of \cite{madau99}:
\begin{equation}
  \frac{dQ}{dt} = \frac{\dot n_{\rm ion}}{n_{\rm H,0}} - \frac{Q}{\bar t_{\rm rec}},
  \label{eq:dqdt}
\end{equation}
where $\bar t_{\rm rec}$ is an ``effective'' recombination timescale and $n_{\rm H,0} = 1.9\times 10^{-7}\,\cc$ is the mean comoving cosmic density of hydrogen atoms. Here, we have written the comoving volume-averaged emission rate into the IGM of ionizing photons as 
\begin{equation}
{\dot n}_{\rm ion} = \frac{1}{V} \langle \dot{N}_{\gamma}\rangle
=\fesc \xi_{\rm ion} \rho_{\rm SFR}.
\label{eq:dotnion}
\end{equation}
In reionization studies, it is convention to include in the ionization balance equation only a fraction of the hydrogen photoionizations and radiative recombinations, i.e. those that take place in the diffuse, low-density IGM. Photons that are absorbed locally, within the immediate high-density vicinity (the ISM and the CGM) of ionizing sources, are accounted for by a reduction of the source term via the escape fraction parameter $f_{\rm esc}$ (see also Section 2.5). 
The third equality in Equation (\ref{eq:dotnion}) assumes  a galaxy-dominated ionizing emissivity, with  $\rho_{\rm SFR}(t)$ being the cosmic star formation rate density,
$\xi_{\rm ion}$ the appropriately-averaged number of LyC photons emitted per unit SFR (a ``yield'' determined by stellar physics and the shape of the initial mass function), and 
$f_{\rm esc}$ the fraction of such photons that leaks into the IGM.

Some care must also be exercised with the photon sink term in 
Equation (\ref{eq:dqdt}). Since the recombination rate is quadratic in density, it depends on the actual
gas density distribution within the volume and cannot be expressed via globally averaged quantities only. It is 
customary then to introduce a ``recombination clumping factor'' $C_\HII$ and write
\begin{equation}
    \frac{1}{\bar t_{\rm rec}} \equiv \frac{\langle n_\HII n_e\alpha^A_\HII \rangle}{\langle n_\HII\rangle} = 
    \langle \alpha^A_\HII \rangle \langle n_e\rangle C_\HII,
\end{equation}
where 
\begin{equation}
    C_\HII \equiv \frac{\langle n_\HII n_e\alpha^A_\HII \rangle}{\langle n_\HII\rangle \langle n_e\rangle \langle \alpha^A_\HII \rangle}
    \label{eq:crec}
\end{equation}
is in general time-dependent. 
Neglecting density-dependent temperature fluctuations, which can impact the recombination coefficient, and assuming 
$n_e\propto n_\HII$, one can rewrite the clumping factor in the more familiar form
$C_\HII=\langle n_\HII^2\rangle/\langle n_\HII\rangle^2$. 

The clumping factor
is an external parameter that must be computed with the help of cosmological hydrodynamics simulations \citep{kohler07,pawlik09,finlator12,shull12,kaurov14,so14}. It has to be chosen appropriately to avoid {\it double counting}, since the impact of recombinations within the host halos of ionizing sources is already implicitly incorporated as a reduction in the source term through the escape fraction $f_{\rm esc}$. For example, all volume elements denser than a given gas overdensity threshold -- say $\Delta >100$ --  should be removed from calculations of the clumping factor if ionizing photons were counted as escaped into the IGM once they entered regions of gas overdensities $\Delta <100$. This means that the escape fraction and the clumping factor are not independent, but as long as the definitions of escape fraction and clumping factor refer to the same decomposition of the gas into IGM and ISM/CGM, the value of the overdensity threshold separating these different components can be chosen arbitrarily. Note, however, that the exclusion in the ionization balance equation of gas above a  certain density 
implies that the quantity $n_{\rm H,0}$ in Equation (\ref{eq:dqdt}) should not be, strictly speaking, the mean hydrogen density of the Universe but rather the redshift-dependent mean density of its diffuse component.

Equation (\ref{eq:dqdt}) above statistically describes the transition from a neutral  Universe to a fully ionized one. Extensively used in the literature \citep[e.g.,][]{haardt12,kuhlen12,bouwens15,Robertson2015,khaire16,ishigaki17,sharma17} as it allows an estimate of the photon budget required to achieve reionization with a fast exploration of parameter space, this simple ODE has been shown to provide an acceptable description of the results of radiative transfer simulations \citep{gnedin00,gnedin16}. The main limitation of this approach is that it assumes that all LyC photons escaping from individual galaxies are absorbed by the diffuse IGM, mathematically permitting unphysical values of $Q$ above unity when reionization is completed and the Universe becomes transparent, because our master equation (\ref{eq:master}) is only valid until the patch is fully ionized.\footnote{See \citet{madau17}, however, for a revised formulation where the source term in Equation (\ref{eq:dqdt}) is modified to explicitly account for the presence of the optically thick LLSs that trace the CGM \citep{crighton19} and determine the mean free path of ionizing radiation after overlap.} Another limitation of the reionization equation is that it ignores collisional ionizations. As long as these take place inside virialized halos, they can be accounted for by appropriately choosing the parameter $f_{\rm esc}$; however, collisional ionizations in filaments outside virialized halos are not included in this approach.

\begin{figure}
\centering
\includegraphics[width=\hsize]{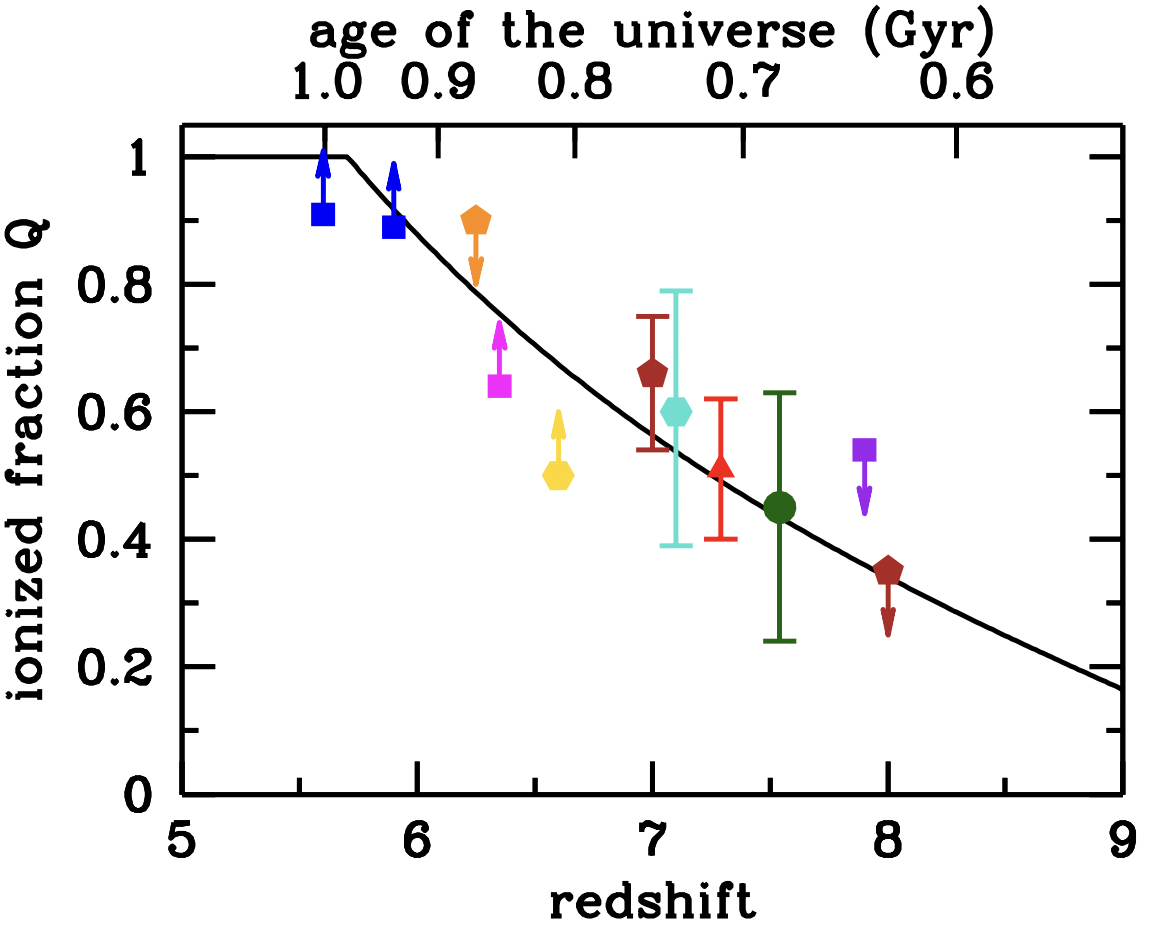}
\vskip 0.2cm
\caption{A toy model of reionization history obtained by integrating Equation (\ref{eq:dqdt}) with a constant emission rate of ionizing photons per hydrogen
atom, $\dot n_{\rm ion}/n_{\rm H,0}=2.7$ Gyr$^{-1}$.
The average ionized fraction $Q$ is shown for an IGM that recombines according to Equation (\ref{eq:trec})). The data points represent constraints from the dark pixel statistics at $z=5.6$ and $z=5.9$ (blue squares, \citealt{mcgreer15}),
the gap/peak statistics at $z=6.32$ (magenta square, \citealt{gallerani08}),
the damping wing absorption profiles in the spectra of quasars at $z=6.3$ (orange pentagon, \citealt{schroeder13})
$z=7.1$ (turquoise hexagon, \citealt{greig17,mortlock11}), $z=7.29$ (red triangle, \citealt{bradley22}),
and $z=7.54$ (green circle, \citealt{banados18}), the redshift-dependent prevalence of \Lya\ emitters (LAEs) at  $7<z<8$ (firebrick pentagons, \citealt{schenker14},
and purple square, \citealt{mason19}), and their clustering properties at $z=6.6$ (gold hexagon, \citealt{ouchi10}).
}
\label{fig:QHII}
\vspace{-0.cm}
\end{figure}

To provide an illustrative description of how the global reionization history of the Universe may proceed over cosmic time, we have numerically integrated Equation (\ref{eq:dqdt}) from $z=10$ onwards, assuming a constant emission rate of LyC photons per hydrogen atom of $\dot n_{\rm ion}/n_{\rm H,0}=2.7$ Gyr$^{-1}$. In this toy model, the IGM is defined as all gas at an overdensity threshold of $\Delta < 100$. Its effective recombination timescale is 
\begin{equation}
\bar t_{\rm rec}=2.3\,[(1+z)/6]^{-4.35}\,{\rm Gyr}, 
\label{eq:trec}
\end{equation}
a fitting formula provided by \citet{so14} and based on the analysis of a radiation hydrodynamical simulation of reionization that begins at $z=10$ and completes at $z\simeq 5.8$.
The resulting average ionized fraction, shown in Figure \ref{fig:QHII}, appears to be consistent with a number of observational constraints on the state of the $z>5$ IGM, and shows that the process of reionization is quite extended: even if it completes at redshift $\lesssim 6$, 40\% of hydrogen atoms in the Universe are ionized already at $z\sim 7.5$. Note that we have artificially set $Q=1$ when reionization is completed, 
the neutral fraction in the IGM drops abruptly, and photoionizations come into balance with radiative recombinations.

\subsection{PDF-based models (``outside-in'')}

A different treatment was presented by \citet{miralda00}, who used the volume-weighted PDF of IGM gas $P_V(\Delta)$ already discussed in Section \ref{sec:PDF}
to describe the ionization state of an inhomogeneous Universe. They assumed that inside each ionized bubble (and for the Universe as a whole after all the low density IGM is ionized) reionization proceeds gradually and ``outside-in'', first into the underdense voids and then more gradually into overdense regions. At any given epoch, all the gas with density above a critical overdensity $\Delta_i(t)$ remains neutral, self-shielded from the ionizing background, while all lower density material is completely ionized. This ansatz cannot clearly be exact, since the degree of ionization of gas at a given density will depend on the environment, but it may provide a useful approximation for understanding the reionization history of cosmic structure.
The overlap of \HII\ regions occurs then first through the lowest density ``tunnels'' found between sources, both because fewer atoms need to be ionized per unit volume and because fewer recombinations take place, while the denser, neutral gas with 
$\Delta>\Delta_i(t)$ limits the mean free path of LyC radiation to sub-horizon scales.

Given a PDF, the mass-averaged ionized fraction is
\begin{equation}
    F_M(\Delta_i)= \int_0^{\Delta_i} \Delta P_V(\Delta) d\Delta. 
\end{equation}
The comoving volume-averaged recombination rate can then be written as 
\begin{equation}
a^3\langle \alpha^A_\nHII n_e n_\nHI \rangle= \frac{n_{\rm H,0}}{t_{\rm rec,0}}\,\int_0^{\Delta_i} 
\Delta^2 P_V(\Delta) d\Delta,
\end{equation}
since only ionized low-density regions contribute to the sink term. In the last expression, $t_{\rm rec,0}$ is the recombination time at the cosmic mean density and we have again neglected the effect of the temperature-density relation on the recombination coefficient. In this formalism, the clumping factor is simply
\begin{equation}
C_\HII = F_M^{-2} \int_0^{\Delta_i} \Delta^2 P_V(\Delta) d\Delta.
\end{equation}
The critical overdensity $\Delta_i$ for a particular ionized patch of volume $V$ (which may be different in different patches) is then determined implicitly by our master equation (\ref{eq:master}), 
\begin{equation}
N_\gamma(t) = 
F_M\left(\Delta_i(t)\right)
+ \int\limits_0^t \frac{dt}{t_{\rm rec,0}} \int\limits_0^{\Delta_i(t)} \Delta^2 P_V(\Delta,t) d\Delta.
   \label{eq:pdfmodel}
\end{equation}
Equation (3) from \citet{miralda00} is then obtained by differentiating this equation with respect to time and dividing it by $V$ and by the Hubble parameter $H$. 
For typical $P_V(\Delta)$, the second term in Equation (\ref{eq:pdfmodel}) dominates over the first term for sufficiently large $\Delta_i$ (typically $\Delta_i\gtrsim10$ at  $z\sim6$). Hence, this class of models is commonly called ``outside-in'', with ``out'' and ``in'' referring to low and high density gas, respectively. The former gets
ionized first, while the latter (excluding the ISM in the very vicinity of UV sources) is ionized later, as $N_{\gamma}$ and $\Delta_i$ increase with time.  In the late stages of reionization, when $N_{\gamma}$ is sufficiently large, it is  recombinations in high density regions that consume most of the LyC photons and determine the rate at which reionization proceeds.

There are two main limitations of the PDF-based approach: 1) the gas density distribution is a fit to the results of cosmological hydrodynamical simulations that are sensitive to resolution, box size, as well as thermal and photoionization history. The PDF choosen by \citet{miralda00}, for example,
\begin{equation}
P_V(\Delta)=A\exp \left[\frac{(\Delta^{-2/3}-C_0)^2}{2(2\delta_0/3)^2}\right]\Delta^{-\beta},
\label{eq:pdf}
\end{equation}
was derived assuming that the initial density fluctuations form a Gaussian random field, the gas in voids is expanding at a constant velocity, and the density field is smoothed on the Jeans scale of the
photoionized gas. Here,  $A, C_0, \delta_0$ and $\beta$ are redshift-dependent best-fit parameters. This PDF has been shown to be inadequate at describing the results of more modern simulations for $\log\Delta>1$ \citep{pawlik09,bolton09}; 
and 2) only recombinations leading to the consumption of LyC
photons that escaped from the high-density ISM of star-forming regions contribute to the balance in Equation (\ref{eq:pdfmodel}).
The number of ionizing photons leaking from the ISM must still be parameterized by an escape fraction $\fesc$ which, contrary to the clumping factor, is not a derived quantity. Hence, the  ``outside-in'' model is incomplete, and should be supplemented by  an equation for $\fesc(\Delta_i)$.

PDF-based models have found wide applicability in studies 
of the post-reionization IGM as probed by observations of hydrogen absorption in the spectra of luminous high-redshift quasars. 
Ignoring the effects of peculiar velocities and of thermal broadening, the 
Gunn-Peterson \citep{gunn65} optical depth to \Lya\ scattering at redshift $z$ is given by
\begin{equation}
\tau=\frac{\pi e^2}{m_ec H(z)}f_\alpha \lambda_\alpha n_\HI, 
\label{eq:tauGP}
\end{equation}
where $f_\alpha$ is the oscillator strength of the \Lya\ transition,
$\lambda_\alpha=1216\,$\AA, $H(z)$ is the Hubble parameter, and
$n_\HI$ is the proper local density of neutral hydrogen in the IGM. In photoionization equilibrium, the neutral hydrogen fraction and therefore the optical depth depend on the local density of the IGM. For a highly ionized region with overdensity $\Delta$ in photoionization equilibrium one derives
\begin{equation}
\tau(\Delta) \simeq 9\times 10^{-4}\, (1+\chi) \frac{T_4^{-0.7} (1+z)^6}{E(z)\Gamma_{-12}} \Delta^2,
\end{equation}
where $\Gamma_{-12}$ is the hydrogen photoionization rate in units of $10^{-12}$ s$^{-1}$, $E(z)\equiv [\Omega_M(1+z)^3+\Omega_\Lambda]^{1/2}$, $\chi$ accounts for the extra electrons liberated by helium reionization, and the temperature is approximately related to the gas density via a power-law density-temperature relation of the form $T=T_0\Delta^{\gamma-1}$,
which should be valid for low-density gas \citep{hui97}. The mean normalized transmitted flux  through the IGM (which is the observed quantity) is then
\begin{equation}
\langle F \rangle=\langle e^{-\tau}\rangle=\int e^{-\tau(\Delta)} P_V(\Delta)d\Delta.
\label{eq:Flux}
\end{equation}
Thus, as long as the PDF and the temperature-density relation are  
sufficiently well understood from theory and simulations, the observed mean transmitted flux can be used to determine the 
hydrogen photoionization rate $\Gamma_\nHI$ \citep{mcdonald01,fan02}. Note that, at high redshift,  the transmitted 
flux distribution $e^{-\tau(\Delta)} P_V(\Delta)\Delta$ peaks in underdense regions with $\Delta\sim 0.3$, as denser 
gas produces saturated \Lya\ absorption with zero transmitted flux. 

\begin{figure}
\centering
\includegraphics[width=\hsize]{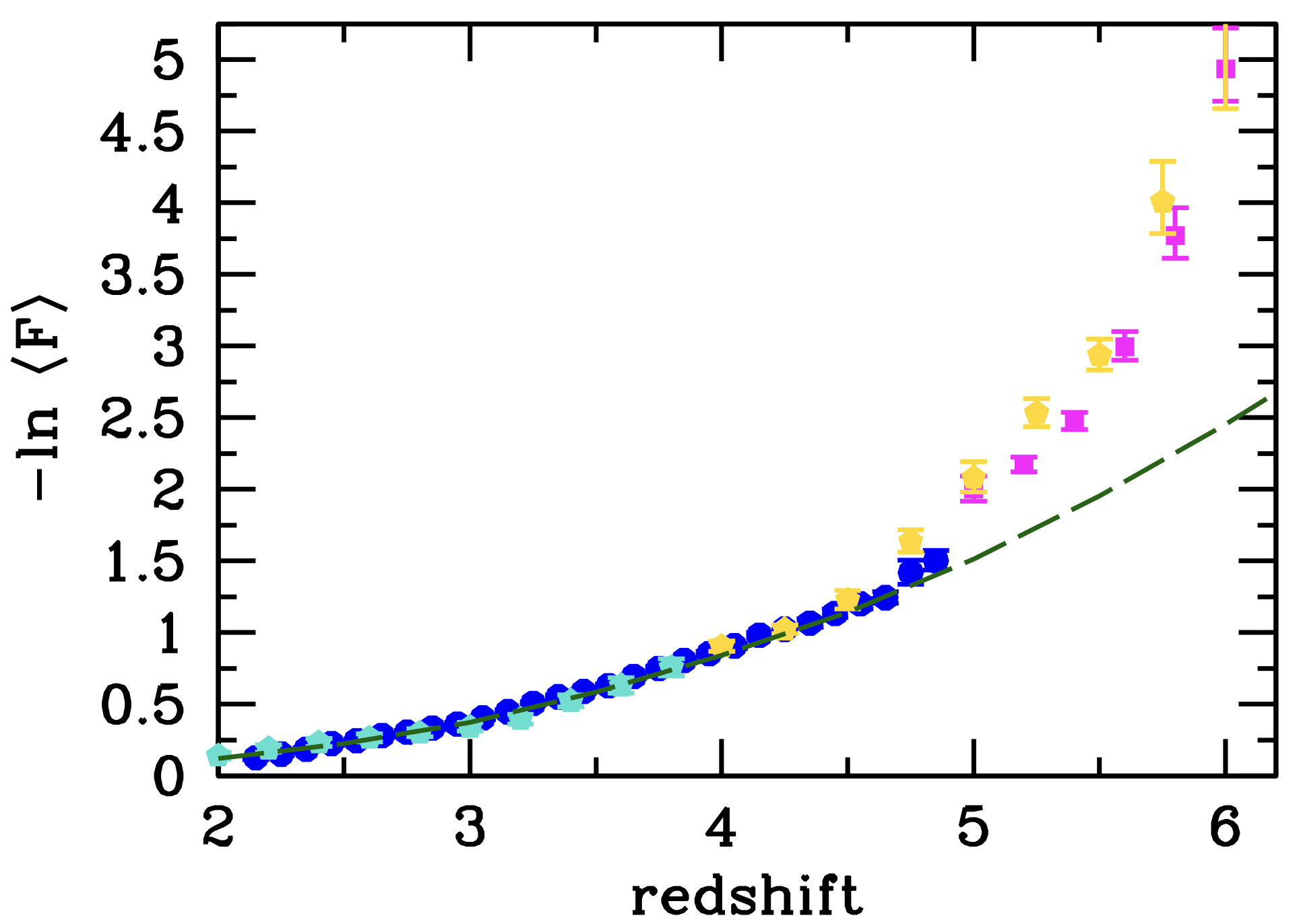}
\caption{Redshift evolution of the observed mean 
transmitted flux from \citet{becker13} (blue dots), \citet{gaikwad21} (turquoise pentagons), \citet{bosman18} (magenta squares) and \citep{eilers18} (yellow pentagons).
The dashed curve shows the prediction of a PDF-based model with constant
$T_0=10^4\,$K, $\gamma=1.3$, and $\Gamma_\nHI=10^{-12}\,$s$^{-1}$ (see text for details).
The data suggest a rapid change in the ionization state of the IGM at $z>5$ produced by a 
decrease in the ionizing ultraviolet background, an indication of the end stages of 
hydrogen reionization. 
}
\label{fig:tau}
\end{figure}

For illustrative purposes, we have calculated here the mean transmitted flux as a function of redshift using the polynomial 
fit to the PDF from \citet{bolton09}, and assuming constant $T_0=10^4\,$K, $\gamma=1.3$, and $\Gamma_\nHI=10^{-12}\,$s$^{-1}$ \citep[c.f.][]{fan06,becker13,gaikwad17,boera19,walther19,gaikwad20,villasenor22}.
Figure \ref{fig:tau} compares the effective \Lya\ optical depth, $-\ln \langle F\rangle$, predicted in the redshift  range $2\lesssim z\lesssim 6$ with the observations. Clearly, a constant hydrogen photoionization rate is unable to reproduce  the rapid increase of the IGM opacity observed at $z>5$,  suggesting  a swift transition in the ionization state of  the Universe. Both the evolution of the observed \Lya\ opacity as well as the increased scatter of the measurements \citep{becker15} are the most compelling constraints to date on the end of the epoch of reionization.

\subsection{Excursion set-based models (``inside-out'')}
\label{sec:excurs}

\begin{figure*}[th]
\centering
\includegraphics[height=0.34\textheight]{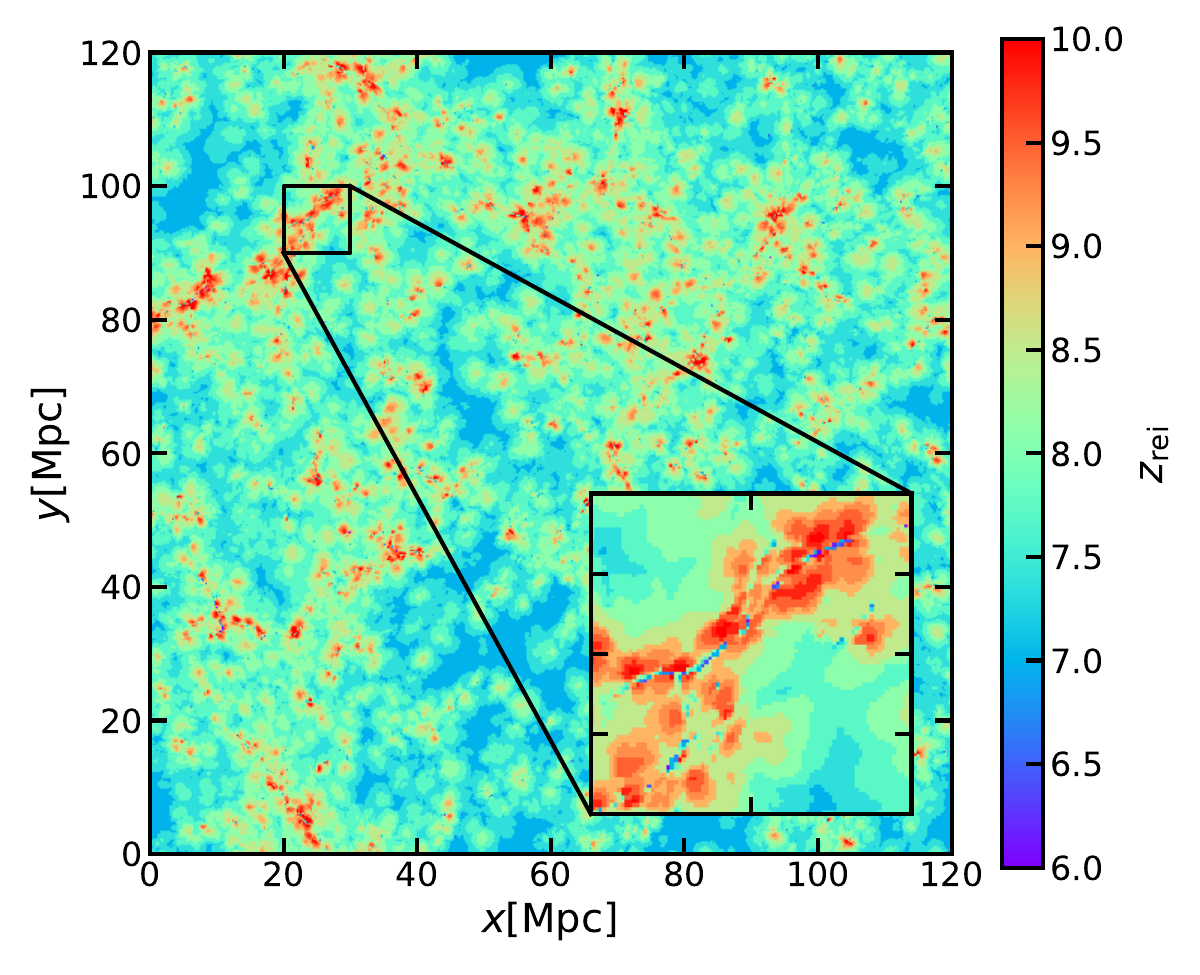}%
\includegraphics[height=0.33\textheight]{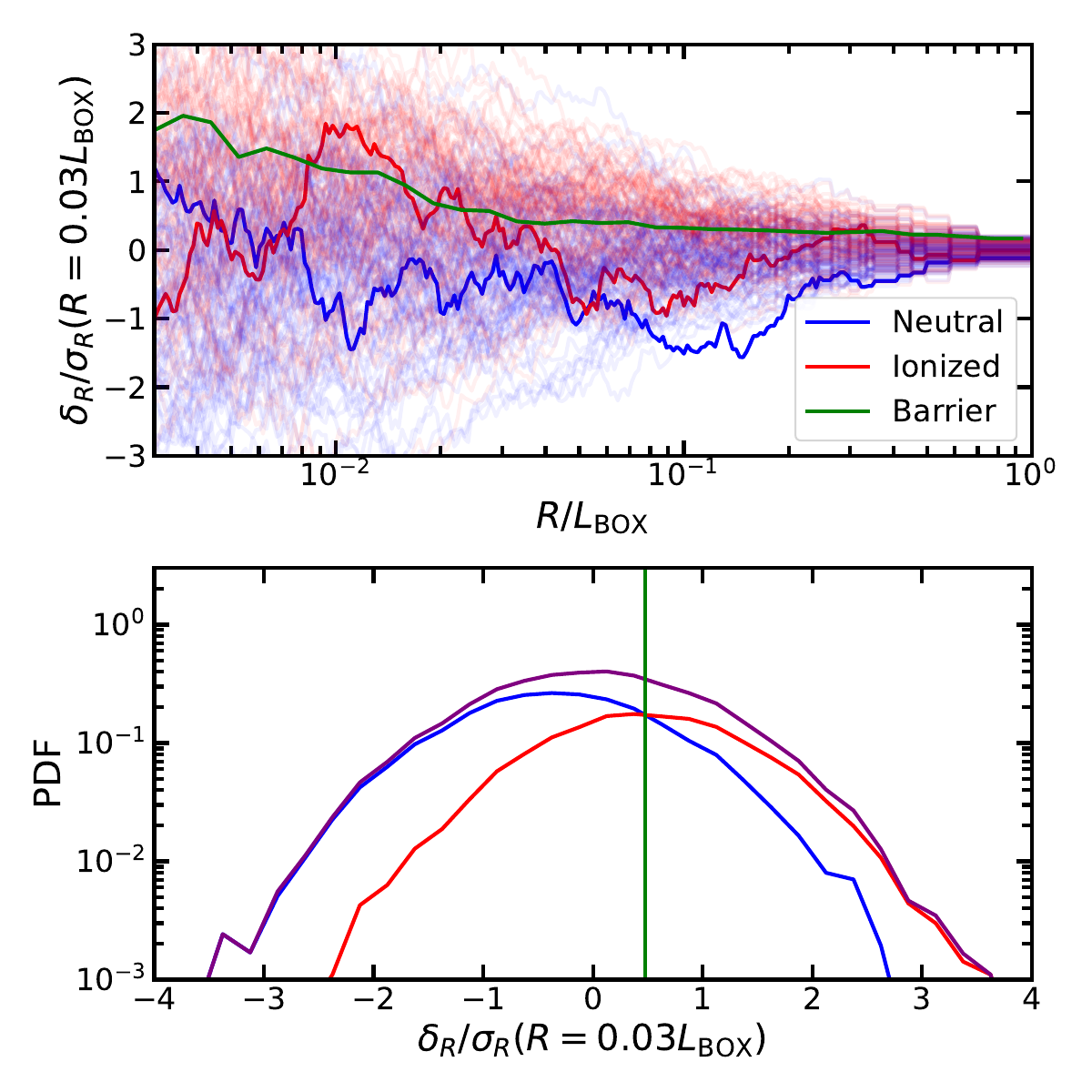}
\caption{Left panel: A thin slice through a 120 comoving Mpc simulation from the CROC suite. 
Each pixel in the image is color-coded by the value of redshift at which gas at that spatial location was 90\% ionized, with the color-bar giving the actual values for each color. The inset zooms on a region that contains both early ionized bubbles and high density gas ionized later. Right top: Trajectories for the simulation box on the left at $z=8$; blue lines (one solid, the other highly translucent) mark trajectories that are not yet ionized, read lines indicate ionized trajectories. The green line shows the reionization barrier. Right bottom: Distribution of initial overdensities on scales $R/L_{\rm BOX}=0.03$. Again blue indicate neutral gas and red ionized; purple is the sum of the two. Both distributions are broad, so the barrier is ``fuzzy''. One plausible definition of a reionization barrier is the overdensity at which the PDFs of neutral and ionized trajectories cross each other (vertical green line).}
\label{fig:traj}
\end{figure*}

While able to characterize an inhomogeneous reionization process where overdense regions are photoionized  after underdense voids, PDF-based models implicitly assume a uniform UV background throughout the Universe and do not take into account large-scale fluctuations in the density field. The success of the 
extended Press-Schechter formalism for describing the correlation of peaks in the primordial mass  distribution \citep{Press1974,Bond1991,Lacey1993} 
has prompted efforts to develop a similar approach for studies of the topology of reionization on large scales. Pioneered by \citet{Furlanetto2004} and \citet{Furlanetto2005}, such modeling has been subsequently 
used in a number of applications, most notably in the 21cmFAST code\footnote{\url{https://github.com/21cmFAST/21cmFAST}.}
for estimating the 21cm signal from neutral hydrogen during the epoch of reionization \citep{Mesinger2007,Mesinger2011}.  The key idea here is to provide a statistical representation  of the fluctuating ionization pattern on the scales 
that are most relevant to observations by connecting the number of ionizing photons present in a spherical region of  comoving radius $R$ or, equivalently, linear mass scale $M\equiv (4\pi/3) \bar \rho R^3$ (where $ \bar \rho$ is the mean density of the Universe) to the collapsed mass  fraction $f_{\rm coll}$ in that region. According to the extended Press-Schechter formalism, this is given by
\begin{equation}
    f_{\rm coll} = \mbox{erfc}\left[\frac{\delta_c-\delta_R(z)}{\sqrt{2[\sigma_{\rm min}^2-\sigma_R(z)^2}]}\right]
    \label{eq:fcoll}
\end{equation}
for Gaussian fluctuations on the scale $R$, where $\delta_R(z)$ and $\sigma_R(z)$ are the linear overdensity in a region of size $R$ and the rms density fluctuation in all regions of size $R$ respectively,
$\delta_c$ is the linear-theory density threshold for collapse (equal to $\simeq 1.686$ for spherical collapse), and $\sigma_{\rm min}$ is the 
rms density fluctuation on the smallest scale that can harbor an ionized source. It can be interpreted as the spatial scale $R_{\rm min}$ corresponding to the mass scale $M_{\rm min}$ for the minimum mass of an ionizing source, which, in turn, may correspond to atomic-cooling halos of virial temperature $10^4$ K.\footnote{A pre-factor of 2 commonly added to this equation to achieve $f_{\rm coll}\rightarrow 1$ at $t\rightarrow\infty$ can always be incorporated into the $\zeta$ factor of Equation (\ref{eq:zeta}).} \citet{Furlanetto2004} postulated that the number of hydrogen ionizing photons per hydrogen atom in the region is
\begin{equation}
    \frac{N_\gamma}{N_\H} = \zeta f_{\rm coll}, 
    \label{eq:zeta}
\end{equation}
where the factor $\zeta$ depends on uncertain source properties like the efficiency of LyC photon 
production, the escape fraction of these photons from the host galaxy, and the star formation efficiency, and  can be a function of mass $M$. Then, from our master equation (\ref{eq:master}), the region under investigation is fully ``self-ionized'' when it contains enough mass in luminous sources, i.e. when 
\begin{equation}
  \zeta f_{\rm coll} = 1 + \int_0^t x_{\HII}\frac{dt}{t_{\rm rec}}.
  \label{eq:barrier}
\end{equation}
This, together with equation (\ref{eq:fcoll}), places a condition $\delta_R>\delta_b(R,z)$ on the mean overdensity within an \HII\ region of size $R$, where $\delta_b(R,z)$ is the threshold or ``ionization barrier''. At any given (randomly chosen) position in space the evolution of $\delta_R$ as a function of $R$ resembles a random trajectory. \citet{Furlanetto2004} used the first-crossing distribution of random walks with a ``moving'' (because 
it is a function of the spatial scale $R$) barrier to model the  size distribution of \HII\ bubbles during reionization.

A practical implementation of the barrier-crossing technique is as follows. Consider a spatially variable field of density fluctuations $\delta(\vec{x})$ (it can be either a linear Gaussian field or an actual nonlinear density field) at a given time in cosmic history, and an arbitrary position $\vec{x}_0$. Let $\delta_R(\vec{x})$ be the density field smoothed with a particular spatial filter of size $R$ (the specific shape of the spatial filter is usually not very important). A sequence of values $\delta_R(\vec{x}_0)$ as a function of $R$ forms a ``trajectory'' in the $\delta_R-R$ plane. The largest value of $R$ for which $\delta_R(\vec{x}_0)=\delta_b(R)$ is called the ``first barrier crossing''. The \citet{Furlanetto2004} model then considers the sphere of size $R$ centered on $\vec{x}_0$ as fully ionized, since the number of ionizing photons per baryon inside that sphere exceeds the threshold value in Equation (\ref{eq:zeta}). A location for which $\delta_R<\delta_b(R)$ for any $R$ is considered to be neutral at the cosmic time under consideration. Note that, if the recombination term in Equation (\ref{eq:barrier}) is omitted (as in \citealt{Furlanetto2004}), and $\zeta$ is assumed to be a constant, then every atom in the Universe eventually becomes ionized -- $f_{\rm coll}$ is an increasing function of time, and thus eventually $f_{\rm coll}$ exceeds $1/\zeta$ at every location and reionization 
ends. In reality recombinations are not negligible at high enough densities, so as $f_{\rm coll}$ increases so does the second term in Equation (\ref{eq:barrier}). Eventually recombinations win, and a few percent of the gas mass remains neutral even after reionization, locked in the LLSs.

\begin{figure}[t]
\includegraphics[width=\hsize]{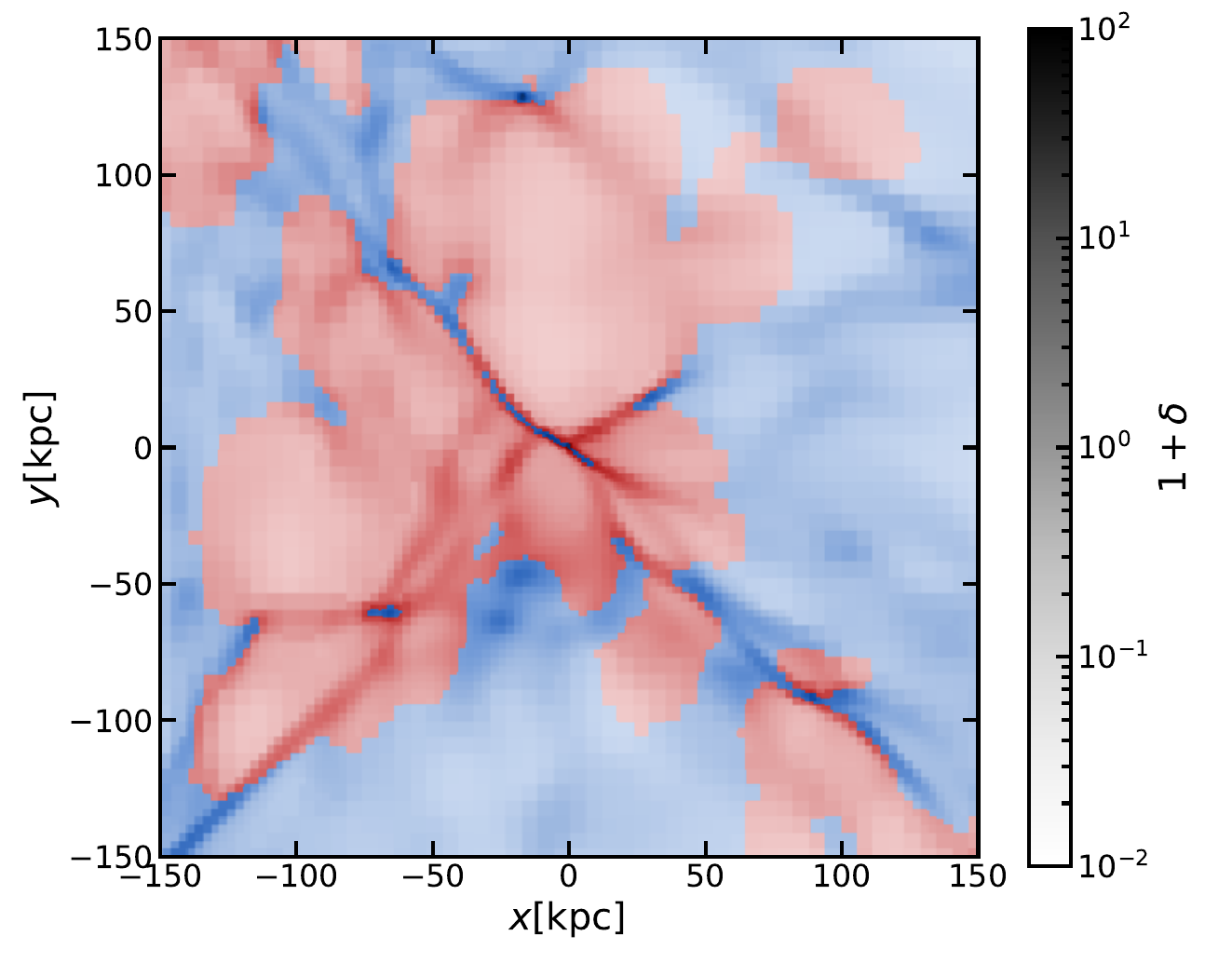}
\caption{Density map shown as intensity (the values of overdensity in the color bar range from $10^{-2}$ to 100) around a massive galaxy in one of the CROC simulations at $z=9$. The red and blue colors show ionized and neutral gas, respectively. Filaments remain neutral long after nearby voids become fully ionized.\label{fig:ifront}}
\end{figure}

\begin{figure*}[t]
\centering
\includegraphics[width=0.5\hsize]{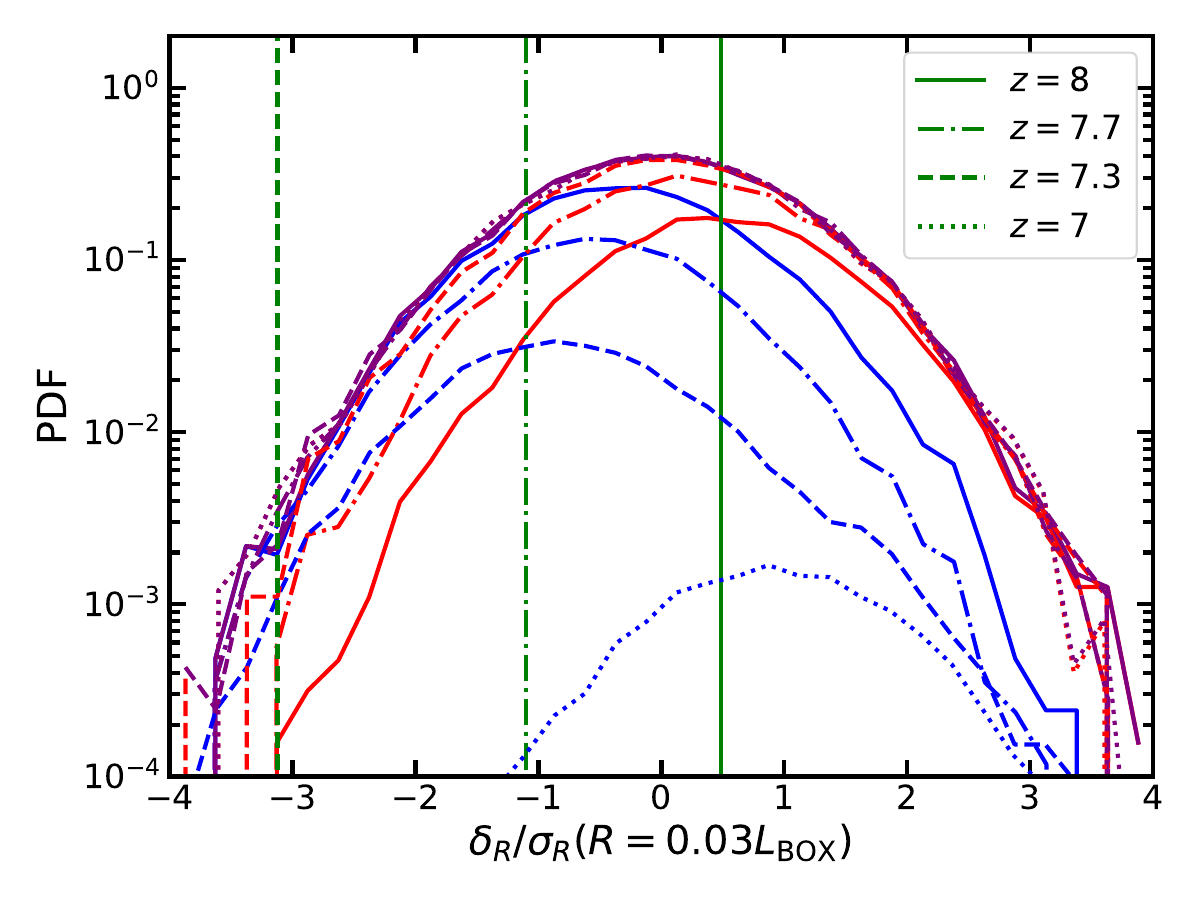}%
\includegraphics[width=0.5\hsize]{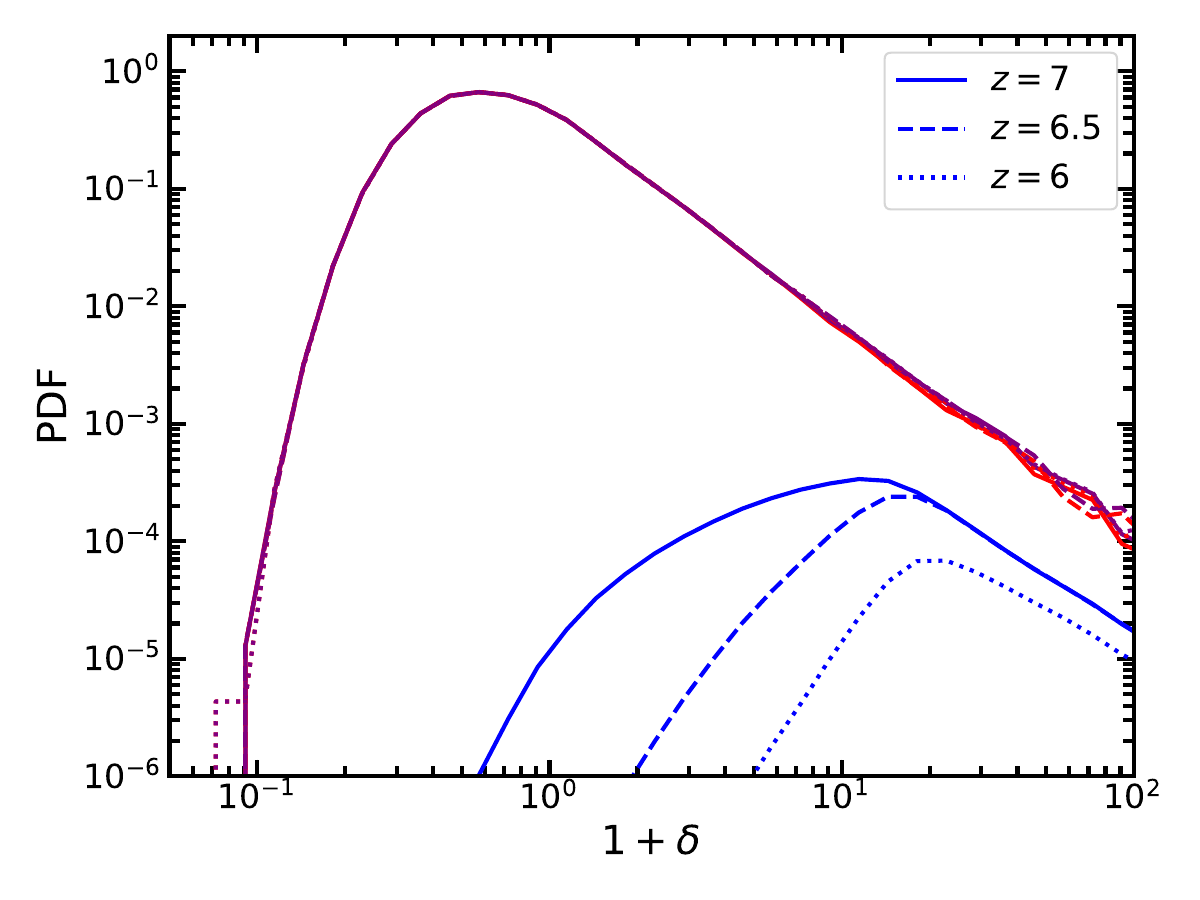}
\caption{Left: Same as the bottom right panel of Fig.\ \ref{fig:traj} at four different redshifts. In this particular realization,  the lowest density gas is ionized at $z=7.3$, and at lower redshifts the reionization barrier cannot be meaningfully defined. Right: Same PDFs at lower redshifts, but now as function of the total gas density. The distributions of neutral gas have well defined peaks ($\Delta_i$ from Eq. \ref{eq:pdfmodel}), although they are not cut off sharply at lower densities, so  $\Delta_i$ is also somewhat ``fuzzy''.}
\label{fig:traj2}
\end{figure*}

Figure \ref{fig:traj} illustrates this model in practice, following the original analysis of \citet{Kaurov2016}. The left panel shows the field of reionization redshifts from a numerical simulation of the ``Cosmic Reionization On Computers'' (CROC) suite (described below). For each pixel, we calculated the moment it reached a 90\% ionization level weighted by mass -- the vast majority of the cells cross this threshold only once. The top right panel depicts a sample of trajectories (or random walks) at $z=8$ (about the midpoint of reionization) from the Gaussian random field initial conditions for this simulation: those corresponding to the locations of gas that remains neutral ($z_{\rm REI}<8$) are colored in blue, while the trajectories of cells that become ionized are shown as red lines. The PDF for neutral gas trajectories at a given value of the smoothing scale $R$ (using a sharp-$k$-space filter) is plotted in the right bottom panel. The distribution is broad and exhibits no sharp features, implying that the barrier is ``fuzzy''. The ratio between the PDF of neutral trajectories and the PDF of all trajectories (i.e.\ the PDF of all overdensities $\delta$) is the cumulative barrier -- if the barrier was infinitely sharp, that ratio would be a Heaviside function $\theta[\delta_b(R)-\delta]$. Hence, choosing a value of $\delta$ where this ratio is $1/2$ serves as a proxy for the barrier. Thus defined barrier is shown in the top right panel as the green line.

The fact that the barrier is ``fuzzy'' should not come as a surprise. Figure \ref{fig:ifront} shows the ionized bubble around a massive galaxy from one of the CROC simulations at $z=9$. Since ionization fronts propagate much faster into  voids than along dense filaments (the I-front speed is proportional to $n_\H^{-1/3}$), the spatial distribution of ionized gas would not correlate well with the density distribution on all but the largest spatial scales. Hence, using density as a ``predictor'' for the ionization state of the gas is going to be inaccurate.

Figure \ref{fig:traj2} shows how the reionization barrier evolves towards later epochs in this particular realization. By $z=7.3$ the barrier reaches the lowest gas densities, and ceases to be meaningfully defined. At even lower redshifts the distribution of neutral gas shifts toward the highest densities, which are poorly represented by the corresponding values in the linear Gaussian  field. The right panel in the figure shows that the actual non-linear PDF of neutral gas has a characteristic overdensity value -- $\Delta_i$ from Eq.~(\ref{eq:pdfmodel}), and therefore the outside-in model provides an approximate description for the evolution of the remaining neutral gas. It does fail, however, in its main assumption that all gas above $\Delta_i$ is neutral -- in fact, most of the gas at any density is ionized, because radiation sources are highly clustered, and hence disproportionately populate the highest density regions.

\section{Numerical Models of Reionization}

\subsection{Semi-numerical models}

In this category we place techniques that do not require accurate knowledge of the full nonlinear matter distribution from numerical simulations. Information about the spatial distribution of \HII\ regions is obtained directly from the Gaussian random field of the initial conditions or their direct derivative, such as the second order perturbation theory.

\subsubsection{21cmFAST}

\begin{figure*}
\centering
\includegraphics[width=0.99\hsize]{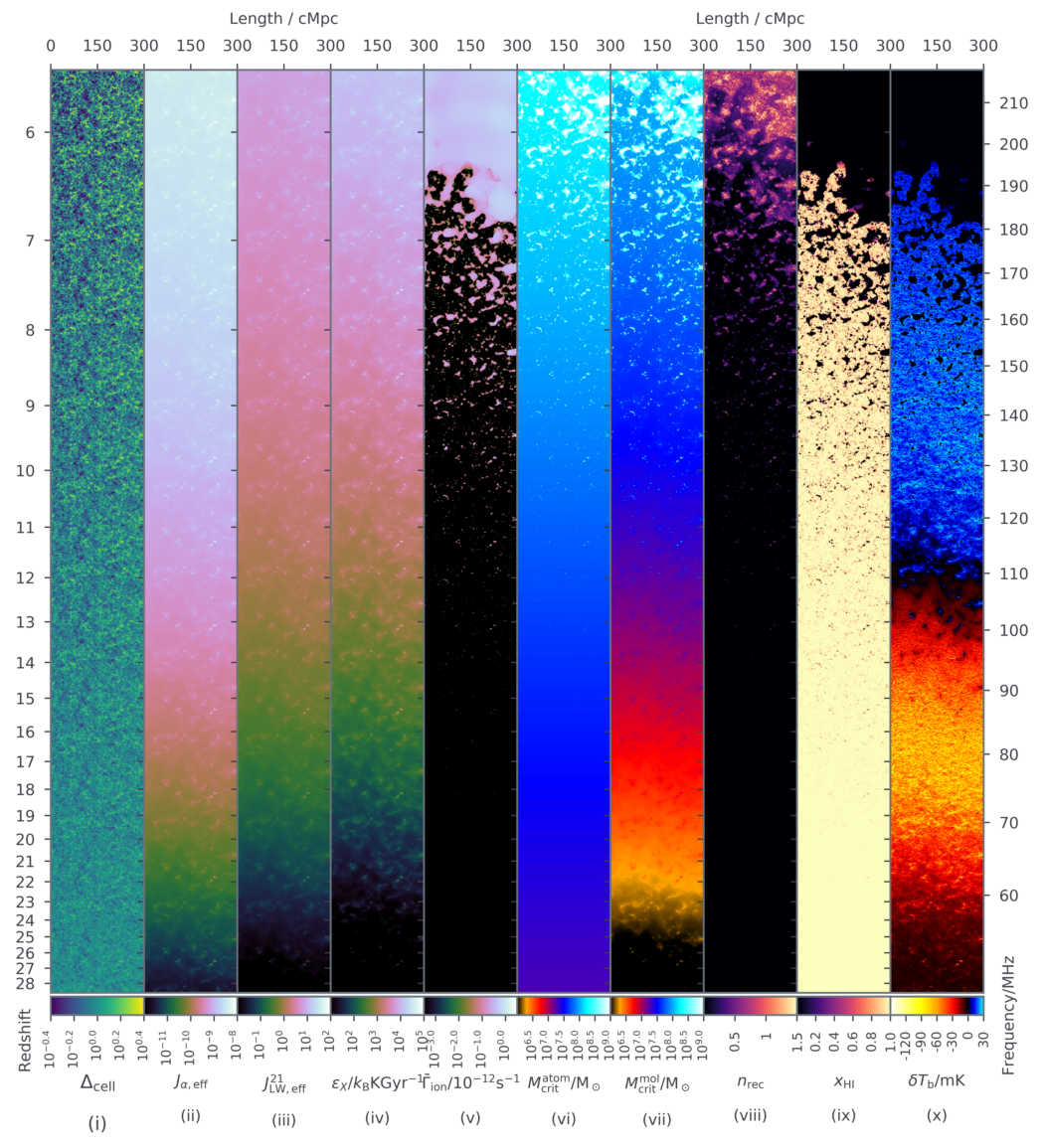}
\caption{\small An example of outputs from the latest version of the 21cmFAST code. Maps of several physical quantities (such as gas overdensity, neutral fraction, number of recombinations, etc) can be generated along the lightcone. (Used by permission from A.\ Mesinger, \url{http://homepage.sns.it/mesinger/Media/lightcones_minihalo.png}).
\label{fig:21cmfast}}
\end{figure*}

The most widely used in this class of models is perhaps the 21cmFAST code \citep{Mesinger2011}. In its original formulation it uses the excursion set-based formalism of \citet{Furlanetto2004} to create a realization of the distribution of ionized gas at any given redshift from equations (\ref{eq:fcoll}) and (\ref{eq:zeta}). The actual density field used in equation (\ref{eq:fcoll}) is obtained from the Zel'dovich approximation (a historical alias to first order Lagrangian perturbation theory). 

One of the most undesirable features of excursion set-based  
algorithms is that they do not explicitly conserve photons -- when two \HII\ bubbles overlap, LyC photons from both bubbles are double-counted in setting the ionization field \citep{zahn11,Paranjape2016,Hutter2018}. The 21cmFAST code includes a special correction to overcome this deficiency \citep{Park2021}.
The latest version of the code is described in \citet{Murray2020}. It is currently the de-facto standard for making large-scale maps of various reionization-related fields (shown in Fig. \ref{fig:21cmfast}), including creating mock maps for current and future 21cm cosmology experiments.

The excursion set formalism can also be used in ``reverse'', to model the distribution of neutral gas -- in particular, the distribution of neutral islands at the end of reionization. Since we showed above that the excursion set formalism loses its accuracy close to the overlap, capturing neutral islands at this epoch is more involved then just ``flipping the sign'' in the 21cmFAST predictions. Such an extension for 21cmFAST has been implemented in the ``IslandFAST'' code \citep{Xu2017}, which agrees well with the results of fully coupled simulations.

\subsubsection{ARTIST-like schemes}

The lack of explicit photon conservation in the excursion set-based models has motivated the development of alternative approaches that explicitly enforce such conservation. One example  is the ``Asymmetric Radiative Transfer In Shells Technique'' \citep[ARTIST,][]{ARTIST}, which solves the actual radiative transfer equation by ray-tracing (see Section \ref{sec:ray}) on a  density field obtained with the excursion set formalism. To make the calculation computationally feasible, ARTIST propagates the radiation emitted by each source in only a small number of directions, and explicitly achieves photon conservation in finite-size segments of the solid angle. 

The comparison between ARTIST and excursion set-based  models 
with the same radiation sources shows significant differences in the ionization history or in the 21cm power spectrum. This unfortunately implies that these approaches are not robust, and contain internal unidentified biases \citep{ARTIST}. On the other hand, one can adjust source parameters such as $f_{\rm esc}$, treating 
them as fudge factors rather than physical quantities, in order to match more closely the results of the different algorithms.

\begin{figure*}[th]
\includegraphics[width=\hsize]{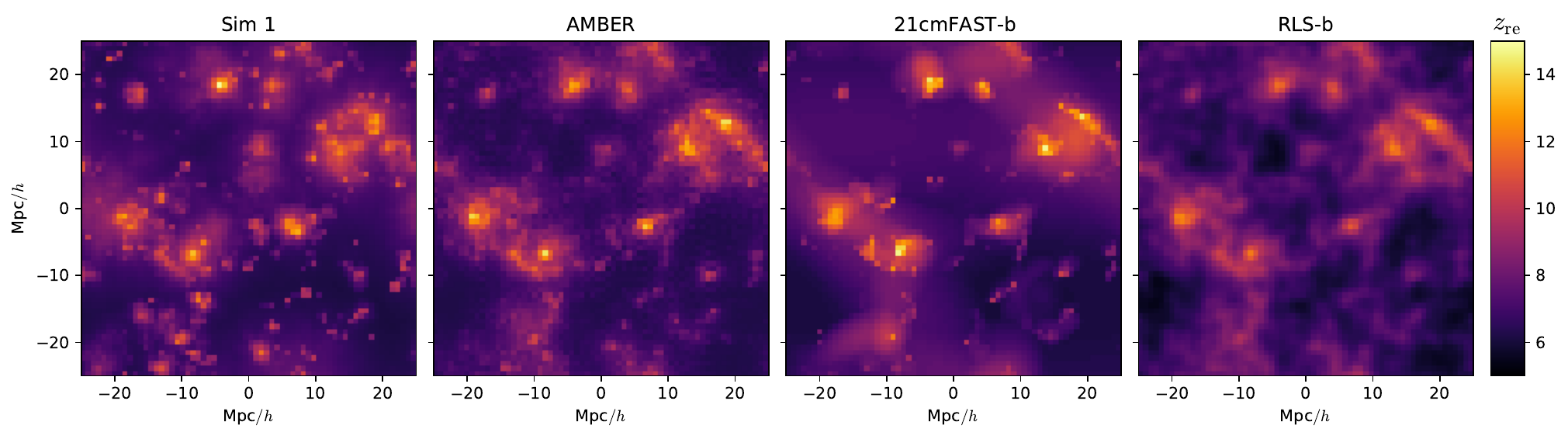}
\caption{Visualization of the reionization-redshift fields from a full radiative transfer simulation, AMBER, 21cmFAST, and ``The Reionization on Large Scales'' (RLS) scheme from \citet{Battaglia2013}. AMBER offers the best approximation to the full numerical simulation out of these three approximate techniques. 
In the units on the $x,y$ axes, the symbol $h$ denotes the Hubble constant in units of 100 km s$^{-1}$ Mpc$^{-1}$.
(Used by permission from Hy Trac.)}
\label{fig:amber}
\end{figure*}

The ``Semi-numerical Code for ReionIzation with PhoTon 
Conservation'' \citep[SCRIPT,][]{SCRIPT} is an intermediate approach between what we call a ``semi-numeric technique'' and a ``decoupled simulation''. It can use either the density field from a dark matter-only (DMO) simulation or the one generated from Gaussian random initial conditions with, say, perturbation theory. The ionization state is modeled with a quasi-ray-tracing algorithm, in which space is divided into cells and cells that contain sources are ionized first, followed by their neighbors, then neighbors-of-neighbors, etc. We list SCRIPT in this section because it is conceptually very similar to ARTIST when used with the density field generated from the perturbation theory. 

\subsubsection{AMBER}

The ``Abundance Matching Box for the Epoch of Reionization'' \citep[AMBER,][]{AMBER} is an entirely different approach for
modeling cosmic dawn. It is based on an ``abundance matching'' hypothesis that there exists a monotonic relation between a base field $\beta(\vec{x})$ and the reionization-redshift field at a given location $z_{\rm rei}(\vec{x})$ (see, e.g.,  example, 
the left panel of Fig.\ \ref{fig:traj}).
The functional form of $z_{\rm rei}(\beta)$ can be deduced from the assumed ionization history. 

The accuracy of such an ansatz depends critically on the choice for the base field $\beta(\vec{x})$. \citet{AMBER} showed that using density as the base field is a poor choice, a conclusion that is generally consistent with our discussion in Section \ref{sec:excurs}.
A much better choice for $\beta$ is an approximation to the ionizing radiation field. The formal solution to the radiative transfer equation for the radiation angle-averaged intensity $J_\nu$ is 
\begin{equation}
    J_\nu(\vec{x}) = \frac{a}{4\pi}\int d^3x_1 \frac{S_\nu(\vec{x}_1)}{(\vec{x}-\vec{x}_1)^2} e^{-\tau_\nu(\vec{x},\vec{x}_1)},
\label{eq:rtsol}
\end{equation}
where $S_\nu$ is the source function and 
the $1/(4\pi r^2)$ term is the inverse-square law for the flux.
Since this equation is not a convolution, it cannot be easily evaluated on an arbitrary mesh or particle distribution and would require using some of the methods for solving the radiative transfer equation discussed in Section \ref{sec:rtalgos}.

There is, however, an \emph{ansatz} that permits  a fast -- $O(N)$ or $O(N\log N)$ -- evaluation for an arbitrary set of $N$ target points $\vec{x}$, and which, on average, gives the correct value for the radiation field:
\[
  \tau_\nu(\vec{x},\vec{x}_1) \rightarrow \frac{1}{\lambda_{\rm mfp}(\nu)}\|\vec{x}-\vec{x}_1\|.
\]
Equation (\ref{eq:rtsol}) then becomes a convolution,
\begin{equation}
    J_\nu(\vec{x}) = \frac{a}{4\pi}\int d^3x_1 \frac{S_\nu(\vec{x}_1) }{(\vec{x}-\vec{x}_1)^2} e^{-\frac{\|\vec{x}-\vec{x}_1\|}{\lambda_{\rm mfp}(\nu)}},
\end{equation}
which can be solved in $O(N)$ operations for $N$ spatial locations.

Figure~\ref{fig:amber} replicates Fig.~17 from the AMBER method paper \citep{AMBER}, and shows a comparison between a full radiative transfer simulation, AMBER, 21cmFAST, and an older approximate scheme called ``The Reionization on Large Scales'' \citep[RLS,][]{Battaglia2013}. Out of the three schemes, AMBER appears to perform the best, with 21cmFAST missing some small scale features but otherwise offering a similarly good approximation on large scales.

\subsection{DMO+SAM}

In this category we place techniques that rely on a dark-matter-only (DMO) simulation combined with a semi-analytical model (SAM). Such an approach is widely used in modeling galaxy formation, and our focus here is on the small subset of such schemes that are specifically designed for modeling reionization  and, therefore, include an algorithm for  computing the spatial distribution of ionized hydrogen. In the two models discussed below this is done with an excursion set-based method.

\subsubsection{ASTRAEUS}
 
\begin{figure}
\includegraphics[width=\hsize,trim=0 0 0 0.86in,clip]{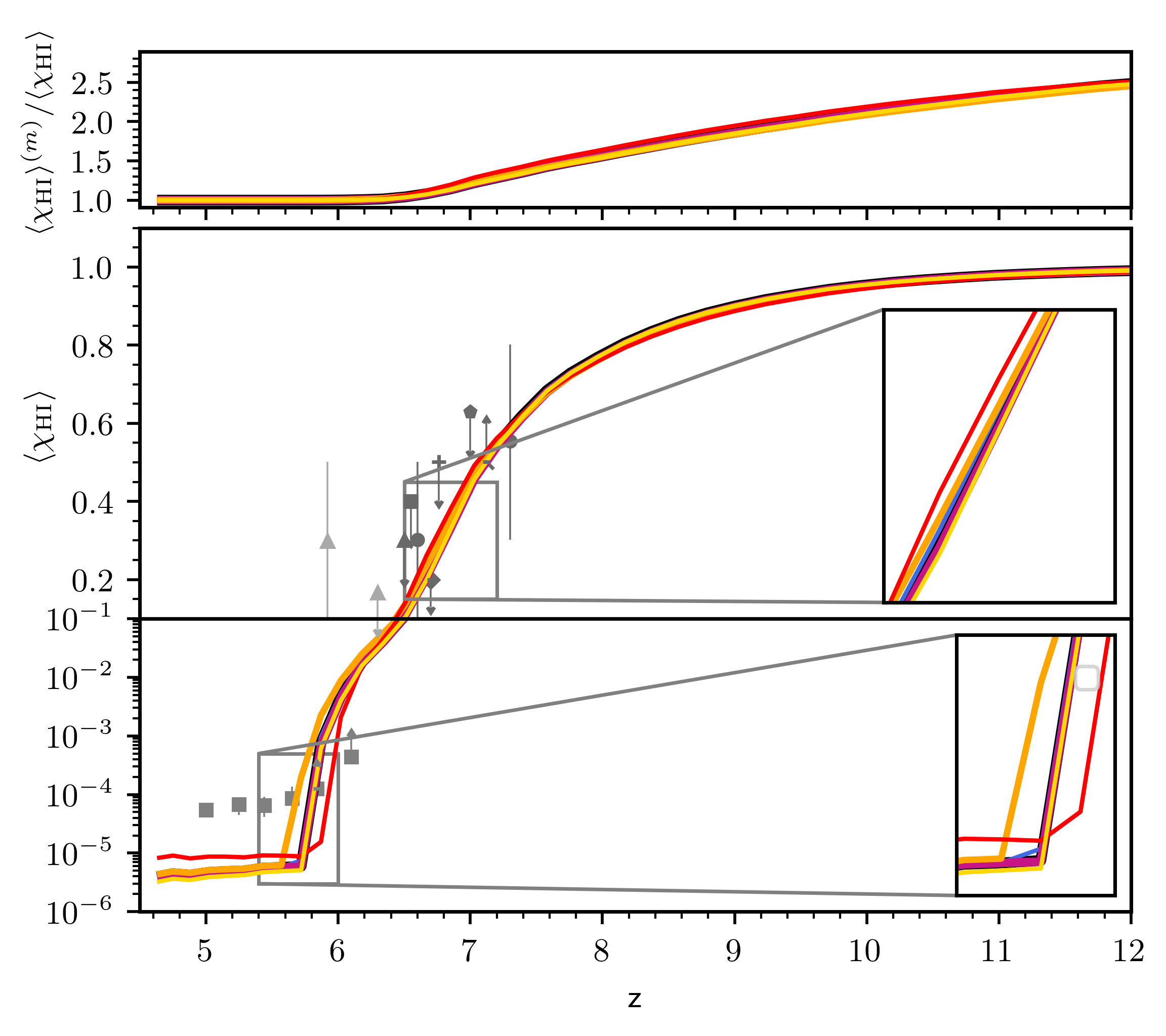}
\caption{Figure 4 from \citet{Hutter2021} showing the evolution of the volume-averaged neutral hydrogen fraction from several ASTRAEUS models and their comparison with observational constraints (grey points) -- see \citet{Hutter2021} for details on the models and the data. The good agreement between the different models illustrates the robustness of the method, but the discrepancy with the observations at $z\lesssim5.8$ suggests that ASTRAEUS does not correctly capture the evolution of the mean free path at these epochs. (Used by permission from P.\ Dayal.)}
\label{fig:aest}
\end{figure}

ASTRAEUS \citep[``seminumerical rAdiative tranSfer coupling of galaxy formaTion and Reionization in N-body dArk mattEr simUlationS'',][]{Hutter2021} is a semi-analytical model \citep[``DELPHI'',][]{Dayal2014} built on 
the ``Very Small MultiDark Planck''
 DMO simulation from the MultiDark simulation suite \citep{MultiDark}. The key parameters of the simulation are listed in Table \ref{tab:sims}. The ionization state of the IGM is modeled by an independent excursion set code developed by \citet{Hutter2018}. With a volume size of $\approx 230$ Mpc, ASTRAEUS simulations are large enough to produce a converged distribution of ionized bubbles during reionization \citep{Iliev2014}.

Figure~\ref{fig:aest} shows an example from ASTRAEUS modeling. The global reionization history in ASTRAEUS depends only weakly on the specifics of the adopted radiative feedback model, in agreement with many other studies. On a negative side, ASTRAEUS fails to model the ionization state of the IGM in the post-reionization era at $z<6$, and thus cannot yet be used for making synthetic quasar spectra for studying transmission spikes, dark gaps, and the 
\Lya\ forest. On a positive side, the large volumes simulated by ASTRAEUS allow one to explore  sample variance in current and future observational surveys. For example, \citet{Ucci2021} found large variations in the galaxy luminosity functions at $M_{\rm UV}<-17$ that will soon be measured by the {\it JWST} Advanced Extragalactic Survey (JADES).

\subsubsection{DRAGONS}

\begin{figure}
\includegraphics[width=\hsize]{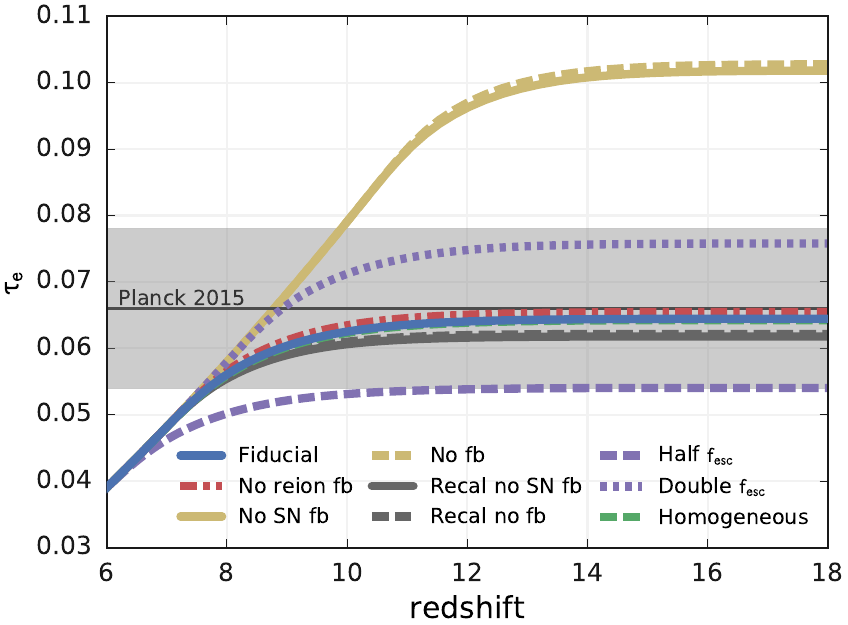}
\caption{Figure 4 from \citet{dragons3} showing the electron scattering optical depth to reionization, $\tau_e$, as a function of redshift for several DRAGONS models - see \citet{dragons3} for details on the simulations and observational constraints. (Used by permission from S.\ Mutch.)}
\vspace{-0.2cm}
\label{fig:drag}
\end{figure}

DRAGONS \citep[``Dark-ages Reionization And Galaxy Observables from Numerical Simulations'',][]{dragons1,dragons2,dragons3} adopt a 
similar scheme to ASTRAEUS. It uses the DMO simulation
``Tiamat'' that was specifically designed for the DRAGONS project. While 12 times smaller in volume and about 6 times smaller in   particle number than the ``Very Small MultiDark Planck'' simulation used in ASTRAEUS, Tiamat has more than 10 times better spatial resolution. 

The power of semi-numerical schemes like ASTRAEUS and DRAGONS is in their flexibility. After the backbone DMO simulation is completed, running the semi-analytical prescriptions on the halo catalogs is straightforward and fast, and can therefore be used for an efficient exploration of a large parameter space. As an example, we show in Fig.~\ref{fig:drag} several DRAGONS models with different stellar feedback recipes (not to be confused with the radiative feedback models shown in Fig.~\ref{fig:aest} for ASTRAEUS). Models without the stellar feedback predict too early reionization,
but, of course, the feedback is well established to be the key in modeling galaxy formation, so models without feedback are of only academic interest. With strong enough feedback, star formation in galaxies ``self-regulates'' \citep{selfreg1,selfreg2,selfreg3,selfreg4,selfreg5,selfreg6,selfreg7,selfreg8,selfreg9,selfreg10,selfreg11,selfreg12,selfreg13,selfreg14,selfreg15,selfreg16}, and the main parameter controlling the timing of reionization is the escape fraction.

\subsubsection{1D radiative transfer}

An interesting simplified approach to implementing radiative transfer on top of a DMO simulation was pioneered by \citet{Thomas2008} and later developed into the GRIZZLY (not an abbreviation) code \citep{Ghara2015,Ghara2018}. Instead of utilizing the full ray-tracing, only one ray is used for each source, with the density profile along the ray being the spherical average over all directions. The radiative transfer is solved along that ray in ``1D", so all ionized regions around sources are spherical. This approach has been shown to be a useful approximation for modeling the 21 cm emission during reionization \citep{Ghara2018}. It should be clear from our discussion in Section 5, however, that such approach fails to capture the ``outside-in" stage of reionization.

\subsection{Partially coupled simulations}

Cosmological simulations vary widely in their setups, input physics, and numerical approaches. In this review we deliberately limit ourselves to simulations that model reionization, i.e.\ include cosmological radiative transfer as one of their main physics packages.
A number of recent computational projects have focused on modeling high redshift galaxy formation without accounting for the thermal and ionization history of their environment, such as FLARES \citep{flares1,flares2,flares3}, FIRE \citep{Ma2020,Liang2021}, or FirstLight \citep{firstlight}. Because they do not address the reionization process itself, they are somewhat outside the scope of the present review. An excellent recent review of galaxy formation simulations is given by \citet{Vogelsberger2020}.

An intermediate stage between a ``semi-numerical'' technique like DMO+SAM and a fully coupled cosmological hydrodynamic simulation is one where 
not all the physics components of a simulation are  actually fully coupled. The classical example of such an approach are C2-Ray simulations \citep{c2ray1,c2ray2}, which include full radiative transfer but assume 
that the hydrogen density tracks the dark matter field of a large DMO run. In this method, the gas dynamics does not respond to heating by UV radiation, and physical effects such as the suppression of gas accretion and condensation in sufficiently low-mass halos cannot be captured directly, although they may be modeled with additional, approximate schemes. These simulations also do not resolve the scales that are relevant for star formation, and model radiation sources with a semi-analytic approach.

A different example of partially coupled simulations can be found in \citet{illust-rei}, where two different radiative transfer solvers are run in post-processing on outputs from the Illustris simulation \citep{illust}. Illustris aims at studying the processes of galaxy formation and evolution in the Universe with a comprehensive physical model that
includes feedback from supernova explosions and
accreting supermassive black holes, and radiative transfer in post-processing complements the original simulation with maps of ionized/neutral IGM gas throughout the reionization process. The limitation of this approach is that the gas dynamics does not respond to the spatially-varying and time-dependent radiation field. An analogous approach using the MassiveBlack-II simulation \citep{MBII} and the radiative transfer code CRASH was presented by \citet{eide18,Eide2020,Ma2022}. Here, the key advantage is the adopted frequency range, which is wider than modeled by most other approaches and  extend to soft X-rays and the associated secondary ionizations. The CRASH+MassiveBlack-II set of simulation also systematically explored the impacts of multiple types of radiation sources -- in addition to normal stars and quasars, they also considered binary stars, shock-heated ISM, and X-ray bursts. The radiative transfer solver has also been used in  post-processing for the modeling, for example, of the escape fraction from individual halos in cosmological simulations
\citep[c.f.][for recent studies]{Ma2020,Kostyuk2022}. As we mentioned at the beginning of this subsection, such models are beyond the scope of this review.

Partially coupled simulations with a  primary focus on galaxy formation at high redshift and not on the process of reionization per se may also include an approximate radiative transfer algorithm rather than an actual numerical solver. The poster child examples of such schemes are BlueTides \citep{bluetides} and Astrid \citep{astrid}. These
simulations combine the physics package of the MassiveBlack-II run \citep{massiveblack} with the ``The Reionization on Large Scales'' approach from \citet{Battaglia2013} that served as a precursor to the AMBER scheme already described. 
They are able to account, albeit approximately, for the ionization history of the galaxy environment, while offering a huge boost to computational efficiency.

Yet another flavor of a partially decoupled scheme can be envisioned as an efficient approach for a multi-fold extension of the simulation size. Most of the methods described here as well as the fully coupled simulations discussed below resolve the low density IGM on scales of order 100 comoving kpc or better. Such resolution, comparable to the Doppler smoothing scale of gas at $10^4\dim{K}$, is the minimum required to capture the IGM density fluctuations that give origin to the \Lya\ forest in the spectra of distant quasars.
A uniform grid simulation with, say, $50\,\dim{kpc}$ resolution in a $500\,\dim{Mpc}$ box would require $10{,}000^3$ cells, a size that is currently achievable on modern 100 petaflops-level platforms 
with, e.g., the GPU-native cosmological hydrodynamics code Cholla \citep{cholla}. 
At $50\dim{kpc}$ resolution, however, no plausible model for the physics of galaxy formation can be constructed. This is similar, for example, to  C2-Ray simulations that have to rely on  semi-analytical schemes for including radiation sources.
\begin{figure}
\includegraphics[width=\hsize,trim=0 0 0 0.68in, clip]{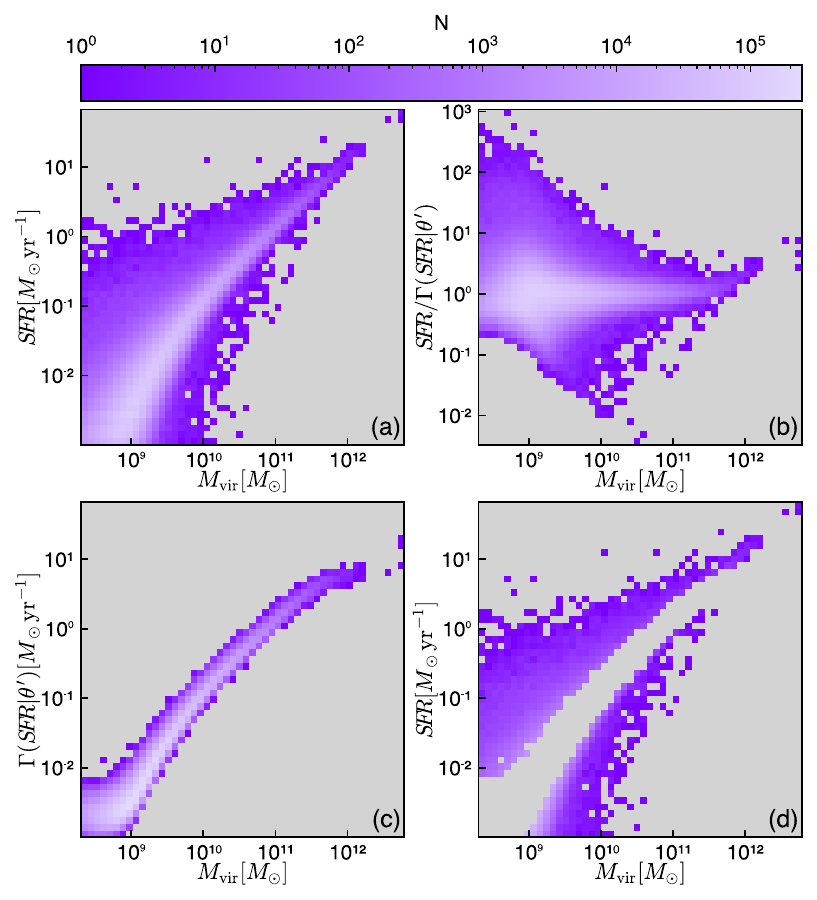}
\caption{The EBM model for the star formation rate (SFR) as a function of virial mass $M_{\rm vir}$. The upper left panel shows the two-dimensional distribution of SFRs with $M_{\rm vir}$ for all galaxies in a subset of CROC simulations, with the color scale showing the number of simulated galaxies at each [SFR, $M_{\rm vir}$] location. The lower left panel shows the EBM model prediction $\Gamma({\rm SFR};\theta^\prime)$ for the distribution of SFR with $M_{\rm vir}$ given the EBM internal parameters $\theta^\prime$. The upper and lower right panels show the residuals and the outliers between the simulated CROC galaxies and the EBM model predictions. These outliers represent less than 5\% of the simulated CROC galaxies.
(Used by permission from Ryan Hausen.)}
\label{fig:ebm}
\end{figure}
A ``two-tier'' simulation approach may be desirable, where a high-resolution small-box simulation is used to inform a coarse-resolution large-box one. Large volume simulations -- which cannot track the interior structure of dark matter halos -- 
may then implement the physics of galaxy formation with a approximate model that recovers the mean trends for galaxy baryonic properties predicted by more detailed calculations. A recent example of  this technique is offered by \citet{Hausen2022}, who trained the Explainable Boosting Machines (EBM) machine learning algorithm to assign stellar masses and star formation rates (SFRs) to the host dark matter halos based on nearly 6 million galaxies simulated by the fully coupled ``Cosmic Reionization On Computers'' (CROC) project. Figure \ref{fig:ebm} shows the main result from that work: a comparison of the simulated SFRs in the original CROC simulations versus the values predicted by machine learning as a function of halo virial mass
$M_{\rm vir}$. The EBM model is highly predictive, failing to capture only the most extreme outliers in SFR at a fixed $M_{\rm vir}$.
Through this approach, the physics of baryonic galaxy formation can be connected to the properties
of dark matter halos and implemented as a ``sub-grid'' prescription in  cosmological hydrodynamics simulations that do not resolve the small scale details of star formation and feedback, while at the same time capturing the variations of SFR in halos of the same mass due to environmental effects and different prior accretion histories.

\begin{table*}[ht]
\setlength\extrarowheight{5pt}
\caption{Simulation-based models and simulations of reionization.}
\begin{minipage}{\hsize}
\begin{tabular}{l|c|c|c|c|l}
     Suite Name &
     N$_{\rm{box}}$\footnote{We list only the number of the largest box runs. Several projects also completed a number of simulations with 8 times smaller N$_{\rm{part}}$.} &
     N$_{\rm{part}}$\footnote{Equivalent number of dark matter particles in the largest simulation.} &
     Box size \footnote{In comoving units.} & Resolution\footnote{Spatial resolution is given in proper units as an effective grid code cell size; for particle codes the conversion between the gravitational softening and the effective cell size is given in \citet{Mansfield2021}. For simulations that maintained their resolution in comoving units, the resolution is quoted at $z=6$.} & Code
     \footnote{AMR=Adaptive Mesh Refinement; SPH=Smooth Particle Hydrodynamics; MM=Moving Mesh; RT=Radiative Transfer.}
     \\
     \hline
     \multicolumn{4}{c}{SAM+DMO} & & \\
     ASTRAEUS  & 1 & $3840^3$ & 230 Mpc& 530 pc& ART (DMO) \\
     DRAGONS   & 1 & $2160^3$ & 100 Mpc& 50 pc& GADGET-2 (DMO) \\
     \multicolumn{4}{c}{Partially Coupled Simulations} & & \\
     Astrid    & 1 & $5500^3$ & 370 Mpc& 115 pc& GADGET-3 (SPH) + semi-analytical RT \\
     MassiveBlack-II & 1 & $1792^3$ & 142 cMpc& 440 pc& GADGET-3 (SPH) + RT in post-processing \\
     C2-Ray  &   1 & $6912^3$ & 714 Mpc& 6.3 kpc& P$^3$M (DMO) + C2-Ray (RT) \\
     Illustris & 1 & $1820^3$ & 107 Mpc& 72   pc& AREPO (MM) + RT in post-processing \\
     \multicolumn{4}{c}{Fully Coupled Simulations} & & \\
     CoDa      & 1  & $8192^3$ & 94~Mpc & 1.7 kpc & RAMSES-CUDATON (uniform grid) \\
     CROC      & 6  & $2048^3$ & 117~Mpc & 100~pc & ART (AMR) \\
     SPHINX    & 1  & $1024^3$ & 20~Mpc & 10 pc & RAMSES-RT (AMR) \\
     Thesan    & 1  & $2100^3$ & 96~Mpc & 300 pc & AREPO-RT (MM)\\
\end{tabular}
\end{minipage}
\label{tab:sims}
\end{table*}

\subsection{Fully coupled simulations}

Self-consistent, fully coupled simulations, often considered the ultimate theoretical model for a given process, can only be trusted as long as they include all the relevant physics (such as gravity, gasdynamics, star formation, stellar and AGN feedback, radiation transport) with sufficient precision and maintain numerical effects under control. 

Table~\ref{tab:sims} summarizes some of the most recent fully coupled simulations (together with several of the approximate methods discussed above).
CosmicDawn (CoDa) \citep{ocvirk16,coda1,coda2,coda3,Lewis2022}, CROC, and Thesan are similar in their computational volume ($\sim100\,\dim{Mpc}$). They differ somewhat in spatial and mass resolution, but all fall into the class of simulations that do not resolve the scale heights of galactic disks (and therefore model galaxies in ``2D''), and can be directly compared to each other. SPHINX simulations 
\citep{sphinx1,sphinx2,sphinx3}, by contrast, focus on resolving the actual vertical structure of star-forming galaxies, achieving much higher spatial resolution at the expense of being unable to model the global reionization history (a 20 comoving Mpc box contains only 4 $L_\star$ galaxies on average). 

\begin{figure*}
\centering
\includegraphics[width=0.5\hsize]{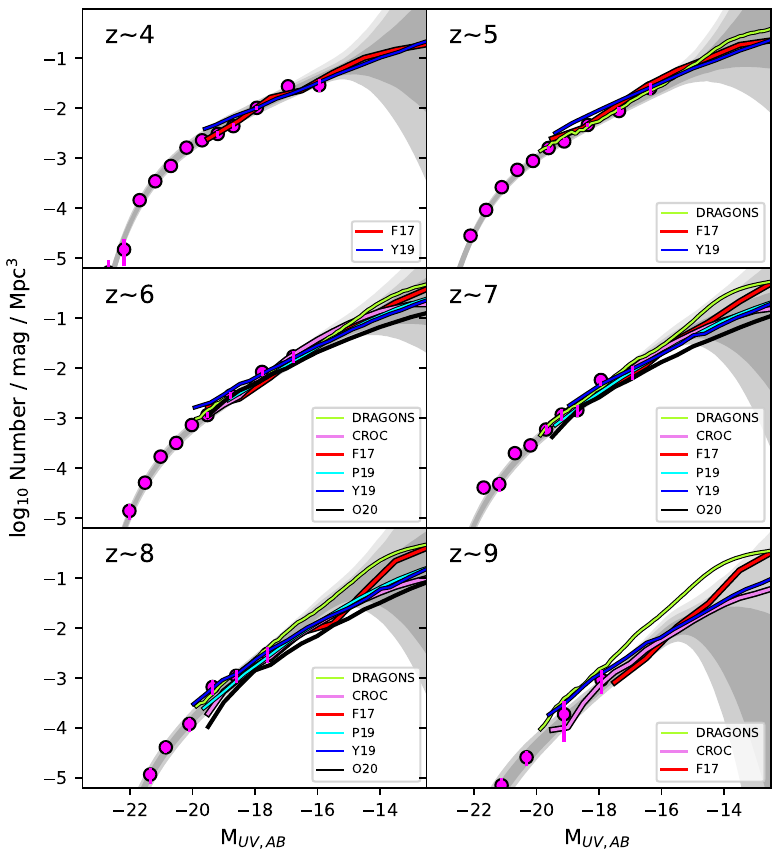}%
\includegraphics[width=0.5\hsize]{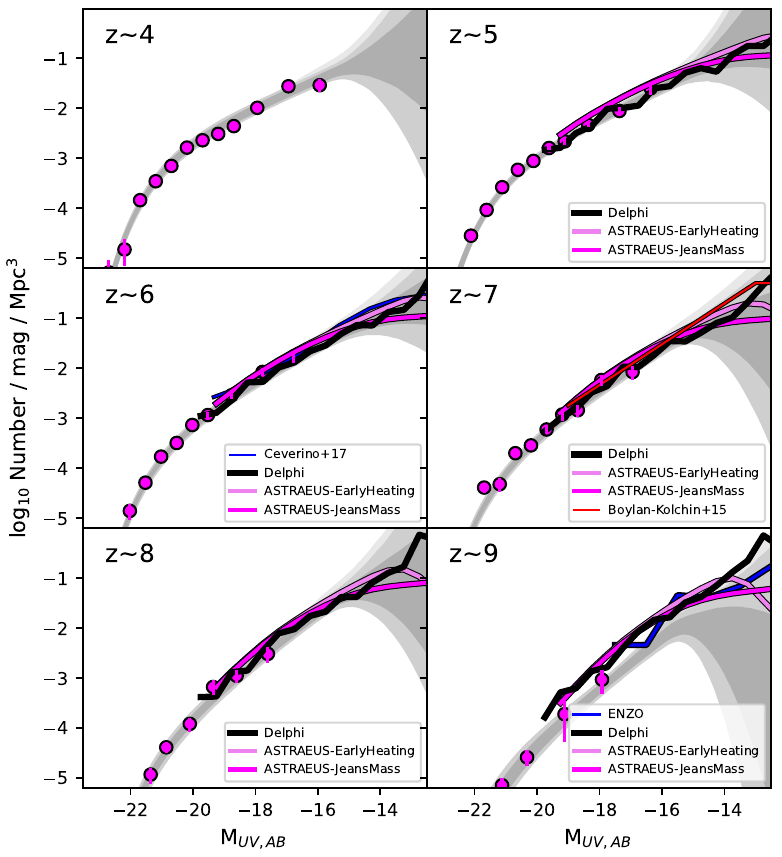}
\vspace{+0.0cm}
\caption{\footnotesize Comparison 
\citep{Bouwens2022} of the observed galaxy UV luminosity functions (gray bands) with model predictions from DRAGONS, CROC, ``Technicolor Dawn'' \citep[F17,][]{finlator18}, 21cmFast \citep[P19,][]{Park2019}, a semi-analytical model \citep[Y19,][]{Yung2019}, ``Cosmic Dawn'' \citep[O20,][]{coda2}, 
ASTRAEUS, Delphi (an earlier version of ASTRAEUS), and FirstLight \citep{firstlight}. (Used by permission from R.\ Bouwens.)}
\label{fig:lfs}
\end{figure*}

The target goal of reaching a $\sim100\,$Mpc simulation box is dictated by the desire to replicate a representative region of the Universe. At $z\sim7$, the correlation length of galaxies is around $10\,$Mpc \citep{Barone-Nugent2014} and is weakly luminosity dependent. The number density of $L_\star$ galaxies is also about 1 per $10\,$Mpc. Hence, a volume of $\sim 100\,$Mpc on a side contains around 1000 $L_\star$ galaxies and has the rms correlation function at half the box size of $(50/10)^{-1.6} \approx 0.08$. Whether $\sim100\,\dim{Mpc}$ simulations actually converge on some of the key features of the reionization process, such as the size distribution of ionized bubbles, is presently unclear. Some earlier C2-Ray simulations \citep{Iliev2014} found convergence only in $\gtrsim 250\,$Mpc boxes, while in CROC simulations (which explicitly match at $z<6$ the LyC mean free path determined by the abundance of LLSs) the bubble size distribution appears to have converged by $z\gtrsim7$ \citep{Gnedin2014b}.

Figure \ref{fig:lfs} from \citet{Bouwens2022} shows a comparison between the galaxy UV luminosity function observed at different redshifts  and several theoretical predictions. The observational data come primarily from the Hubble Frontier Fields, and are subject to systematic uncertainties from lensing modeling \citep{Bouwens2017,Bouwens2022} that rapidly increase for galaxies fainter than $M_{\rm UV}>-15$. At brighter magnitudes most theoretical models match the data well,\footnote{Notice, however, that CROC results are shown for earlier, smaller boxes. CROC underpredicts luminosities and stellar masses of super-$L_*$ galaxies in its largest, 117 Mpc boxes \citep{Zhu2020}.}\, but they differ
widely at $M_{\rm UV}\gtrsim-14$ mag, 
a faint luminosity regime that will soon be probed by {\it  JWST} observations. Theorists are eagerly waiting for {\it JWST} deep field data as few, if any, of the current models will survive these ground-breaking sensitive measurements.

\begin{figure*}
\centering
\includegraphics[width=0.57\hsize]{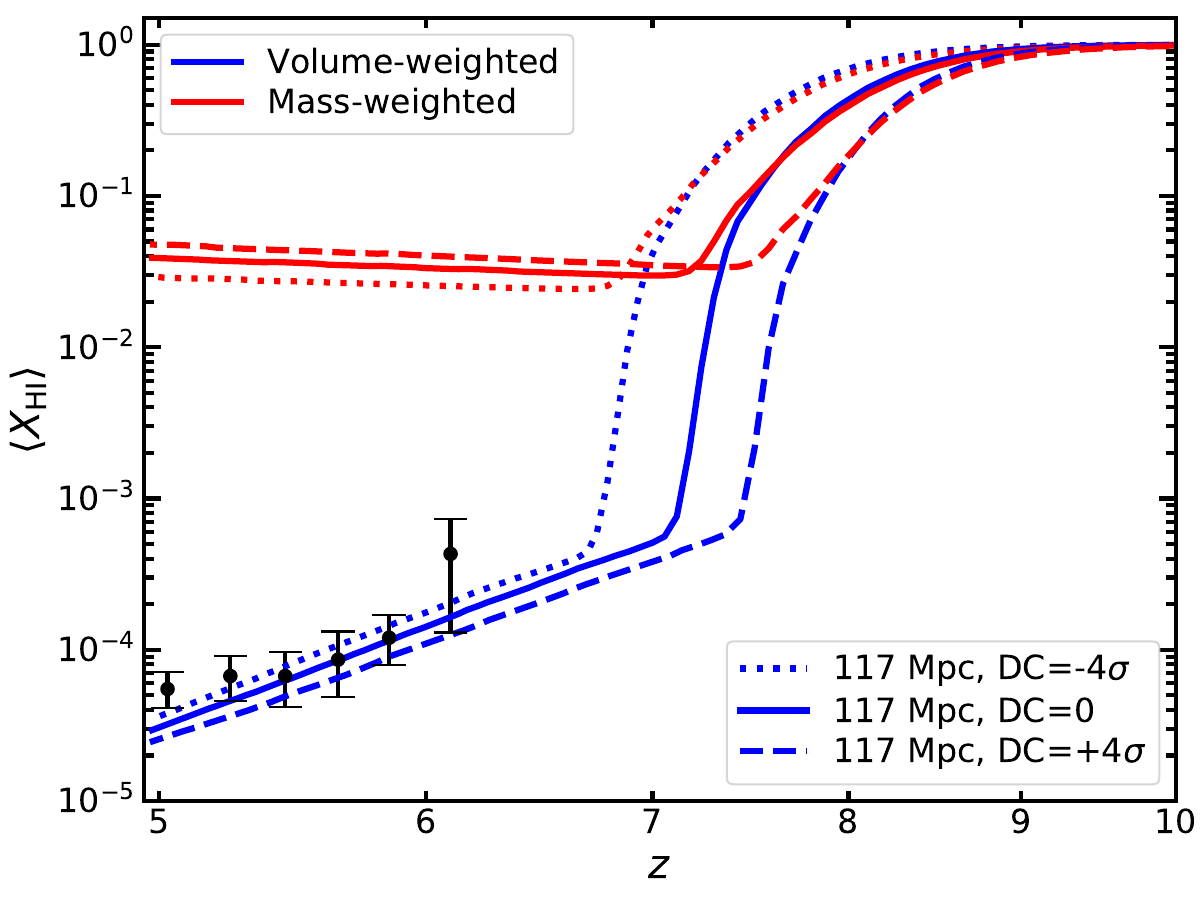}%
\includegraphics[width=0.43\hsize]{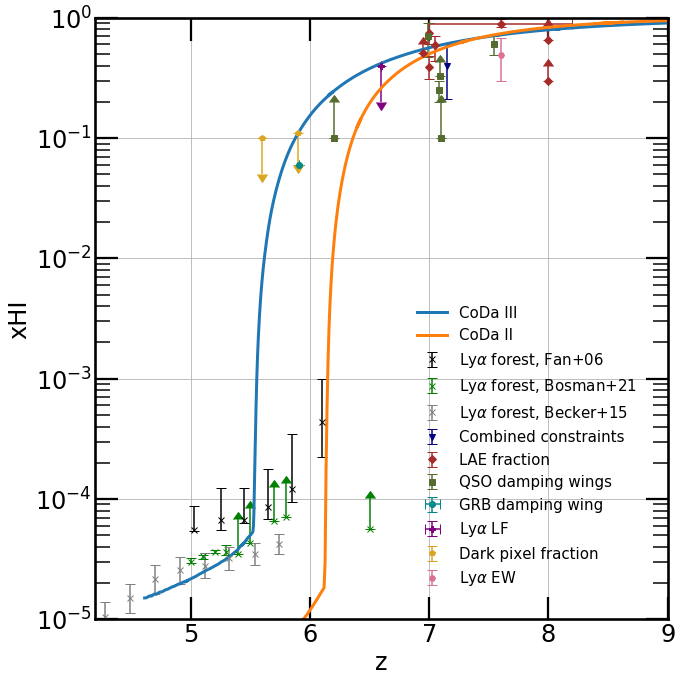}
\includegraphics[width=0.62\hsize,trim=0 0 50 0,clip]{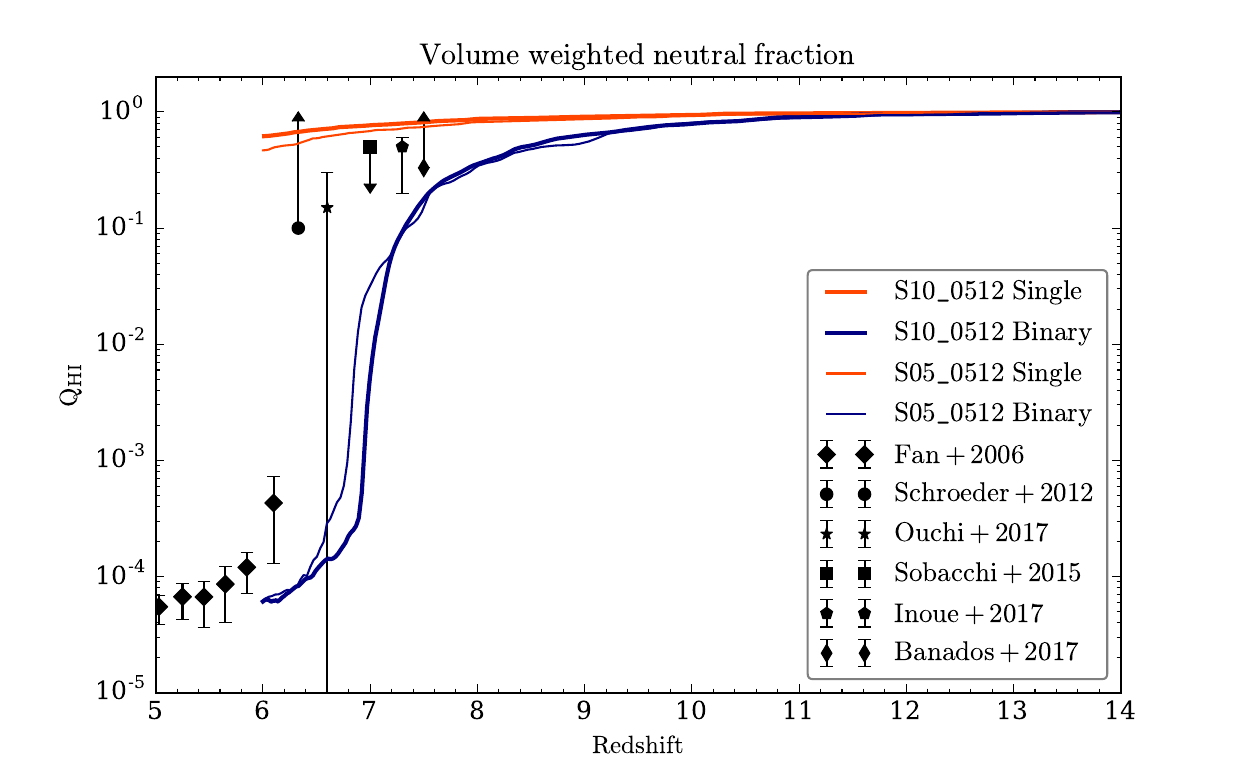}%
\includegraphics[width=0.38\hsize]{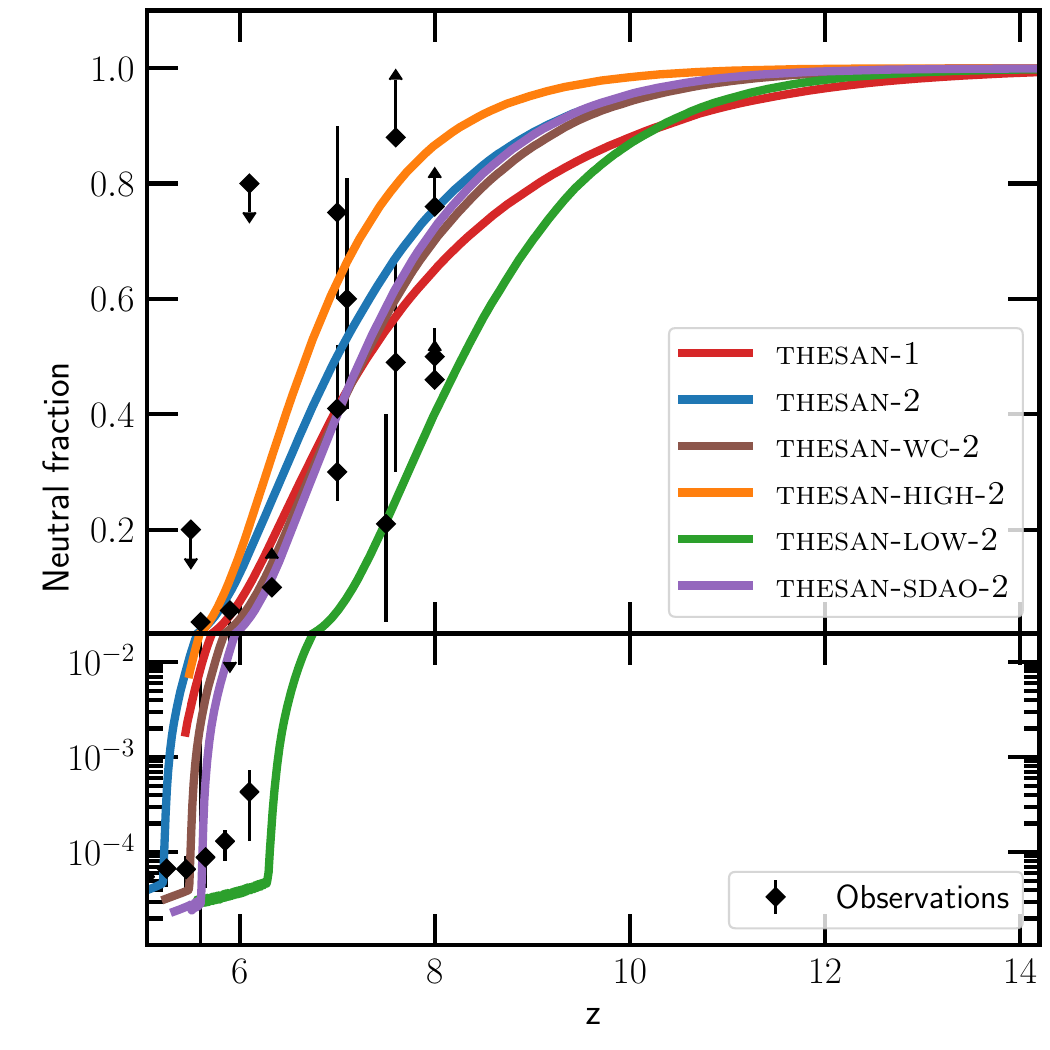}
\caption{\footnotesize Evolution of the neutral hydrogen fraction in four recent fully coupled simulations of the reionization process: CROC (upper left), CoDa \citep[upper right,][]{Lewis2022}, SPHINX \citep[lower right,][]{sphinx1}, and Thesan \citep[lower right,][]{thesan1}. Notice that 6 data points from \citet{fan06} shown in the CROC panel are also shown in the other 3 panels with different symbols. (Used by permission from J.\ Lewis, J.\ Rosdahl, R.\ Kannan.)}
\label{fig:xhz}
\end{figure*}

Since the goal of fully coupled simulations is to actually model the reionization process, their performance must be measured against that metric. In Figure \ref{fig:xhz} we show the most basic prediction for any simulation of the epoch of reionization, the evolution of the globally volume-averaged neutral hydrogen fraction.\footnote{Predicting the mass-weighted neutral hydrogen fraction is much harder, as after reionization this is dominated by the residual neutral component locked in the Damped Lyman-$\alpha$ Systems.} While the latest fully coupled simulations do produce neutral fractions similar to the observed values, they do not do so at the level of precision required 
by the observations. The only simulation project that does match the observations accurately is CROC, but this is \emph{by construction}, as the data measurements were actually used to calibrate the effective escape fraction from the radiation sources. The unsatisfactory level of agreement 
between data and theory points to the direction where efforts in designing the next generation of cosmological simulations should be aimed at -- i.e. on significantly improving the modeling of intergalactic gas. A number of recent observations have provided measurements of various properties of the post-reionization IGM, such as the distribution of mean \Lya\ opacities along skewers of fixed length \citep{becker15,bosman18,Eilers2018,Yang2020}, the distribution of ``dark gaps'' --  continuous spectral regions in distant quasar spectra where the trasmitted flux is below a specified threshold \citep{Zhu2020,Zhu2021}, and the cross-correlation between quasar absorption spectra and properties of galaxies along the same sightline \citep{Meyer2019,Meyer2020}. None of the existing fully coupled simulations are able to match all of these observational constraints.


\section{Concluding Remarks}

The epoch of cosmic reionization is a vast science frontier discovery area where detailed modeling and complex computational tools are required to support the anticipated observational breakthroughs. Since the very first paper on reionization by \citet{arons70}, this field has remained heavily theory dominated until the very recent times. The prospects for improving our understanding of this era over the next few years are outstanding, and a number of massive datasets from new ground- and space-based instruments and facilities, like the Dark Energy Spectroscopic Instrument ({\it DESI}), the Subaru Hyper Suprime-Cam ({\it HSC}), the Extremely Large Telescope ({\it ELT}), the Atacama Large Millimeter Array ({\it ALMA}), the Hydrogen Epoch of Reionization Array ({\it HERA}), the Square Kilometre Array ({\it SKA}), {\it JWST} and {\it Euclid}, are poised to revolutionize our understanding of the dawn of galaxies, the cosmic ionizing photon budget, the physics of the early IGM and galaxy-IGM interactions, and promise an era of precision reionization studies. 

It is generally expected that, by performing throughout analyses that comprise the latest observational results, elaborate theoretical modeling, state-of-the-art simulations that incorporate realistic baryonic physics over a large range of spatial scales, and statistical inference, astrophysicists will be able over the next decade to: 1) unravel the nature of the first astrophysical sources of radiation and heating, and investigate how they interacted with their environment; 2) trace the overall progression of the cosmic reionization process and establish its impact on observations of the early Universe -- from the large-scale signal of neutral hydrogen in the redshifted 21cm line at $z\gtrsim 10$ to the fluctuating \Lya\ opacity of the $z\sim 5.5$ IGM; 3) reveal the thermal history of cosmic gas through the imprint left by hydrogen and helium reionization heating on the power spectrum of the \Lya\ forest and other observables; and 4) model the range of cosmological physics that influence the small-scale properties of the cosmic web.

The precision of future observational data will 
exceed the precision of current theoretical models 
(this is already happening, as can be seen in Fig.\ \ref{fig:xhz}). Theorists must therefore face the challenge of significantly increasing the physical fidelity and numerical precision of their models, and enhance their predictive power. The world of theoretical modeling of reionization is very diverse, as we strove to demonstrate above. Such a diversity is useful during the formative years of any field, but as the field matures one expects many of the simpler, less accurate methods and tools to become obsolete. This is very likely going to be the case in reionization studies as well. It is our belief then that this ``living" review  will 
lead a very active life indeed, with many of the schemes described above falling out of use. Hopefully, this article will be of help to the next generation of theorists and computational scientists, charting the directions where a beginning graduate student or a postdoc switching fields may make a large and visible impact.  


\paragraph{Acknowledgments}

We thank the anonymous referees for catching a number of critical omissions and mistakes in the first version of this review. We also greatly benefited from comments by Benedetta Ciardi, Linhua Jiang, Dylan Nelson, Michael Shull, Hy Trac, and Yidong Xu.
This work was supported in part by the NASA Theoretical
and Computational Astrophysics Network (TCAN) grant
80NSSC21K0271. The authors would like to express their 
appreciation to the TCAN team for many useful inputs, 
valuable discussions, and inspiring ideas. Fermilab is operated by Fermi Research Alliance, LLC, under Contract No. DE-AC02-07CH11359 with the United States Department of Energy.



\bibliography{main}   

\end{document}